\documentclass[structabstract]{aa}
\usepackage[dvips]{graphicx}
\usepackage{caption}
\usepackage{subcaption}
\usepackage{txfonts}
\usepackage{color}
\usepackage{natbib}
\bibpunct{(}{)}{;}{a}{}{,} 
\usepackage{psfrag}
\usepackage{version}

\begin{document}

\title{The evolution of amorphous hydrocarbons in the ISM: \\
        dust modelling from a new vantage point}
\authorrunning{}
\titlerunning{Dust modelling from a new vantage point}

\author{A.P. Jones$^{1,2}$, 
             L. Fanciullo$^{1,2}$, 
             M. K\"ohler$^{1,2}$, 
             L. Verstraete$^{1,2}$, 
             V. Guillet$^{1,2}$, 
             M. Bocchio$^{1,2}$,
             N. Ysard$^{1,2}$ }  
           
    \institute{CNRS, Institut d'Astrophysique Spatiale, UMR8617, Orsay F-91405, France\\[-0.3cm]
         \and  Universit\'e Paris Sud, Institut d'Astrophysique Spatiale, UMR8617, Orsay F-91405, France\\[0.1cm]
    \email{Anthony.Jones@ias.u-psud.fr} }

    \date{Received 12 April 2013 / Accepted 15 July 2013}

   \abstract
{The evolution of amorphous hydrocarbon materials, a-C(:H), principally resulting from ultraviolet (UV) photon absorption-induced processing, are likely at the heart of the variations in the observed properties of dust in the interstellar medium.}
{The consequences of the size-dependent and compositional variations in a-C(:H), from aliphatic-rich a-C:H to aromatic-rich a-C, are studied within the context of the interstellar dust extinction and emission.}
{Newly-derived 
optical property data for a-C(:H) materials, combined with that for an amorphous forsterite-type silicate with iron nano-particle inclusions, a-Sil$_{\rm Fe}$, are used to explore dust evolution in the interstellar medium.}
{We present a new dust model that consists of  a power-law distribution of small a-C grains and log-normal distributions of large a-Sil$_{\rm Fe}$ and a-C(:H) grains. The model, which is firmly anchored by laboratory-data, is shown to quite naturally explain the variations in the infrared (IR) to far-ultraviolet (FUV) extinction, the 217\,nm UV bump, the IR absorption and emission bands  and the IR-mm dust emission.}
{The major strengths of the new model are its inherent simplicity and built-in capacity to follow dust evolution in interstellar media. We show that mantle accretion in molecular clouds and UV photo-processing in photo-dominated regions are likely the major drivers of dust evolution.} 


\maketitle

\section{Introduction}

The evolution of interstellar dust  is a key but complex issue that is now receiving some well-focussed attention. In particular, hydrocarbon solids ({\it i.e.}, a-C:H and a-C, also known as HAC) present a particular challenge because of their inherent complexity and also because they appear to be rather vulnerable to destruction \cite[{\it e.g.},][]{2008A&A...492..127S,2011A&A...530A..44J}. Further, they appear to undergo rather complex, size-dependent evolution arising, principally, from ultraviolet (UV) photon absorption leading to photo- or thermal-processing \cite[{\it e.g.},][]{2009ASPC..414..473J,2012aA&A...540A...1J,2012bA&A...540A...2J,2012cA&A...542A..98J}  and incident ion and electron collisions in shock waves and in a hot gas \cite[{\it e.g.},][]{2010A&A...510A..36M,2010A&A...510A..37M,2012A&A...545A.124B}. The evolution of their properties have received quite some interest as a model for the solid carbonaceous matter in the interstellar medium \citep[ISM, {\it e.g.},][]{1990QJRAS..31..567J,1995ApJ...445..240D,1997ApJ...482..866D,2004A&A...423..549D,2004A&A...423L..33D,2005A&A...432..895D,2008A&A...490..665P,2008A&A...492..127S,2009ASPC..414..473J,2010A&A...519A..39G,2011A&A...529A.146G,2011A&A...525A.103C}, in circumstellar media \citep[{\it e.g.},][]{2003ApJ...589..419G,2007ApJ...664.1144S} and in the Solar System \citep[{\it e.g.},][]{2011A&A...533A..98D}. 

By dust {\em evolution} we here mean {\em a change in the observable properties of the dust arising from processes such as accretion, coagulation, (photo-)fragmentation, erosion and grain charge effects.}  
This evolution may be constructive, as for accretion and coagulation and lead to an increase in the total dust mass and/or its mean size, or destructive, as for (photo-)fragmentation and erosion and lead to a decrease in the total dust mass and/or its mean size. 

The evolution of hydrocarbon solids was recently elucidated in detail in the series of preceding papers and their corrigenda \citep[][hereafter called papers~I, II and III]{2012aA&A...540A...1J,2012bA&A...540A...2J,2012cA&A...542A..98J,2012dA&A...545C...2J,2012eA&A...545C...3J} and this work builds upon these foundations in order to explore some of the consequences arising from the use of the optEC$_{\rm (s)}$(a) optical property data to explain the observed interstellar dust extinction and emission properties. As described in papers~I to III, amorphous hydrocarbon particles are macroscopically-structured ({\it i.e.}, a contiguous network of atoms), solid-state materials consisting of only carbon and hydrogen atoms.  For reference, a-C:H materials are H-rich ($\sim 15-60$ at. \% H), aliphatic-rich and wide band gap ($> 1$\,eV), whereas a-C materials are H-poor ($\lesssim 15$ at. \% H), aromatic-rich and narrow band gap ($\sim -0.1$ to $1$\,eV). The designation a-C(:H) is used here to cover the whole family of H-rich a-C:H to H-poor a-C carbonaceous solids, whose properties have been well-studied within both the physics and astrophysics communities \citep[{\it e.g.},][]{1979PhRvL..42.1151P,1980JNS...42...87D,1983JNCS...57..355T,1986AdPhy..35..317R,1987PhRvB..35.2946R,1988JVST....6.1778A,1988Sci...241..913A,1988PMagL..57..143R,1990JAP....67.1007T,1991PSSC..21..199R,1995ApJS..100..149M,1996ApJ...464L.191M,2000PhRvB..6114095F,2001PSSAR.186.1521R,2002MatSciEng..37..129R,2003ApJ...587..727M,2004PhilTransRSocLondA..362.2477F,2007DiamondaRM...16.1813K,2007Carbon.45.1542L,2008ApJ...682L.101M,2011A&A...528A..56G}.

This paper uses a slightly-modified version of the laboratory-constrained optEC$_{\rm (s)}$(a) data presented in paper III, which naturally explain many of the observed properties attributed to carbonaceous dust in the ISM. 
Here we insert these data into the DustEM dust extinction and emission calculator \citep{2011A&A...525A.103C} in order to quantitatively examine and explore their viability as a tool in explaining the range of interstellar hydrocarbon dust extinction and emission observables and their associated (non-)correlations. 

This work is a departure from the `traditional' interstellar dust modelling methodology because it adopts a `holistic' or `global' approach, in which we are able to coherently vary the dust properties over extreme-ultraviolet (EUV) to cm wavelengths. 
The result is a dust model that self-consistently explains almost all dust observables.
This is important because it paves the way for an investigation of the interdependencies between the dust observables and, hence, their variations, correlations and non- or anti-correlations.  

This paper is organised as follows:
Sections \ref{sect_proc_timescales} and \ref{sect_reform_acc} summarise a-C(:H) dust photo-processing and its re-formation and re-accretion in the ISM, 
Section \ref{sect_constraints} briefly summarises the constraints imposed on dust models by experimental data and diffuse ISM dust observations, 
Section \ref{sect_dust_model} presents our new evolutionary dust model, 
Section \ref{sect_dust_ext_sed} discusses the astrophysical implications of the new model, 
Section \ref{sect_dust_variations} gives a schematic view of the a-C(:H) dust life-cycle and explores its evolution and variation in the ISM and  
Section \ref{sect_conclusions} concludes this work.

\section{Carbon dust photo-processing}
\label{sect_proc_timescales}

Carbonaceous dust will be processed (evolve) as it traverses the ISM from its sites of formation to its  demise in energetic regions and its eventual re-birth in some dense phase of the ISM \citep[{\it e.g.},][]{2011A&A...530A..44J}. For small hydrocarbon grains and polycyclic aromatic hydrocarbons (PAHs), species of molecular dimensions, this processing can be catastrophic because high-energy photon absorption and electron collisions will result in highly electronically-excited particles that undergo dissociation before they can relax radiatively \citep[{\it e.g.},][]{2010A&A...510A..36M,2010A&A...510A..37M,2012A&A...545A.124B}.

Principal among the processes that affect the structure and composition of a-C(:H) dust is the photolytic and/or thermal processing arising from UV-EUV photon absorption ({\it e.g.}, papers II and III). As shown in papers II and III the relevant time-scales for a-C(:H) photo-processing will be composition-dependent ({\it i.e.},  depend on the H content, $X_{\rm H}$, which is directly proportional to the optical band gap, $E_{\rm g}$)  
and also size-dependent. The characteristic time-scale for the direct photo-processing (photo-darkening or aromatisation) of a-C(:H) grains in the diffuse ISM appears to be relatively short and of the order of a million years for a-C(:H) nano-particles (see papers II and III and their corrigenda). Thus, it is to be expected that small grains and thin mantles, of a-C(:H), ought to be maximally-aromatised in the diffuse ISM\footnote{Complete aromatisation is expected to occur to a depth of $\simeq 20$\,nm, which is where the optical depth for EUV-UV photons ($E_{h\nu} \gtrsim 10$\,eV) reaches unity for all a-C(:H) materials (see paper~II).\label{footnote_20nm}}. Large a-C(:H) grains ($a \geq 200$\,nm) will, however, remain predominantly aliphatic-rich because their cores, which  are not UV photo-processed, make up $\geq 70$\% of the grain volume. This appears to be qualitatively consistent with observations, {\it i.e.}, that the smaller carbonaceous particles are ``aromatic-rich'' while the larger carbonaceous grains are ``aliphatic-rich'' and that the carbon mantle/coating on  amorphous silicate dust must be rather aromatic \citep[{\it e.g.},][]{1999ApJ...517..883B}. 

\cite{1999ApJ...512..224A} showed that the 3.4\,$\mu$m absorption band is not polarised along the  line of sight towards the Galactic Centre source \object{Sagittarius A IRS~7}. At the time there were no spectropolarimetric observations of the 9.7\,$\mu$m silicate band along this line of sight. This led \cite{2002ApJ...577..789L} to conclude that there was, at that time, no reason to reject a silicate core-carbonaceous organic mantle interstellar dust model. The modelling work by \cite{2002ApJ...577..789L} showed that, although both the 3.4\,$\mu$m CH and 9.7\,$\mu$m silicate bands are expected to be polarised, they are unlikely to be polarised to the same degree. However, \cite{2006ApJ...651..268C} found that the 3.4\,$\mu$m band is essentially unpolarised along the same Galactic Centre lines of sight towards \object{GCS 3-II} and \object{GCS 3-IV} where the 9.7\,$\mu$m silicate band is polarised. Thus, it is clear that any carbonaceous material associated with the silicate dust cannot be aliphatic-rich because of the  lack of polarisation of the 3.4\,$\mu$m absorption band. This then implies that there must be a population of large ($a \gg 20$\,nm) carbonaceous grains in the ISM that contain a significant aliphatic-rich component that is the origin of the 3.4\,$\mu$m absorption band \citep[{\it e.g.},][]{2012aA&A...540A...1J,2012cA&A...542A..98J} and that this dust component is separate from the large, amorphous silicate grain population. 

However, the exact nature of, and time-scale for, UV photon-induced processing is rather difficult to quantify because of uncertainties in the wavelength-dependence of the photo-dissociation cross-sections, and the importance of  competing channels, such as UV photon absorption leading to heating or fluorescence ({\it e.g.}, paper~III). In this work we therefore keep an open mind on this subject, construct a viable model for interstellar carbonaceous dust, using the optEC$_{\rm (s)}$(a) data, that best fits the observations and the variations in those observations.

\section{a-C:H re-formation/accretion in the ISM}
\label{sect_reform_acc}

The work presented in paper~II indicates that enhanced large particle ($a \geq 100$\,nm) scattering at near-IR to IR wavelengths ($1-5\,\mu$m) occurs in wide band gap a-C:H materials ({\it i.e.}, aliphatic-rich a-C:H with $E_{\rm g} > 1.25 $\,eV). Carbon accretion from the gas phase in a moderately extinguished medium ($A_{\rm V} \gtrsim 1$) will tend to form wide band gap a-C:H mantles on all grains. Such mantles, accreted in cloud cores, could be an explanation for the ``cloudshine'' observed predominantly in the $H$ ($1.6\,\mu$m) and $K_s$ ($2.2\,\mu$m) bands and explained as starlight scattered by the dust in cloud interiors \citep{2006ApJ...636L.105F}. Possibly also related to a-C:H mantling is the observed ``coreshine'' effect observed in the Spitzer IRAC $3.6$ and $4.5\,\mu$m bands, which is assumed to arise from cloud core dust emission \citep{2010Sci...329.1622P,2010A&A...511A...9S}. It is possible that the observed ``coreshine'' could also be due to the scattering of starlight by a-C:H grain mantles or it could be due to luminescence from H-rich, a-C:H nano-particles recently released from carbonaceous mantles accreted within molecular clouds. The latter scenario is similar to the recently-proposed mechanism for fullerene formation around planetary nebul\ae\ via the photo-processing and vibrational excitation of ``arophatic" clusters derived from a-C:H particles formed in denser regions \citep{2012ApJ...757...41B,2012ApJ...761...35M}. 

The observed ``cloudshine'' and ``coreshine'' can therefore be re-interpreted in terms of the scattering and/or  emission by H-rich, carbonaceous grains formed by the accretion of gas phase carbon into aliphatic-rich mantles within molecular clouds. An initially-accreted a-C(:H) mantle will probably be rather H-poor ({\it i.e.}, a-C) where the ambient radiation field is only weakly attenuated because of UV photon-driven dehydrogenation. However, with increasing density and extinction the accreting mantle will be H-rich ({\it i.e.}, a-C:H).  The maximum thickness of an a-C(:H) mantle is likely to be of the order of a few nm \cite[{\it e.g.},][]{1990QJRAS..31..567J}. This carbonaceous mantle accretion-transformation scenario could quite naturally explain the observed variations in the far-infrared (FIR) to mm dust emission in the transition between high-density/low-excitation molecular clouds and low-density/high-excitation photon-dominated regions (PDRs, see Section~\ref{sect_dust_construct_accn}).

\section{Diffuse ISM dust constraints}
\label{sect_constraints}

Observations, complimented by fundamental solid-state physics, optical property modelling and experimental data, impose stringent constraints on interstellar dust analogue materials and the models that use these data. Any viable dust model must therefore be carefully constructed within the framework of these constraints. \\[0.05cm]

\noindent {\bf Observational constraints} include the: 
\begin{enumerate}
  \item $\lambda$-dependent EUV to FIR extinction,
  \item $\lambda$-dependent albedo and scattering,  
  \item uniformity of the curvature of the far-ultraviolet (FUV) extinction, 
  \item fixed UV bump position and limited variations in width,
  \item FUV, UV bump, visible/near-IR (NIR) and mid-IR (MIR)  (non-)correlations,
  \item FUV and UV bump (non-)correlations with (C/H)$_{\rm dust}$, 
  \item $\lambda$-dependent polarisation; polarisation-to-extinction ratio, 
  \item NIR-FIR absorption and emission bands,  
  \item extended red emission (ERE) and blue luminescence (BL), 
  \item full EUV-cm dust spectral energy distribution (SED), 
  \item cosmic abundance constraints (O, C, Si, Mg, Fe \ldots),
  \item known dust sources and pre-solar grain compositions, 
  \item x-ray halo and absorption data, 
  \item inferable dust (re-)formation processes in the ISM, and 
  \item physical reasonableness and survivability in the ISM.
\end{enumerate}
{\bf Experimental constraints} include the measured behaviours of: 
\begin{enumerate}
  \item material optical properties, 
  \item T-dependence of dust analogue optical properties, 
  \item quantum efficiency of likely ERE and BL carriers, and 
  \item material deposition onto a substrate (accretion).
\end{enumerate}
{\bf Modelling constraints} include the: 
\begin{enumerate}
  \item shape (irregularity) and structure (core-mantle, aggregate),
  \item dust size distribution (power law, log-normal, \ldots), 
  \item small size ($a \lesssim 10$\,nm) of FUV and UV bump carriers, and 
  \item a de-coupling of the FUV, UV bump and visible extinction.   
\end{enumerate}

As shown by \cite{1983ApJ...272..563G}, particles that produce the UV bump at 217\,nm, apparently, cannot contribute significantly to the FUV extinction and, further, the particles that are responsible for the UV bump can only make a small contribution to the extinction longwards of $\sim 170$\,nm. They also infer that the physical characteristics of the FUV extinction carriers {\it ``remain fairly stable once the grains have emerged from the molecular cloud phase of their evolution.''} \cite{1983ApJ...272..563G} also convincingly show that any dust model for which the FUV extinction is a sum of carbonaceous/graphite and silicate contributions is inconsistent with observations. The \cite{1977ApJ...217..425M} and \cite{1984ApJ...285...89D} models, and also the more recent \cite{2001ApJ...554..778L} and \cite{2011A&A...525A.103C} models, do not satisfy this constraint. 

\cite{2004ASPC..309...33F} and \cite{2007ApJ...663..320F,2009ApJ...699.1209F} have undertaken the most recent and detailed series of analyses of the extinction curve component (non-)correlations and we will use these studies as the basis for comparison of the optEC$_{\rm (s)}$(a) model data with observations. 
From these works we note that the major observed trends and variations that need to be explained by any viable dust model are: 
\begin{enumerate}
  \item The FUV extinction rise (or curvature) and the intercept of an underlying linear component at infinite wavelength, in an $E{\rm (\lambda-V)}/E{\rm (B-V)}$ {\it vs.} $1/\lambda$ plot, are very well correlated. 
  \item The 217\,nm UV bump characteristics: 
  \begin{itemize}
    \item small peak position variations: $4.5903\pm0.0085\,\mu$m$^{-1}$ ($217.85\pm0.91$\,nm).
    \item a range of UV bump widths: FWHM~$1.00\pm0.15\,\mu$m$^{-1}$.
    \item no correlation between peak position and width. 
    \item broader where the FUV curvature is greater. 
    \item strongest for intermediate levels of FUV extinction.
    \item weaker for high or low FUV extinction. 
    \item weaker for  low or high values of $R_{\rm V}$.
  \end{itemize}
  \item The FUV extinction rise and UV bump do not appear to correlate with the abundance of the mid-IR (12 and 25\,$\mu$m) emitters \citep{1994A&A...284..956B}. 
  \item The IR extinction, for $\lambda > 1\,\mu$m, exhibit a power-law-like behaviour that is a function of $R_{\rm V}$ \citep{2009ApJ...699.1209F}.
\end{enumerate}

\section{Towards a more realistic interstellar dust model}
\label{sect_dust_model}
  
\begin{table*}
\caption{The model dust compositions based on the optEC$_{(s)}$(a) data and the dust observables.}
\begin{center}
\begin{tabular}{lccccccccc}
\hline
\hline
               &        &     &       &             \\[-0.2cm]
material & typical radius [nm] & $E_{\rm g}$(bulk)  [ eV ]    &   $E_{\rm g}$(eff.)  [ eV ]    & FUV & UV bump & vis.-NIR & EBs &  3.4\,$\mu$m abs.  &  IR-mm  \\[0.05cm]
 \hline
               &        &     &       &             \\[-0.2cm]
 a-C           &  $< 1$    &  0.1        &  $> 0.8$      & $\bullet$    & $\bullet$    & $\times$ & $\bullet$ & $\times$ & $\times$   \\
 a-C           &  $1-5$    &  0.1        &  $0.1-0.8$ & $\bullet$ & $\bullet$ & $\circ$ & $\circ$    & $\times$ & $\times$  \\
 a-C           &   $5-20$ &  0.1        &  0.1           & $\circ$    & $\circ$    & $\circ$ & $\times$ & $\times$ & $\circ$  \\    
 a-C:H/a-C &  $100-200$  &  2.5/0.1  &  2.5/0.1     & $\times$ & $\times$ & $\circ$ & $\times$ & $\bullet$ & $\bullet$   \\
  a-Sil$_{\rm Fe}$/a-C     &    $100-200$    &  $\sim$$8/0.1$   &  $\sim$$8/0.1$  & $\times$ & $\times$ & $\bullet$ & $\times$ & $\times$ & $\bullet$ \\
                &        &     &       &    & & & & &           \\[-0.2cm]
\hline
\hline
\end{tabular}
\begin{list}{}{}
Notes:
\item[] 1. The materials for core/mantle particles are indicated as such in column 1. 
\item[] 2. The symbols $\bullet$, $\circ$ and $\times$ indicate major, minor and no contributions to the observed, FUV extinction, UV bump at 217\,nm, visible-NIR extinction, emission bands (EBs),  3.4\,$\mu$m absorption and IR-mm continuum emission.
\item[] 3. The outer surfaces of the a-C:H grains are UV photo-processed (into a-C) to a depth of 20\,nm (see paper~II).  
\item[] 4. The accreted and UV photo-processed a-C mantle on the amorphous silicates is 5\,nm thick. 
\item[] 5. The relative contributions of the small a-C particles to the UV bump and FUV extinction depend on the size distribution (see Sections \ref{sect_FUV_ext} and \ref{sect_UV_bump}). 
\end{list}
\end{center}
\label{table_dust_model}
\end{table*}
  
Based on earlier considerations \citep[{\it e.g.},][]{1990QJRAS..31..567J,2001ASPC..231..171J,2011A&A...530A..44J}, 
the work of \cite{2011A&A...528A..96K,2012A&A...548A..61K}, {\em Planck} satellite observations \citep{2011A&A...536A..24P,2011A&A...536A..25P}, and the optEC$_{\rm(s)}$(a) model data (papers I to III), we suggest that the usual decomposition of the  interstellar dust species into two major populations of amorphous carbonaceous and silicate grains remains the most viable explanation. 
However, the newly-available optEC$_{\rm(s)}$(a) data imply that material composition and size effects likely play an important role. The a-C(:H) component is the detailed subject of this, and papers I--III, and has properties determined by and calibrated against the available laboratory data.  

\subsection{Interstellar dust components}
\label{sect_composite}

\begin{figure} 
 \resizebox{\hsize}{!}{\includegraphics[angle=270]{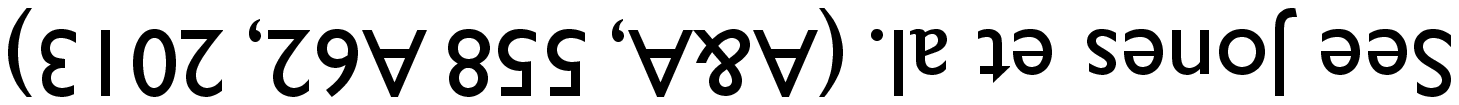}}
 \vspace{-0.5cm}
 \caption{The model dust populations, as seen in cross-section. In the upper part the a-C:H/a-C grains are shown, where black represents aromatic-rich material and white aliphatic-rich material. In the lower part amorphous silicate grains (green) are shown with a 5\,nm thick coagulated/accreted a-C mantle. The particle radii are indicated on a logarithmic scale.}
 \label{fig_dust_model}
\end{figure}

In the ISM it is hard to see how the silicate and carbonaceous dust populations could be completely segregated because mixing, even at some minor (contaminant) level, must occur.   Thus, the amorphous silicates must be mixed, to some degree, with a carbonaceous dust component  \cite[{\it e.g.},][]{1989ApJ...341..808M,1990QJRAS..31..567J}. In Table~\ref{table_dust_model} and Fig. \ref{fig_dust_model} we summarise our assumed dust properties as a function of size and in the following we describe these in more detail: \\
\noindent {\bf 1) a-C(:H) grains with size-dependent properties}:\footnote{For a-C(:H) particles, with fewer than several thousand atoms, surface hydrogenation becomes important. Also, the aromatic cluster sizes are limited by the particle radius and the band gap for small ($a < 30$\,nm) a-C particles can be significantly higher than expected. For example, for a-C the band gap can be as low as $\simeq -0.1$\,eV but for a 1\,nm radius a-C particle the band gap cannot be smaller than 0.7\,eV (for full details see paper III).}  
This population represents a fundamental continuity in composition and size distribution, which is qualitatively consistent with a-C:H dust that has been exposed to the ``equilibriating effects'' of the local ISRF for at least $10^6$\,yr (see Fig.~18 in the Corrigendum to paper~III), {\it i.e.}, long enough for any sub-nm particles to be aromatised to a-C.\footnote{Aromatised a-C nano-particles consist of small aromatic clusters, with a few aromatic rings per cluster, that are covalently-linked together by aliphatic and olefinic bridging structures. They are three-dimensional networks that intimately mix predominantly aromatic carbon but which also contain a significant aliphatic and olefinic carbon component in a single, contiguous chemical structure. The term PAH cannot be applied to them because they are not purely aromatic carbon structures.}  Any particles larger than a few tens of nm in radius will be incompletely aromatised or will consist of an aliphatic-rich core surrounded by a more absorbing, aromatised mantle layer. 
(Therefore, carbonaceous mantles $\lesssim 20$\,nm thick will be completely aromatised, see following.) \\
\noindent {\bf 2) a-Sil grains $\pm$ a-C(:H) mantles}: 
The interpretation of recent observations made by the {\em Planck} satellite \citep{2011A&A...536A..24P,2011A&A...536A..25P}, coupled with {\em Herschel} and {\em IRAS} data, shows that the observed dust emission spectral energy distribution (SED) in the diffuse ISM can be empirically, and extremely well, fit with a single temperature ($T_{\rm dust} \sim 18$\,K), emissivity-modified  ($\beta \simeq 1.8$) black-body.  This strongly suggests that the dust emission at long wavelengths is dominated by emission from a single dust population that mixes, predominantly, amorphous silicate (a-Sil) materials \citep[{\it e.g.},][]{2011A&A...535A.124C} with a carbonaceous (a-C or a-C:H) component in the form of mantles or accreted small grains \citep[{\it e.g.},][]{2011A&A...528A..96K,2012A&A...548A..61K}.
The surface a-C(:H), whether accreted as a mantle or formed by the coagulation of small a-C particles, must be $\lesssim 20$\,nm thick (the depth at which the optical depth for the FUV processing photons is unity) otherwise it will be {\em incompletely photolysed} to a-C (see paper II).
An a-C:H mantle would be traceable through the polarisation of the aliphatic C$-$H $3.4\,\mu$m absorption band, which will follow that of the host a-Sil.  
The presence of an a-C:H mantle on a-Sil dust is inconsistent with observations \citep{1999ApJ...512..224A,2006ApJ...651..268C}. 
In our model we assume 5\,nm thick a-C mantles ($E_{\rm g} = 0.1$\,eV) on the a-Sil grains. \\
For the amorphous silicate optical properties we use those for an amorphous forsterite-type silicate with iron incorporated into the material as nano-particle inclusions, a-Sil$_{\rm Fe}$, which is equivalent to incorporating iron directly into the silicate lattice structure. In Appendix \ref{sect_silicate} we discuss the consequences of this form of amorphous silicate, in relation to those of the usually-adopted optical properties for `astronomical silicates' \citep[{\it e.g.},][]{1984ApJ...285...89D,1996A&A...309..258G}. 
We will investigate the optical properties of the range of likely interstellar amorphous silicate dust analogue materials in a future paper. \\
\noindent {\bf 3) Coagulated, a-C(:H)-mantled a-Sil grains}: 
Enhanced dust emissivities in denser regions of the ISM perhaps indicate the onset of dust coagulation into larger inhomogeneous aggregates, which is consistent with recent modelling results \citep[{\it e.g.},][]{2011A&A...528A..96K,2012A&A...548A..61K}. This component therefore represents a coagulated mixing of all dust components, possibly with an additional a-C:H mantle material accreted from the remnant carbon in the gas phase, and also ice mantles in dense molecular clouds. 

Comparing these dust populations with those used in the \cite{2011A&A...525A.103C} DustEM model, we find that there is no direct one-for-one correspondence because of overlapping properties. However, and in general, the small carbonaceous grains ($a \lesssim 20$\,nm) are the carriers of the same observables as the combined DustEM PAH and small amorphous carbon, SamC, grains. (Taken together we note that the combined DustEM PAH+SamC particles have a similar size distribution to our small a-C(:H) particles with a power-law size distribution.) The large a-C:H/a-C grains are equivalent to the DustEM large amorphous carbon, LamC, grains and the large carbon-coated amorphous silicates equivalent to the DustEM amorphous silicate, aSil, grains. Compared to most other dust models, we use a naturally-continuous distribution of carbonaceous grain properties rather than separate and disconnected populations of small carbon grains and PAHs. For the model  proposed here we find it necessary to add an a-C mantle to the Fe-containing amorphous silicate grains in order to increase their emissivity and decrease their temperature (see Appendix \ref{sect_silicate} and Section \ref{sect_MIRmm_emission}). 

The proposed dust model has the advantage of adding a carbonaceous component to the large ($a \sim 10-4000$\,nm) silicate population (see Fig. \ref{fig_dust_model}) that does not exhibit a $3.4\,\mu$m aliphatic carbon band but which ought to show a weak $3.3\,\mu$m aromatic CH band and other associated CH and CC bands. However, as we show later (see Section \ref{sect_dust_starter}), the a-C mantles on the a-Sil grains do not appear to manifest any absorption features in the $3\,\mu$m region. In any event, the IR absorption signatures of the a-C mantles will not be observable in the diffuse ISM because they will be swamped by emission bands at the same wavelengths. Observations of a $3.3\,\mu$m aromatic CH band in absorption would be possible along lines of sight where the emission bands are weak or absent \citep[{\it e.g.}, towards young stellar objects and the Galactic Centre, ][]{1989ApJ...344..413S,1995ApJ...449L..69S,1999ApJ...517..883B,2000ApJ...537..749C}. \cite{1999ApJ...517..883B} find that the optical depth of the $3.25\,\mu$m `aromatic' absorption feature in their spectra correlates better with the optical depth of the silicate feature than with the water ice feature, perhaps indicating  an aromatic carbon component more intimately associated with the amorphous silicate grains than with their accreted ice mantles.

\section{Astrophysical implications}
\label{sect_dust_ext_sed}

The \cite{1997A&A...323..566L} three-component dust model consists of large organic refractory-coated silicates (responsible for the visible extinction and polarisation), empirical UV bump-carrying particles (with a weak contribution to the visible extinction) and PAHs (responsible for the FUV extinction). In this model the extinction can naturally be decomposed into three separate populations and it thus satisfies the observed extinction constraints and (non-)correlations. It also satisfies the observed wavelength-dependent polarisation and extinction-to-polarisation ratio. As we shall see the dust model that we have developed bears some qualitatively-similar traits to the \cite{1997A&A...323..566L} model. However, our model uses only two dust materials, {\it i.e.}, a-Sil$_{\rm Fe}$ and a-C(:H). In the model two of the dust observables, the FUV extinction and the UV bump, are due to the size-dependent properties of a-C(:H).

\begin{figure} 
 \resizebox{\hsize}{!}{\includegraphics{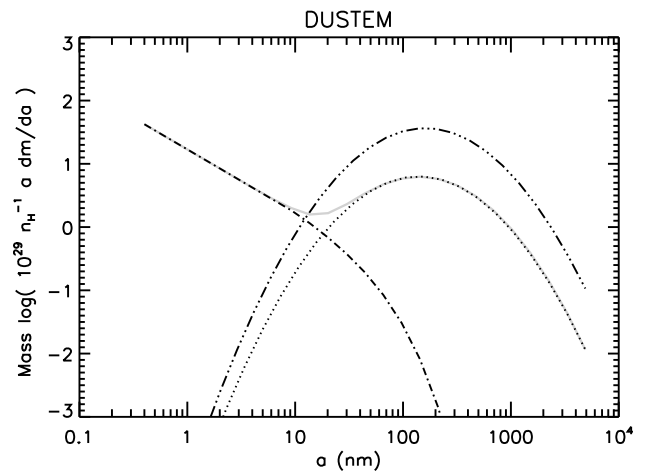}}
 \caption{The standard dust model size distributions. a-C-coated ($E_{\rm g} = 0.1$\,eV, mantle depth 5\,nm) amorphous forsterite-type silicate grains with $\sim 70$\% of the available iron in metallic, nano-particle inclusions (triple dot-dashed). Core/mantle, a-C:H/a-C ($E_{\rm g} = 2.5/0.1$\,eV) particles: large, log-normal (dotted) and small, power-law with an exponential tail (dash-dotted). The grey line shows the overall a-C(:H) grain size distribution, with physically-continuous properties, which is only separated into `large' and 'small' components for modelling convenience.} 
 \label{fig_dust_model_sdist}
\end{figure}

In order to illustrate the utility of the optEC$_{\rm (s)}$(a) data for ISM dust modelling we present a surprisingly-simple dust model, based on the discussion presented in the previous section. This model uses three dust populations but only two dust materials (see Figs.~\ref{fig_dust_model} and \ref{fig_dust_model_sdist}): 
\begin{enumerate}
  \item A  power-law distribution, with an exponential tail\footnote{This exponential cut-off is required in order to limit the size of the largest a-C particles when we explore variations in the power-law index due to dust evolution (see Section \ref{sect_FUV_ext}).}, of small (0.4 to $\simeq 100$\,nm), a-C(:H) particles that have been photo-processed to depths of 20\,nm and that show a continuity of size-dependent optical properties,  
  \item a log-normal size distribution of large ($\simeq 200$\,nm) a-C:H grains ($E_{\rm g} = 2.5$\,eV)\footnote{The dust extinction and emission results are not sensitive to the core a-C:H material band gap as long as it is $\gg 0$\,eV. However, the observed IR absorption bands in the $3-4\,\mu$m region and the derived aliphatic CH$_2$/CH$_3$ ratio both require that the large a-C(:H) grain core material has a wide band gap, {\it i.e.}, $E_{\rm g} \gtrsim 2.5$\,eV (see paper I and paper III and the corrigendum to paper III).\label{fn_Eg2p5}} with an outer, photo-processed, a-C layer (20\,nm thick, $E_{\rm g} = 0.1$\,eV), and  
  \item a log-normal size distribution of large ($\simeq 200$\,nm) amorphous forsterite-type silicate grains, with Fe nano-particle inclusions (containing $\sim 70$\,\% $\equiv 22$\,ppm of the cosmic iron),\footnote{We note that adding 100\,\% of the cosmic iron into the silicates does not significantly modify the results.} that are coated with 5\,nm thick a-C mantles ($E_{\rm g} = 0.1$\,eV).  
\end{enumerate}
With this model we are able to achieve an excellent fit to the dust observables using a coherent set of physically-realistic a-C(:H) and a-Sil$_{\rm Fe}$ data that are based on laboratory measurements. In order to maximise the usefulness of the model we have ``astronomicalised'' the optEC$_{\rm(s)}$(a) data.  All of the required adjustments are rather minor and physically-justified (see Appendix~\ref{astro_fudge}). With these modified data we have re-visited the comparison of the optical properties of nm-sized a-C(:H) particles with those of ``astronomical PAHs'' (as per paper III, {\it i.e.}, see Section \ref{astro_fudge_cfPAH}).

The model results are not particularly sensitive to the large a-C:H/a-C grain size distribution, which contains a relatively small fraction of the dust mass. We find that removing the large a-C:H/a-C dust component and extending the small a-C:H/a-C grain power law to $4900$\,nm gives an equally good fit to the dust observables. However, at mm wavelengths the fit is not so good but we note that this is in a region where the amorphous silicate emission is not yet well constrained. 

In the following sub-sections we present the detailed extinction and emission predictions for our dust model and compare them to the observations of dust in the diffuse ISM.

\subsection{A diffuse ISM dust model}
\label{sect_dust_starter}

\begin{figure*}
        \centering
        \begin{subfigure}[b]{0.5\textwidth}
                \centering
                \includegraphics[width=9.0cm]{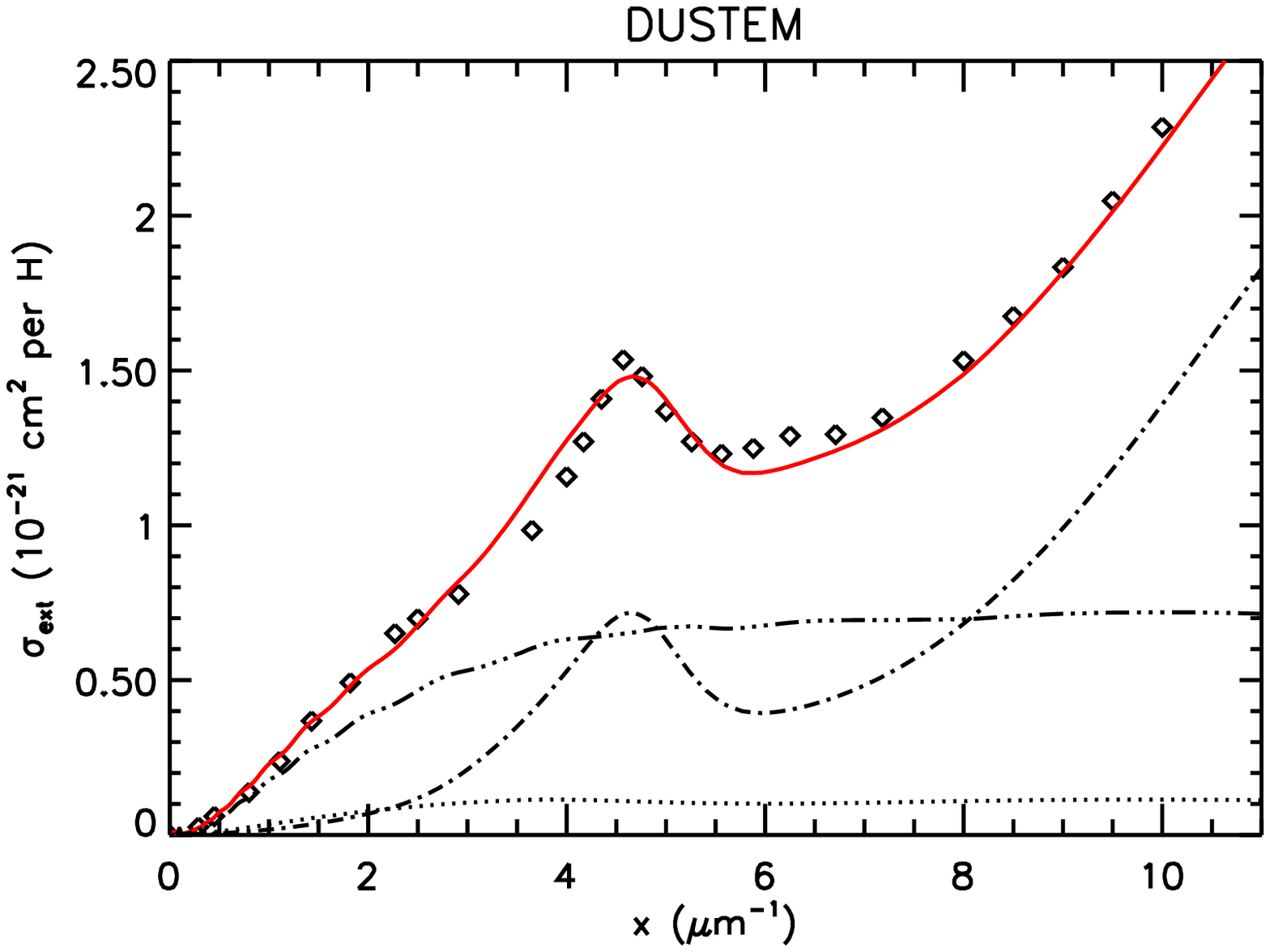}
                \includegraphics[width=9.0cm]{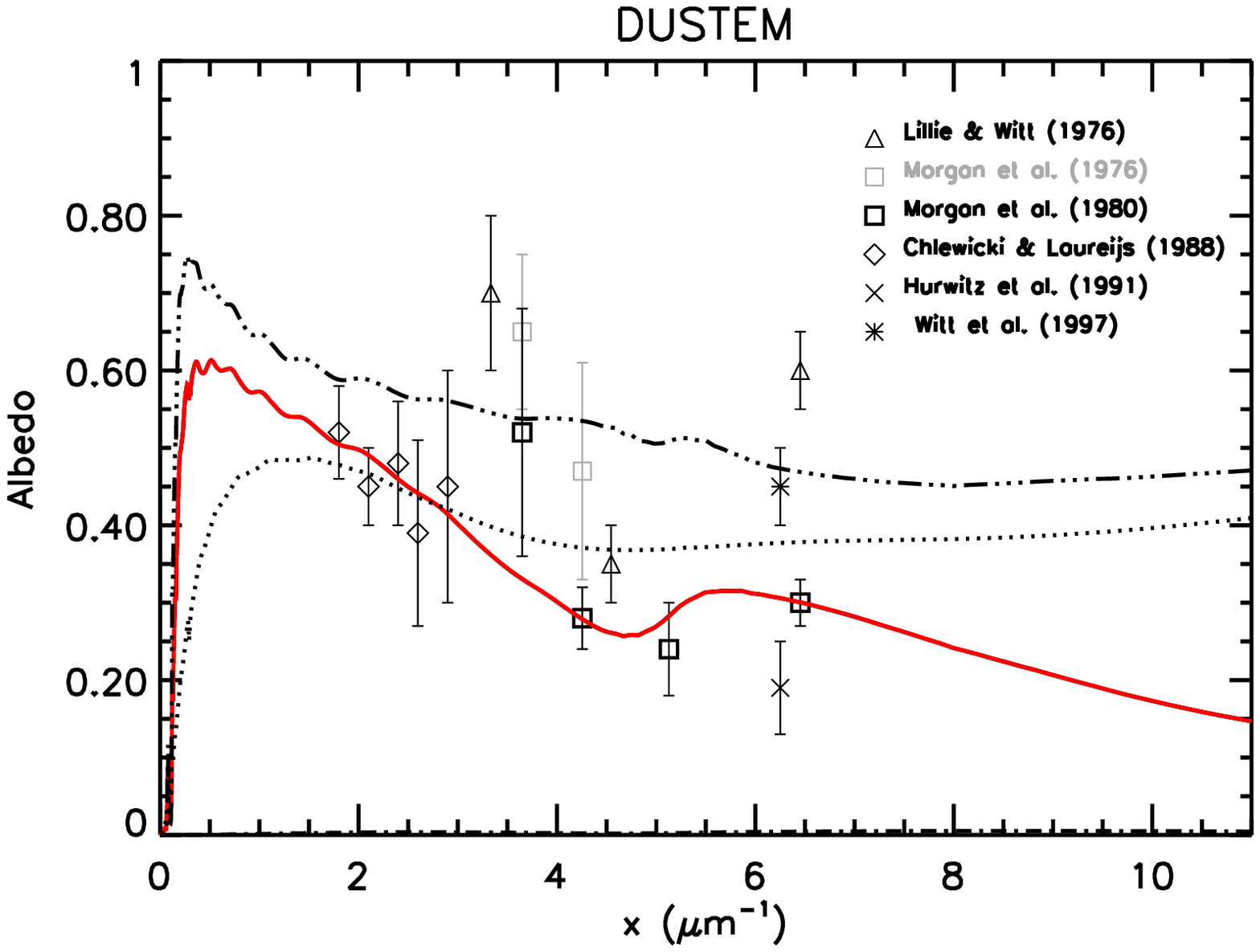}
                \label{fig_dust_model_extuv_alb}
        \end{subfigure}%
        ~ 
        \begin{subfigure}[b]{0.5\textwidth}
                \centering
                \includegraphics[width=9.0cm]{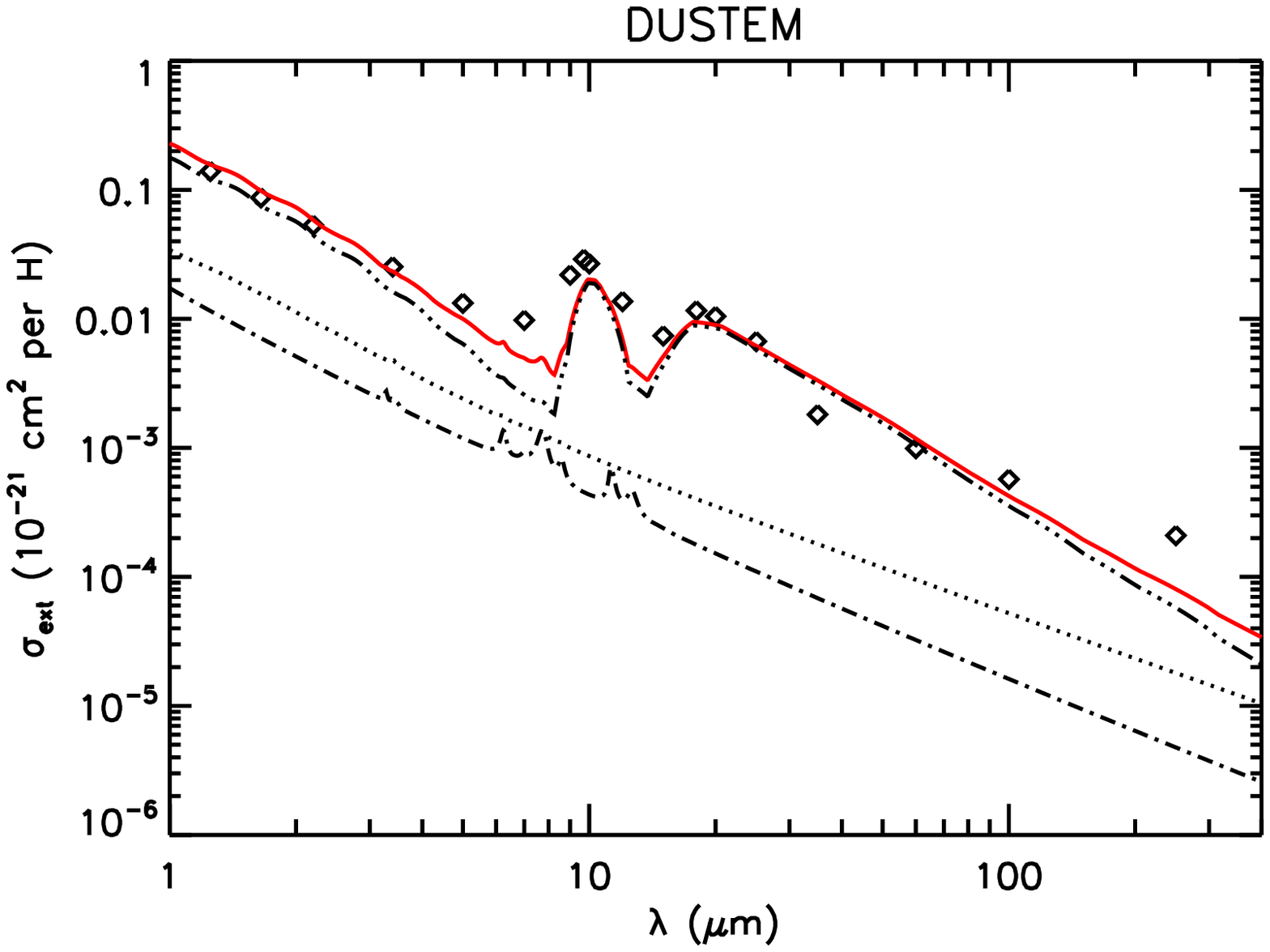}
                \includegraphics[width=9.0cm]{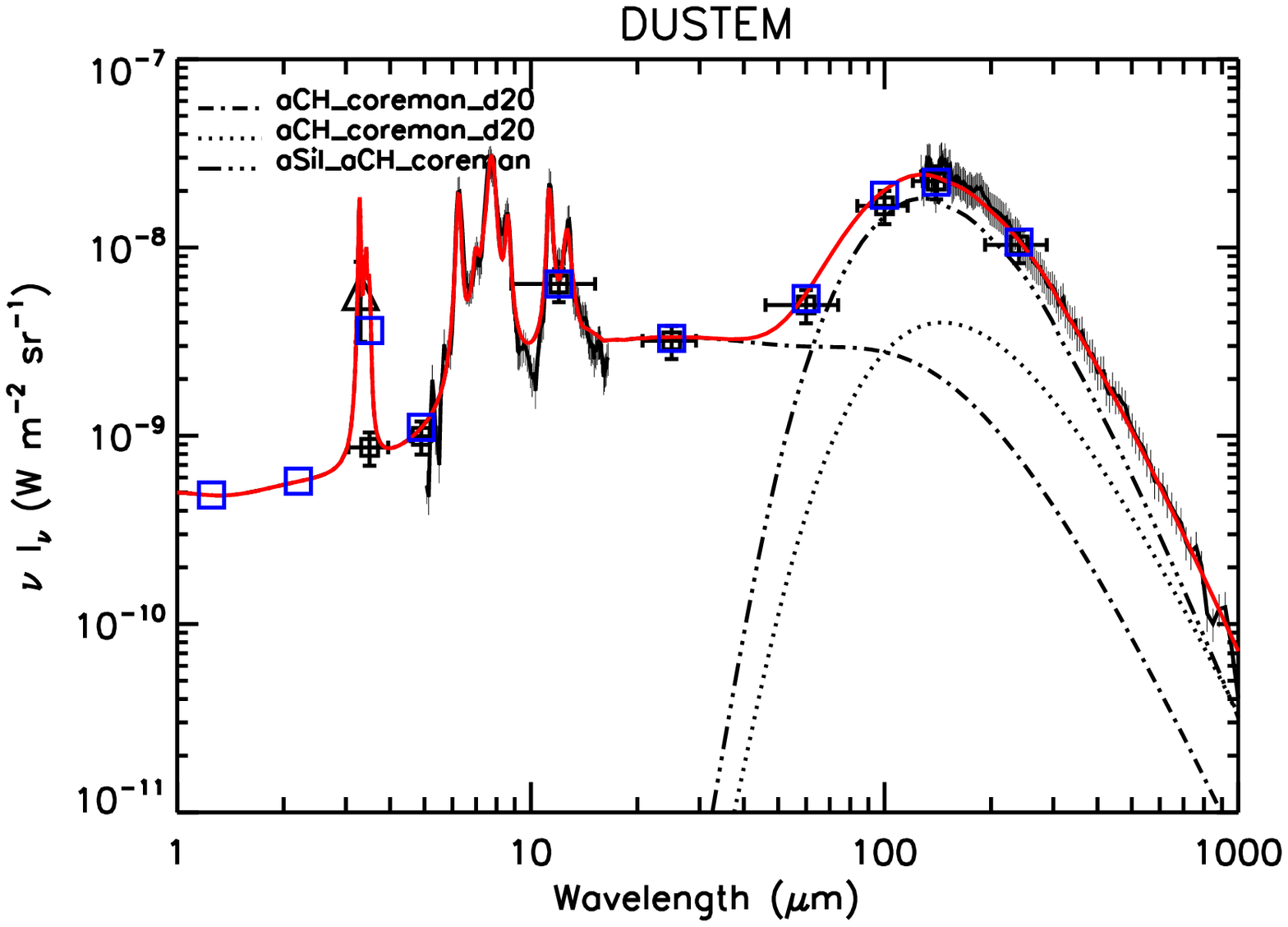}
                \label{fig_dust_model_extir_emis}
        \end{subfigure}
        ~ 
        \caption{The standard dust model for $N_{\rm H} = 10^{20}$\,H\,cm$^{-2}$: 
                     NIR-UV extinction (top left),  
                    IR extinction (top right),  
                    albedo (bottom left) and  
                    the full dust SED (bottom right). 
                    The grain types are: 
                     a-Sil$_{\rm Fe}$/a-C core/mantle grains (triple dot-dashed), 
                     a-C:H/a-C core/mantle grains (dotted) and 
                     small a-C grains (dash-dotted). 
                    In each plot the totals are shown by the solid lines (red). 
                    The extinction data in the upper figures are taken from \cite{1990ARA&A..28...37M}.
                    The observational data in the lower right plot are: $\sim5-15\,\mu$m ISOCAM/CVF ISO spectrum, $\sim100-1000\,\mu$m FIRAS/COBE spectrum, $3.3\,\mu$m AROME narrow band measurement (triangle) and   DIRBE/COBE photometry (squares) \citep[for details see][]{2011A&A...525A.103C}.
                     }
        \label{fig_DustEM_output}
\end{figure*}

\begin{table*}
  \caption{The dust model parameters as input into DustEM.}
\begin{center}
\begin{tabular}{llcccccccc}
\hline
\hline
                &       &      &         &          &         &         &         &        &     \\[-0.25cm]
composition &   size       & $E_{\rm g}$ ( eV ) &  $\rho$ ( g\,cm$^{-3}$ ) & $\alpha$  &  $a_{\rm min}$ / $a_{\rm max}$      &  $a_{\rm c},\ a_{\rm t}$ /  $a_0$ & $\gamma$ / $\sigma$ & $Y$ & $f_{M-{tot}}$ \\
  core / mantle &    distribution  & core / mantle &  core / mantle  &        &    ( nm )                   &    ( nm )                    &                  &  ( $M/M_{\rm H}$ )      &     \\
                &      &       &         &          &         &         &         &        &     \\[-0.25cm]
\hline
                      &   &    &         &          &         &         &         &        &     \\[-0.25cm]
  a-C:H / a-C  \ &   p-law &2.5 / 0.1     &  1.3 / 1.6  &   5.0   &  0.4 / 4900 &  50, 10 / $-$ &  1.0 / $-$     & $1.6\times10^{-3}$ &  18.6\,\%   \\
                   &    &      &         &          &         &         &         &        &     \\[-0.25cm]
  a-C:H / a-C  \ &  log-n & 2.5 / 0.1     &  1.3 / 1.6  &   $-$   &  0.5 / 4900 &  \ \ \ \ $-$, \ $-$ / 7.0 &  \ \ \ $-$\ \  / 1.0   & $0.6\times10^{-3}$ &  \ \,7.0\,\%   \\
                   &    &      &         &          &         &         &         &        &     \\[-0.25cm]
  a-Sil$_{\rm Fe}$ / a-C & log-n &$\sim 8$ / 0.1 &  2.5 / 1.6 &        & 1.0 / 4900 &  \ \ \ \ $-$, \ $-$ / 8.0 &  \ \ \ $-$\ \ / 1.0  & $5.8\times10^{-3}$ &   67.4\,\%  \\
                   &    &      &         &          &         &         &         &        &     \\[-0.25cm]
  \ \ \ \ \ \ \ \ $-$\ / a-C & mantle &\ \ $-$\ \  / 0.1   &   $-$\ \ \,  / 1.6  &  $-$   & $d=5$\,nm & \ \ \ \ $-$, \ $-$ / \ $-$\ \   &  \ \ $-$\ \  / \ $-$\ \   & $0.6\times10^{-3}$  &  \ \,7.0\,\% \\[0.05cm]
\hline
                &       &      &         &          &         &         &         &        &     \\[-0.25cm]
               &        &       &         &          &         &         &   TOTAL  & $8.6\times10^{-3}$  &     \\[0.05cm]
\hline
               &       &        &         &          &         &         &         &        &     \\[-0.25cm]
\end{tabular}  
\begin{list}{}{}
Notes:
\item[]  1. The elemental abundances for this model are, 233\,ppm for C, and 
{\bf 50}-67, 22, 45-{\bf 50} and 180-{\bf 199}\,ppm, for Mg, Fe, Si and O, respectively.  The abundances in boldface type, for Mg and Fe, are for the standard model where the Fe is present in nano-particles, the other values indicate the Mg and Fe abundances when all of the Fe is incorporated into the amorphous silicate phase. 
\item[] 2. $d$ is the a-C mantle depth on the amorphous silicates (5\,nm). 
\item[] 3. p-law indicates a power-law distribution, $dn/da \propto a^{-\alpha}$, with an exponential tail, $D(a) = {\{\rm exp}(-[a-a_t]/a_c)^\gamma\}$ for $a>a_t$ else $D(a) = 1$. 
\item[] 4. log-n indicates a logarithmic normal distribution, $dn/da \propto {\rm exp}(-{\rm log}[a/a_0]^2)/\sigma$. 
\end{list}    
\end{center}    
  \label{table_model_params}
\end{table*}

The use of the optEC$_{\rm(s)}$(a) data, as derived in papers II and III, is quite naturally able to explain the major dust observables. However, the fit to the UV bump, the $6-9\,\mu$m emission bands and the $\sim 10-40\,\mu$m MIR continuum emission is not entirely satisfactory. We note that the latter two deficiencies are related, for a better account for the $6-9\,\mu$m emission bands leads to increased energy-loss via band emission for the larger a-C:H particles, which are then cooler and show weaker continuum emission in the $\sim 10-40\,\mu$m region. Also, an increase in the IR band intensities in the $6-9\,\mu$m region leads to a slight increase in the extinction on the short wavelength side of the $9.7\,\mu$m silicate band (see Fig.~\ref{fig_DustEM_output} top right).  We note that the $9.7\,\mu$m band in our laboratory data-derived a-Sil$_{\rm Fe}$ is  too narrow in comparison with the \cite{1990ARA&A..28...37M} extinction profile (see Fig.~\ref{fig_DustEM_output}, top right, and Appendix \ref{sect_silicate}), even with the addition of carbonaceous mantles. However, the extinction towards the Galactic Centre observed by  \cite{1996A&A...315L.269L} looks rather different in that the short wavelength dip appears to be filled in, tantalisingly exhibiting what could be broad absorption bands in the $3-4$ and $6-8\,\mu$m regions (similar to but much stronger than those of the a-C:H materials seen in Fig.~\ref{fig_DustEM_output} top right). 

As mentioned above, we have adjusted the optEC$_{\rm(s)}$(a) data in a physically-meaningful way in order to get a good fit to the astronomical data.\footnote{The required modifications to the IR bands, and the $\pi-\pi^\star$ and $\sigma-\sigma^\star$ bands are minor and are explained in full detail in Appendix~\ref{astro_fudge}.} 
Following our use of the published optEC$_{\rm(s)}$(a) data we find that it is difficult to interpolate for intermediate particle sizes because of the sharp IR bands, which are composition- and size-dependent. Thus, the optical property data ideally need to be calculated {\it from scratch} for each required particle size.\footnote{This capability will be introduced into a future release of the DustEM code and documentation that integrates the optEC$_{\rm(s)}$(a) data.} 

\begin{figure} 
 \resizebox{\hsize}{!}{\includegraphics{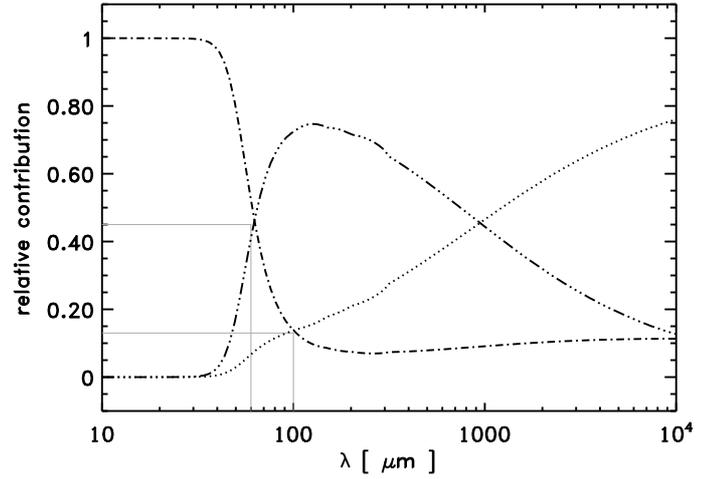}}
 \caption{The dust component relative contributions to the SED: the line designations are as per Fig.~\ref{fig_dust_model_sdist}. Note that the small a-C(:H) grain population contributes $\approx 45$\% of the total emission at 60\,$\mu$m and $\approx 13$\% of the total emission at 100\,$\mu$m.} 
 \label{fig_dust_model_decomp_std}
\end{figure}

Using the DustEM tool  \citep{2011A&A...525A.103C} we have calculated the predicted properties for our dust model.  The small and large a-C(:H) grain size distributions are shown separately in Fig.~\ref{fig_dust_model_sdist} but they do in fact form a single grain population with size-dependent properties (the grey line in Fig.~\ref{fig_dust_model_sdist}), which are important for $a \lesssim 30$\,nm (the separation into two distributions is simply a modelling convenience). Fig.~\ref{fig_DustEM_output} shows the UV-visible extinction curve, the albedo,\footnote{The observational data are from \cite{1976ApJ...208...64L,1976MNRAS.177..531M,1980MNRAS.190..825M,1988A&A...207L..11C,1991ApJ...372..167H,1997ApJ...481..809W}} the IR extinction and the emission spectrum ({\it i.e.}, the SED). In this model the large a-C(:H) grain mass is about one tenth of that of the amorphous silicates, which reflects the lower resilience of carbonaceous dust to ISM dust processing, compared to silicate dust \citep[see later, and also][]{2008A&A...492..127S,2011A&A...530A..44J}. Fig.~\ref{fig_dust_model_decomp_std} gives the relative dust component contributions at each wavelength and shows that for $\lambda \approx 1-60\,\mu$m the emission is dominated by the small (power-law) a-C grains, for $\lambda \approx 60\,\mu$m-1\,mm by the a-Sil$_{\rm Fe}$/a-C grains and for $\lambda  \gtrsim 1$\,mm by the large a-C:H/a-C grains. Fig.~\ref{fig_DustEM_output} indicates that the model gives an extremely good fit to all of the major dust extinction and emission features observed in the diffuse ISM.  

Note that the a-C:H to a-C dust mass ratio, an indicator of the $sp^3/sp^2$ C atom ratio, is $\approx 1/4$ for our model ($\equiv R = X_{sp^3}/(X_{sp^3}+X_{sp2}) \simeq 0.2$ and $E_{\rm g} \simeq 0.9$\,eV)\footnote{{\it N.B.} This is an effective band gap for all of the a-C(:H) dust phases and does not take into account that the a-C:H and a-C dust are segregated into core and mantle components, respectively, each  characterised by a different band gap, {\it i.e.}, $\simeq 2.5$\,eV for a-C:H and $\simeq 0.1$\,eV for a-C.}, indicating that in the diffuse ISM the grains are aromatic-rich in character, in agreement with the recent work by \cite{2012ApJ...760L..35L} and \cite{2013ApJ...770...78C}. In fact, for the a-C nano-particles ($E_{\rm g} \simeq 0.1$\,eV), which are responsible for the IR emission bands in our model, $\sim 90$\% of the carbon atoms are in $sp^2$ aromatic clusters. This is consistent with the finding of \cite{2012ApJ...760L..35L} that the emission band carriers contain $< 15$\% of aliphatic carbon. \cite{2013ApJ...770...78C} interpret the observational evidence for the diffuse ISM carbonaceous dust as resulting from grain processing leading to the a-C:H mantling of aromatic-rich stardust grains. 
 
We note that this dust model requires similar silicate-element depletion (321 $\equiv$ 50, 22, 50, 199\,ppm for Mg, Fe, Si and O, respectively) but more carbon depletion (233\,ppm) than the \cite{2011A&A...525A.103C} model (315 $\equiv$ 45, 45, 45 and 180 for Mg, Fe, Si and O, respectively, and 200\,ppm for C). For {\em bulk, crystalline} olivine minerals the density ranges from 3.2\,g\,cm$^{-3}$ for forsterite (Mg$_2$SiO$_4$) to 4.4\,g\,cm$^{-3}$ for fayalite (Fe$_2$SiO$_4$). However, for sub-$\mu$m, amorphous materials that have been ion-irradiated through the effects of cosmic rays and shocks in the ISM the material densities will be significantly lower. We therefore adopt a density of $2.5$\,g\,cm$^{-3}$ for our interstellar amorphous forsterite-type silicate,  compared to the bulk, crystalline mineral density of 3.2\,g\,cm$^{-3}$. For iron in the form of nano-particles  we assume a density of $7.87$\,g\,cm$^{-3}$.  Adopting Mg:Fe = 2.3:1 for the silicate, the same quantity of iron, as was assumed to be in Fe nano-particles ({\it i.e.}, 22\,ppm), can be incorporated into the silicate structure.\footnote{The optical properties for this silicate are very similar to those of a-Sil$_{\rm Fe}$ (see Apendix \ref{sect_silicate}). Further, as shown by \cite{2006ApJ...637..774C}, a mix of amorphous olivine-type  and pyroxene-type silicate materials can equally well explain the profiles of the interstellar silicate bands. Hence, the Planck extinction and polarisation data will be critical in constraining the optical and physical properties of the large interstellar silicate and carbon dust populations.} In this case all of the required silicate-forming elemental depletions (314\,ppm) are similar to those of the \cite{2011A&A...525A.103C} model, {\it i.e.}, 67, 22, 45, 180\,ppm for Mg, Fe, Si and O, respectively. Nevertheless, the nature of amorphous silicates and the solid material into which iron is bound in the ISM, {\it i.e.}, silicate, oxide or metal, remains something of an open question that awaits new laboratory and observational data. 

\begin{figure} 
 \resizebox{8.5cm}{!}{\includegraphics{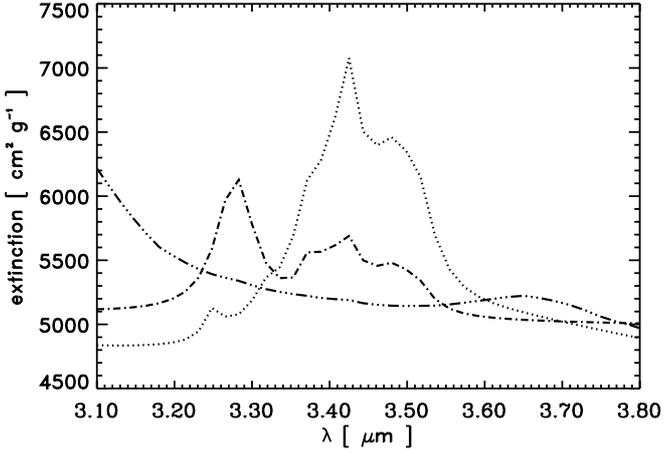}}
 \caption{The scaled dust model extinction in the $3.4\,\mu$m region: small a-C:H/a-C ($\times \lambda^{1.5}$, dash-dotted), large a-C:H/a-C ($\times \lambda^{1.65} - 2.9 \times 10^4$, dotted) and large a-Sil$_{\rm Fe}$/a-C ($\times \lambda^{1.7} - 8.2 \times 10^3$, triple dot-dashed).} 
 \label{fig_dust_3p4_ext}
\end{figure}

In Fig.~\ref{fig_dust_3p4_ext} we show a zoom into the extinction in the $3.4\,\mu$m region for each of the three dust components in our standard model. Note that the data have been scaled for comparison. Fig.~\ref{fig_dust_3p4_ext} shows that the large a-C:H/a-C grains exhibit a predominantly aliphatic $3.4\,\mu$m band with a peak at $\simeq 3.42\,\mu$m, a shoulder at $\simeq 3.5\,\mu$m and weak absorption in the $3.28\,\mu$m region, which resemble the band profile seen towards the Galactic Centre \citep[see][]{2012aA&A...540A...1J,2012cA&A...542A..98J}. Any small a-C:H/a-C grains, seen in absorption, will exhibit a dominant $3.28\,\mu$m aromatic CH band with sub-bands or shoulders at 3.37, 3.42 and $3.5\,\mu$m but all of these are significantly weaker than the large a-C:H/a-C grain absorption bands. Further, the large a-Sil$_{\rm Fe}$/a-C core/mantle grains appear to show no spectroscopic evidence of their a-C mantles in this wavelength region. The non-linear baseline seen in the a-Sil$_{\rm Fe}$/a-C extinction data is due to the narrow size range used to derive the $Q_{ext}$ data.
In our model the carbonaceous $3.4\,\mu$m and the amorphous silicate $9.7\,\mu$m bands are due to distinct dust populations and some variation in their relative band strengths is therefore to be expected \cite[{\it e.g.},][]{2010EP&S...62...63G}. 

From the predicted dust SED in Fig.~\ref{fig_DustEM_output} it is clear that the emission bands in the $3-4\,\mu$m region have the correct spectral signatures, {\it i.e.}, a dominant $3.3\,\mu$m with a side-band at $\simeq 3.4\,\mu$m with a shoulder at $\simeq 3.5\,\mu$m, but that they are too strong. However, we note that this model does not yet take into account the effects of grain charge, which (as in astronomical PAH models) will reduce the band strengths of the predominantly aromatic carriers that are the originators of the emission bands in this wavelength region. In a follow-up paper we will look at the detailed effects of grain charge and the anomalous microwave emission from spinning a-C(:H) grains.

With our proposed model we expect little silicate grain fragmentation, in shock waves and turbulent regions, for two reasons. Firstly, there are too few sufficiently-large grains ($a \simeq 10$\,nm) for catastrophic collisions with silicate grains to be important \citep{1996ApJ...469..740J} and, secondly, the silicate grains are protected from erosion-type, cratering fragmentation by their encasing a-C mantles. In contrast, the less resistant a-C(:H) grains will experience some fragmentation \citep{2008A&A...492..127S} and we propose that our small a-C grain size distribution is the product of fragmentation in the ISM. We will investigate these fragmentation/cratering effects in detail in a follow-up paper. Further, we adopt log-normal size distributions for the large a-C:H/a-C and a-Sil$_{\rm Fe}$ grains because, firstly, they are typical of dust condensation from the vapour phase, {\it e.g.}, in the dust shells around evolved stars. Secondly, they appear to be consistent with the sizes of the pre-solar amorphous and crystalline silicate grains of evolved star origin that have been analysed in meteorites \citep[{\it e.g.},][]{2007ApJ...656.1223N}. Also, a theoretical study of cloud fragmentation and collapse prior to star formation shows that a log-normal distribution is a characteristic product of multiplicative processes acting on a size distribution \citep[{\it e.g.},][]{1973MNRAS.161..133L}. This would appear to be similar to the case for the multiple processing of dust in turbulent and shocked interstellar regions. 

\subsection{The optEC$_{\rm (s)}$(a) model: diffuse ISM dust and beyond}
 
Following on from papers~I to III, we consider the usefulness of the optEC$_{\rm(s)}$(a) data as a tool to interpret interstellar dust observations. In the following sub-sub-sections we compare the predicted properties, derived using the modified optEC$_{\rm(s)}$(a) data (see Appendix \ref{astro_fudge}), with the major constraints imposed by interstellar dust observations. The model-observation comparisons are presented in order of increasing wavelength. 

\begin{figure} 
 \resizebox{\hsize}{!}{\includegraphics{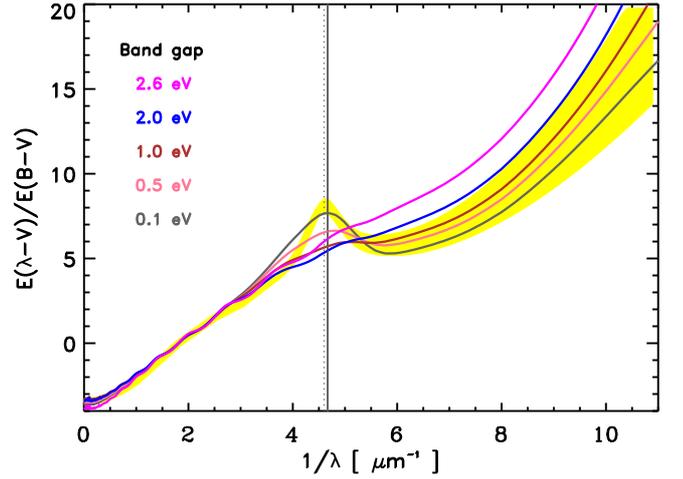}}
  \caption{The dust model normalised extinction, E($\lambda$-V)/E(B-V) for fixed dust mass and for varying band gap, $E_{\rm g}$ [eV]:   0.1 (dark grey),   0.5 (pink), 1.0 (brown),   2.0 (blue) and 2.6 (violet). The solid (dotted) vertical grey lines mark the model (observed) UV bump positions.   The yellow band indicates the observed extinction and its range of variation \citep{2007ApJ...663..320F}.}
 \label{fig_dust_results_1}
\end{figure}
%
\begin{figure} 
\resizebox{\hsize}{!}{\includegraphics{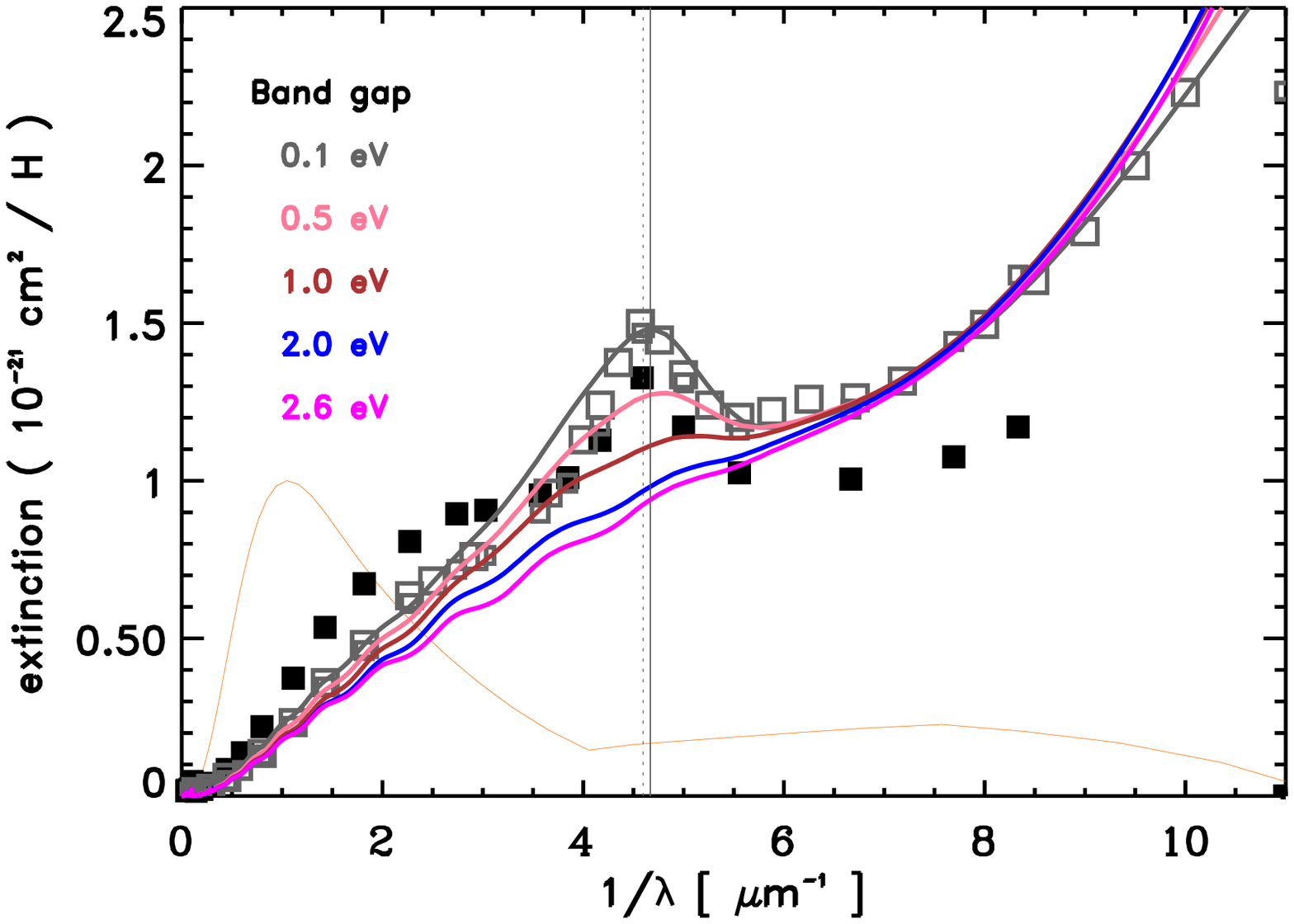}}  
\resizebox{\hsize}{!}{\includegraphics{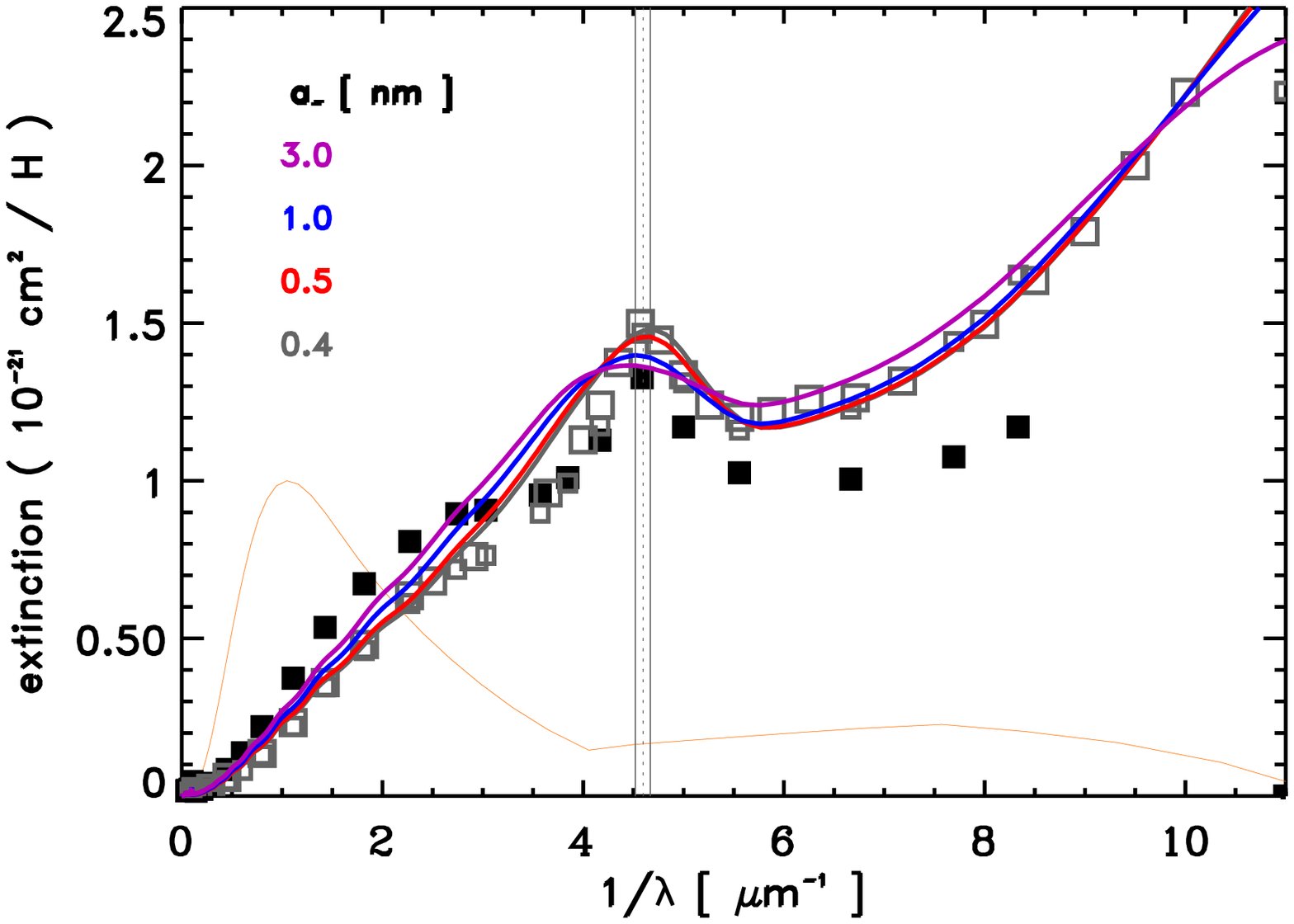}}   
\resizebox{\hsize}{!}{\includegraphics{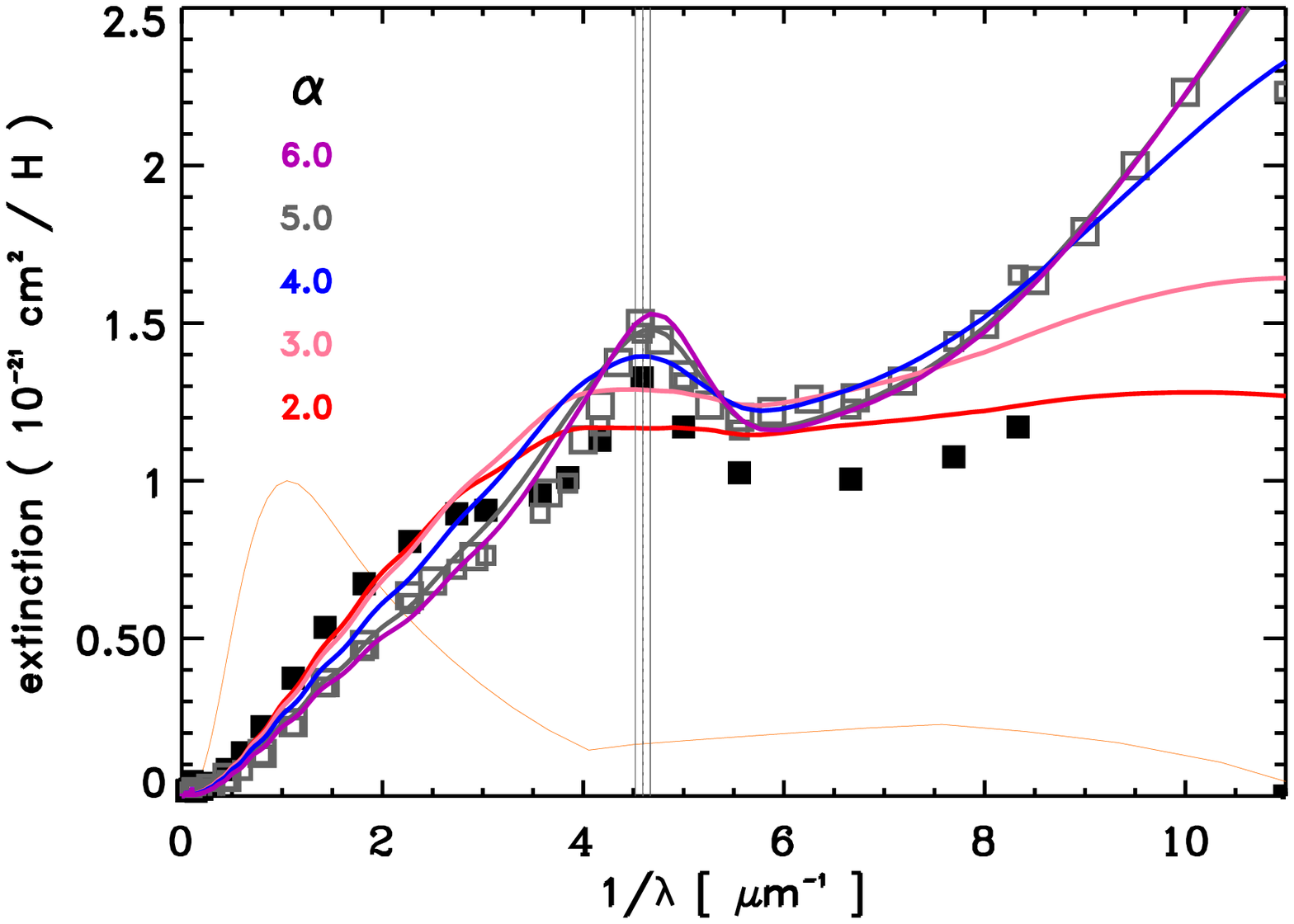}}
  \caption{The dust model extinction for fixed dust mass as a function of $E_{\rm g}$ (top), $a_-$ (middle) and $\alpha$ (bottom). 
  The solid (dotted) vertical grey lines mark the model (observed) UV bump peak positions. The symbols indicate the observed extinction for $R_V = 3.1$ (grey open squares) and 5.1 (black filled squares) \citep{1979ARA&A..17...73S,1990ARA&A..28...37M}.
  The orange line shows the intensity of the adopted ISRF   plotted as $J_\lambda/\lambda$ and normalised to unity.}
 \label{fig_dust_results_2}
\end{figure}
\begin{figure} 
\resizebox{\hsize}{!}{\includegraphics{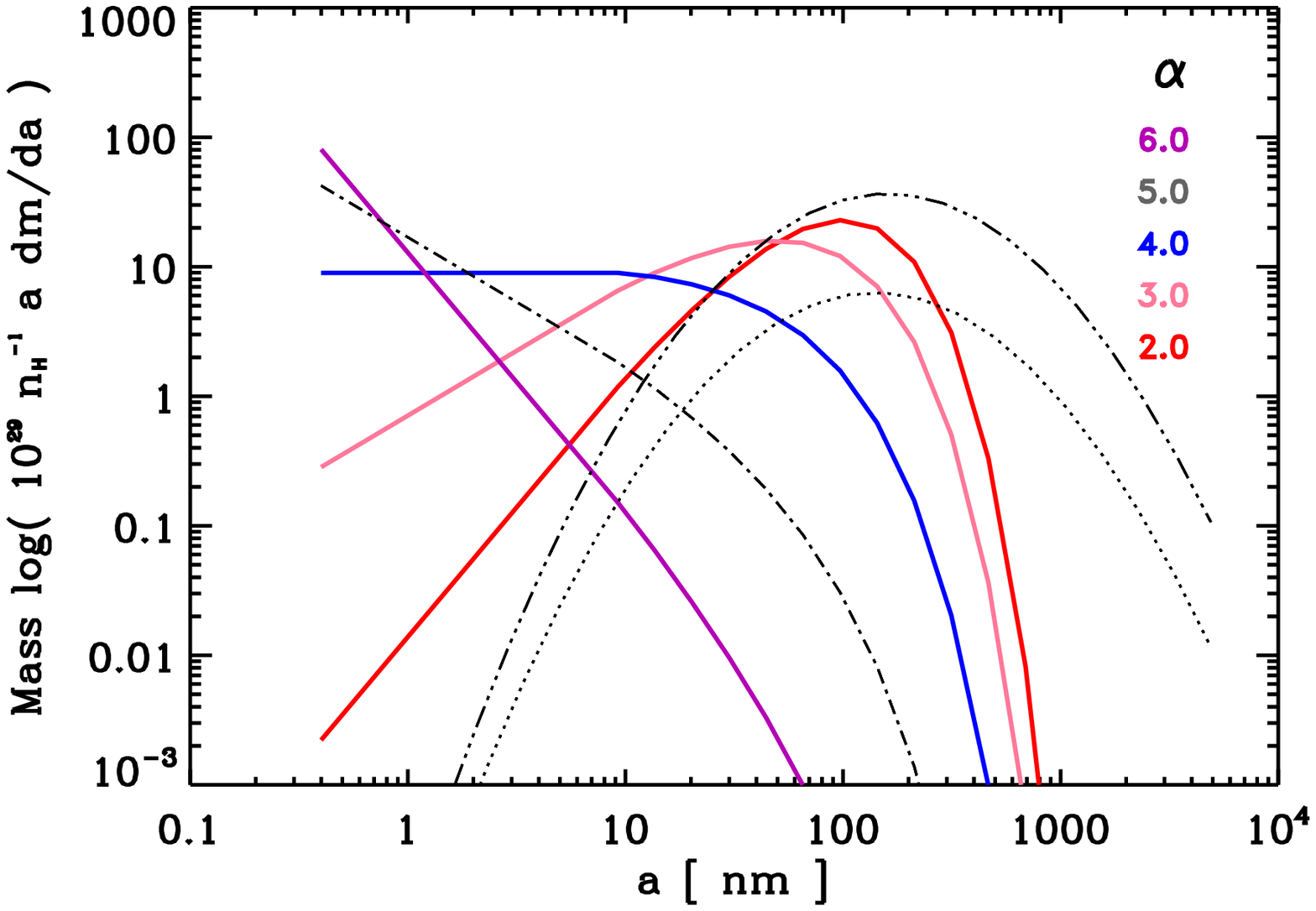}}
  \caption{Dust size distributions for fixed mass in each component: large a-C:H/a-C particles (black dotted), large a-Sil$_{\rm Fe}$/a-C particles (black tripple-dot-dashed), for varying small a-C(:H) grain size distribution power-law index, $\alpha$.} 
 \label{fig_dust_dist}
\end{figure}

\vspace*{0.5cm}
\subsubsection{The FUV extinction rise}
\label{sect_FUV_ext}

In Fig.~\ref{fig_dust_results_1} we show the model  extinction, for fixed dust component masses, normalised in the usual $E(B-V)$ manner, as a function of the `outer' material band gap. These data are compared to the  observed diffuse ISM extinction variations from the work of \cite{2007ApJ...663..320F}. 
Here we vary the band gap of the outer 20\,nm of all of the carbonaceous grains and, therefore, particles with $a < 20$\,nm, which dominate the FUV extinction, are homogeneous.  For particles with $a > 20$\,nm the core material gap is fixed at 2.5\,eV (see Footnote \ref{fn_Eg2p5}) and only the gap of the outer  20\,nm deep layer is varied. 

Fig.~\ref{fig_dust_results_1} indicates that the observed FUV extinction (in normalised form) is only consistent with a-C:H dust with a band gap $\lesssim 2$\,eV.  Nevertheless, care must be exercised in the use of  $E(B-V)$-normalised  extinction data because of the strong leverage by the V and B band data point anchors. Thus, in Fig.~\ref{fig_dust_results_2} we show the {\em un-normalised} FUV extinction for the model as a function of $E_{\rm g}$ (upper plot), $a_-$ (middle plot) and $\alpha$ (lower plot), along with the \cite{1979ARA&A..17...73S} and \cite{1990ARA&A..28...37M} extinction data (arbitrarily-scaled for qualitative comparison). Plotting the data in this way, we note that, for a fixed dust mass, the FUV extinction is intrinsically invariant in the $6-8\,\mu$m$^{-1}$ region, as noted by \cite{1983ApJ...272..563G}. Thus, for fixed dust mass and size distributions, the shape and intensity of the FUV extinction is independent of the band gap and hence the composition of the a-C(:H) dust component. This behaviour arises because the FUV optical properties, which are dominated by the $\sigma-\sigma^\star$ band of small a-C(:H) particles, are material independent. The FUV extinction shown in Fig.~\ref{fig_dust_results_2} is then qualitatively consistent with all of the plotted a-C(:H)  band gap materials ($E_{\rm g} = 0.1-2.6$\,eV). Thus, it appears that the FUV extinction could be a direct tracer of the small particle dust mass. \cite{2012ApJ...760...36P} find a gradual decrease in the FUV extinction with decreasing density, which they interpret as indicating the disruption of small grains in the diffuse ISM.  

In the lower plot in Fig.~\ref{fig_dust_results_2} we show how the extinction is affected by varying the power-law index, $\alpha$, of the small a-C particles (see Fig. \ref{fig_dust_dist}). As $\alpha$ decreases the FUV extinction flattens because of the increasing loss of small particles from the size distribution. The middle plot in Fig.~\ref{fig_dust_results_2} shows the effects of varying the minimum particle size, $a_-$, which must be in the range $0.4-1$\,nm (for $\alpha = 5$) in order to obtain a satisfactory fit to the FUV extinction.\footnote{However, as we show in Section \ref{sect_MIRmm_emission}, increasing $a_-$ leads to a progressive loss, and eventual disappearance, of the IR emission bands.}  \ We find that the form of $6-8\,\mu$m$^{-1}$ ($8-11\,\mu$m$^{-1}$) FUV extinction in the diffuse ISM predicted by our model does not vary as long as $a_- < 5$\,nm ($a_- \lesssim 3$\,nm) as indicated in Fig. \ref{fig_dust_results_2} (middle).  For $a_- > 3$\,nm the FUV extinction turns over for $\lambda^{-1} > 8\,\mu$m$^{-1}$ and for $a_- > 5$\,nm it is flatter in the $\lambda^{-1} = 6-8\,\mu$m$^{-1}$ region. 

The form of the FUV extinction and the observed ``tight'' FUV rise-intercept correlation \citep{2004ASPC..309...33F,2007ApJ...663..320F} seems to imply a ``lever effect'' with a ``pivot point'' at FUV wavelengths. Consistent with this  constraint, we note that the optEC$_{(s)}$(a) $Q_{\rm abs}$ data, for particles smaller than 10\,nm and for all band gap materials, have a ``pivot point'' in the $\approx 7.0-7.5$\,eV region, for small a-C(:H) dust size distributions with $\alpha \geq 3$ (Fig.~\ref{fig_dust_results_2} here, and  Figs. 7 and D.2 to D5 in paper III).

We conclude that the variations in the FUV extinction, predicted by our model, are consistent with the FUV extinction characteristics determined by \cite{1983ApJ...272..563G} and the range of variations noted by \cite{2009ApJ...699.1209F}.

\subsubsection{The 217\,nm UV bump}
\label{sect_UV_bump}

The size dependence inherent in the optEC$_{\rm (s)}$(a) data  naturally leads to a UV bump for small ($a \lesssim 1$\,nm), narrow band gap ($E_{\rm g} = -0.1$ to 0.5\,eV), a-C particles, as can clearly be seen in  
Fig.~\ref{fig_DustEM_output} and the extinction plots in Figs.~\ref{fig_dust_results_1}  and \ref{fig_dust_results_2}. 
This occurs because the UV bump in a-C is carried by a restricted-range of aromatic (or PAH-like) clusters with 1-3 rings.   In large particles ($a \geqslant 10$\,nm) the feature is very broad and peaks in the $\lambda^{-1} < 4\,\mu$m region. We emphasise that the aromatic clusters intrinsic to a-C(:H) materials are not strictly PAHs because they form a cohesive part of a much larger network structure, within which they are bound to other clusters by aliphatic and olefinic bridging groups \citep{2012cA&A...542A..98J,2012ApJ...761...35M}.

It is clear that the derived UV bump at $\approx 4.7\,\mu$m$^{-1}$ is slightly too large with respect to the observed interstellar feature ({\it e.g.}, see Figs.~\ref{fig_DustEM_output}, \ref{fig_dust_results_1} and \ref{fig_dust_results_2}). For varying band gap, but fixed dust mass,  the predicted UV bump position is practically unvarying, which is in agreement with the immovable position of the observed bump \citep[{\it e.g.},][]{2004ASPC..309...33F,2007ApJ...663..320F}. However, it does shift to positions as short as $\approx 4\,\mu$m$^{-1}$, compared to the observed range of $\simeq 4.58-4.60\,\mu$m$^{-1}$ \citep{2007ApJ...663..320F}, but only in the case of extreme size distributions, {\it i.e.}, for large $a_-$ or small-$\alpha$ distributions (see Fig.~\ref{fig_dust_results_2}, middle and lower panels, and Section \ref{sect_dust_variations}). We have investigated the width and postion of the UV bump extinction as a function of $a_-$, as predicted by our diffuse ISM dust model (Fig. \ref{fig_DustEM_output}), and find no significant variations as long as $a_- < 1$\,nm.  For $a_- > 1$\,nm the UV bump is broader and the peak shifts to longer wavelengths (see Fig \ref{fig_dust_results_2}, middle). We then note that, in the absence of extreme size distribution variations,  it is encouraging that the predicted variations in the UV bump position and width are rather small, in accordance with observations.

\cite{2004ASPC..309...33F} and \cite{2007ApJ...663..320F} have undertaken a detailed study of the UV extinction bump and its variations as seen in the $E(B-V)$-normalised extinction data. They find the UV bump peak positions and widths are uncorrelated, broader bumps for greater FUV curvature,  stronger bumps for intermediate levels of FUV extinction ({\it i.e}, weaker for high or low FUV extinction), stronger FUV extinction as the bump weakens, and 
weaker bumps for  low or high values of $R_{\rm V}$.  Using the optEC$_{(s)}$(a) data to model the UV bump we find that the peak position and width are not correlated.  We also find broader bumps for larger band gap materials that exhibit more curvature in their FUV extinction and that the FUV extinction strengthens and the bump weakens with increasing $E_{\rm g}$ (see Fig.~\ref{fig_dust_results_1}).  The weaker bump associated with low FUV extinction could be explained by a `flatter', {\it i.e.}, high $R_{\rm V}$, size distribution, see Figs.~\ref{fig_dust_results_1} and \ref{fig_dust_results_2}. Conversely, a weak UV bump for high  FUV extinction is consistent with large band gap a-C:H materials, which show basically no UV bump (Fig.~\ref{fig_dust_results_1} and \ref{fig_dust_results_2}). 

In the un-normalised extinction plots in Fig.~\ref{fig_dust_results_2} (upper and middle panels) we can see that the FUV extinction shows little or no variation with $E_{\rm g}$ or $a_-$ but does depend, rather sensitively, on the assumed power law, $\alpha$, for the small a-C grains (Fig.~\ref{fig_dust_results_2}, lower panel). 
Thus, the observed UV extinction variations seen in the $E(\lambda-V)/E(B-V)$ plot (Fig.~\ref{fig_dust_results_1}), can be explained by variations in the visible extinction slope (Fig.~\ref{fig_dust_results_2}, upper panel), which biases the $E(B-V)$-normalised data. The visible extinction and the UV bump are determined by the a-C(:H) composition and, for fixed dust mass, both increase in strength as $E_{\rm g}$ decreases, which leads to an anti-correlation between the UV bump strength and the FUV continuum in $E(B-V)$-normalised data \citep[{\it e.g.},][]{2004ASPC..309...33F}. Further, the range of FUV extinction variations predicted by the model are in very good agreement with the observed FUV extinction dispersion seen in the diffuse ISM \citep{2004ASPC..309...33F,2007ApJ...663..320F}. 

For our standard model parameter set (Table \ref{table_model_params}) the UV bump and the FUV extinction rise are due to a-C particles with  $< 250$ C atoms ($\equiv a < 1$\,nm) and $< 7 \times 10^3$ C atoms ($\equiv a < 3$\,nm), respectively.  This could then explain why the UV bump and FUV extinction are partially de-coupled  \citep[{\it e.g.},][]{1983ApJ...272..563G,2004ASPC..309...33F,2007ApJ...663..320F}. As shown in Sections \ref{sect_constraints}, \ref{sect_FUV_ext} and \ref{sect_UV_bump} the UV bump and FUV extinction produced by our model depend on the adopted minimum size for the a-C grains, $a_-$, and their power law index, $\alpha$ (see Fig.~\ref{fig_dust_results_2} middle and lower panels). The upper panel of Fig.~\ref{fig_dust_results_2} shows that the UV bump intensity is sensitive to the material band gap. However, given that the aromatisation time-scale in the diffuse ISM seems to be short \citep[{\it e.g.}, $\simeq 10^5/G_0$\,yr,][]{2012dA&A...545C...2J} most small a-C(:H) particles with radii $\lesssim 20$\,nm will be a-C. The partial de-coupling of the UV bump and FUV extinction in the diffuse ISM would then be due to variations in the a-C dust mass for $a \leqslant 1$\,nm (UV bump carriers) with respect to somewhat larger particles ($a \leqslant 3$\,nm, the FUV extinction carriers),  as shown in Fig. \ref{fig_dust_results_2}. In particular, we propose that the diffuse ISM FUV {\em extinction is principally characterised by variations in} $\alpha$ and that the IR-MIR SED {\em is determined by the minimum particle size}, $a_-$, and  that both are a function of the local ISRF intensity and hardness \citep[{\it e.g.},][]{1994ApJ...420..307J,2012A&A...545A.124B}. This is in agreement with the conclusions of the astronomical PAH model, {\it i.e.}, that the IR emission band profiles are sensitive to the number of C atoms in the smallest PAHs \citep{2001A&A...372..981V}. 

In conclusion, it is evident that the UV extinction bump, which is reasonably well fit with $0-0.22$\,eV band gap materials, appears to provide a stronger constraint on the composition of the small a-C dust particles than does the rather invariant FUV extinction rise. However, as noted by \cite{2012ApJ...760...36P} the intensity of the UV bump extinction appears to be essentially uncorrelated with the degree of carbon depletion into dust, (C/H)$_{\rm dust}$.

\subsubsection{Visible-NIR extinction and variations}

\begin{figure} 
\resizebox{\hsize}{!}{\includegraphics{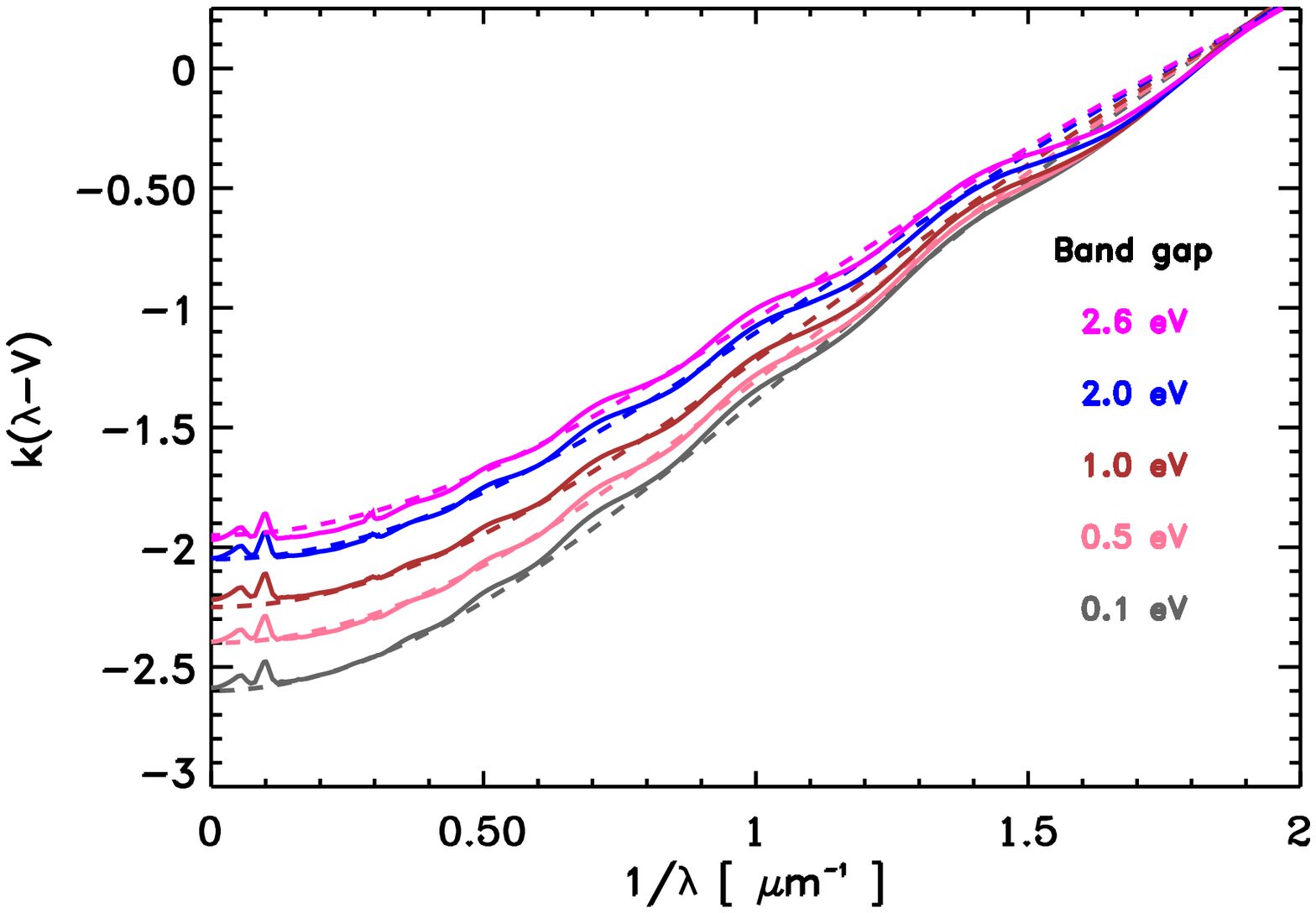}} 
  \caption{The NIR extinction predicted by the model (solid lines) and the parameterised fits \citep[as per][]{2009ApJ...699.1209F} for varying band gap, $E_{\rm g}$ [eV] (dashed lines), for a column density $N_{\rm H} = 5 \times 10^{21}$\,cm$^{-2}$. The model data colour-coding is as per Fig.~\ref{fig_dust_results_2}. The sinusoidal structure is due to the adopted narrow silicate size range and will not be present in real, irregular and non-spherical particles.}
 \label{fig_NIR_ext0}
\end{figure}
\begin{figure} 
\resizebox{\hsize}{!}{\includegraphics{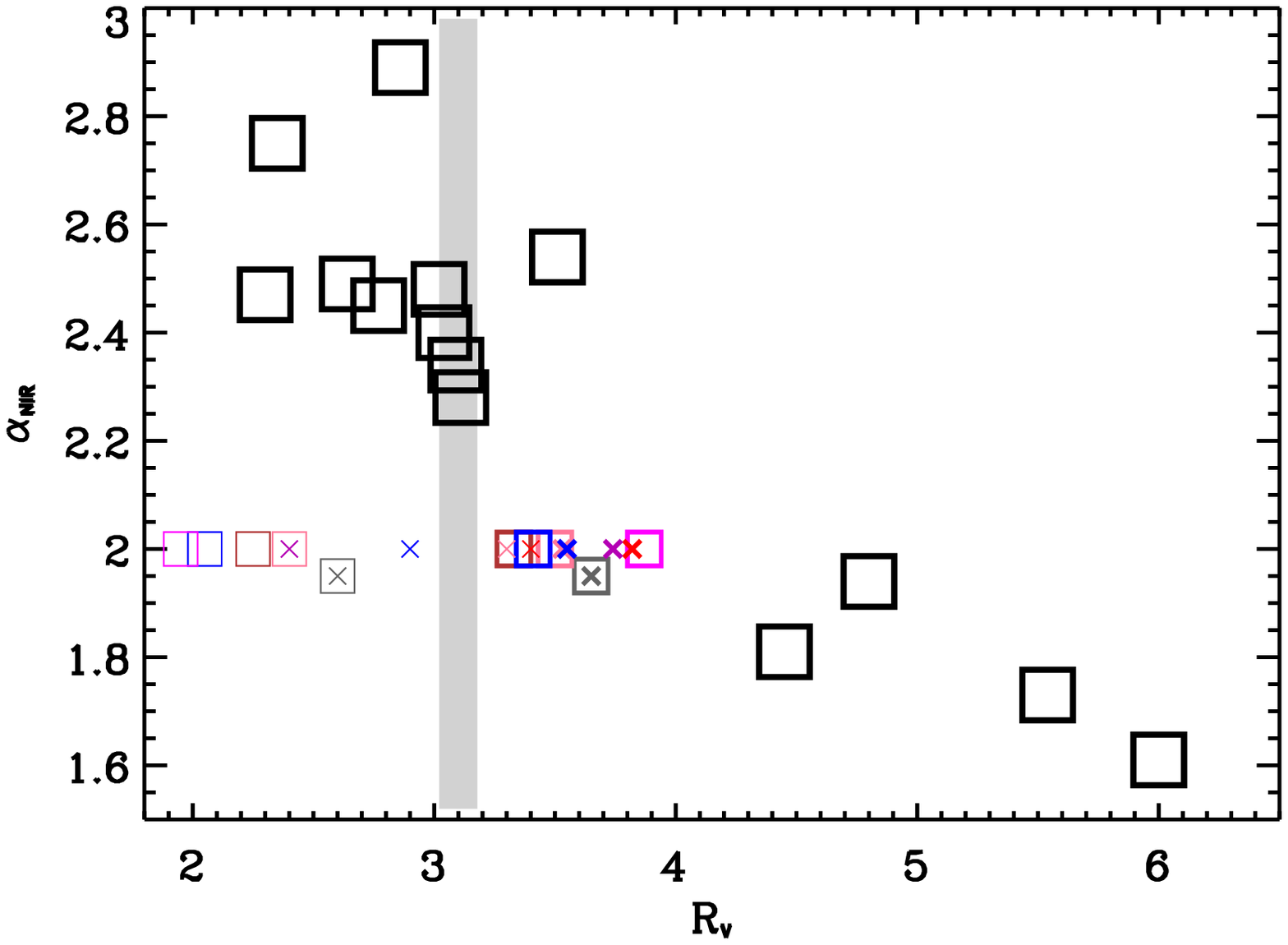}} 
  \caption{The \cite{2009ApJ...699.1209F} NIR extinction parameter $\alpha_{\rm NIR}$ as a function of $R_V$ (large black squares), see Eq. (\ref{eq_NIRext_fit}). The fit parameters for our model as a function of $E_{\rm g}$ ($\alpha$) are shown by the coloured squares (crosses), the colour-coding as per Fig.~\ref{fig_dust_results_2}. The thick squares use the model $R_V$ and the thin squares the best fit $R_V$ values from Eq. (\ref{eq_NIRext_fit}), shown as the dashed lines in Fig. \ref{fig_NIR_ext0}. The vertical grey band shows the typical diffuse ISM value of $R_V$.}
 \label{fig_NIR_ext1}
\end{figure}

The extinction data shown in Fig.~\ref{fig_dust_results_2} indicate that, for fixed dust mass, this model predicts a significant dispersion in the $B$ and $V$ band extinction, which is due to variations in the dust composition ({\it i.e.}, decreasing $(B-V)$ as $E_{\rm g}$ increases) or its size distribution ({\it i.e.}, increasing $(B-V)$ as $\alpha$ decreases or $a_-$ increases). Unfortunately, it is exactly these interesting variations that are `damped out' by the  normalisation of the observed extinction by $E(B-V)$. Thus, based on the above it  appears that normalising the interstellar extinction in the FUV at  $\simeq 7\,\mu$m$^{-1}$ ($\equiv 143$\,nm) would render the data in much more diagnostically-useful manner. 

\cite{2004ASPC..309...33F}, \cite{2009ApJ...699.1209F} and \cite{2011ApJ...737...73F} note that the IR extinction longward of $1\,\mu$m generally has a power-law type of behaviour. As Fig.~\ref{fig_NIR_ext0} demonstrates, the range of a-C(:H) grain material optical properties, with $E_{\rm g} \simeq 0.0-2.6$\,eV ($\equiv X_{\rm H} = 0.0-0.6$), in combination with the adopted amorphous silicate, are consistent with this important observational constraint.  

The NIR extinction can be fit with a two-parameter model, where the free parameters are a power law, $\alpha_{\rm NIR}$, and $R_{\rm V}$, {\it i.e.},
\begin{equation}
k(\lambda-V) = \left[ \frac{0.349+2.087R_{\rm V}}{1 + (\lambda/0.507)^{\alpha_{\rm NIR}}} \right] - R_{\rm V}, 
\label{eq_NIRext_fit}
\end{equation}
as determined by \cite{2009ApJ...699.1209F}. As they noted, and as we also find with our model, the good fits to the data extend out to shorter and longer wavelengths than those used by \cite{2009ApJ...699.1209F} in deriving their NIR extinction law (Fig.~\ref{fig_NIR_ext0}). In fitting the observations \cite{2009ApJ...699.1209F} show that $\alpha_{\rm NIR}$ generally decreases as $R_{\rm V}$  increases. However, for our dust model we find good fits with $\alpha_{\rm NIR} \simeq 2$ for all $R_{\rm V}$. In Fig.~\ref{fig_NIR_ext1} we show $\alpha_{\rm NIR}$ as a function of $R_{\rm V}$, as per \cite{2009ApJ...699.1209F}, and find that the lower $R_{\rm V}$ values derived for our model span the same range as those for the \cite{2009ApJ...699.1209F} fits to their data. Discrepancies between the model and observational data are also apparent in Fig. \ref{fig_DustEM_output} where the model data show slight curvature that is not apparent in the observations. This indicates that the model does not yet include all of the elements necessary to explain the exact form of the NIR extinction. This could be related to an underestimate of the absorption in the $5-10\,\mu$m region compared to observations \citep[{\it e.g.},][]{1996A&A...315L.269L}. The discrepancy could also be due to grain charge effects, which we do not take into account here. For example, aromatic cation species show NIR absorption in the $\sim 0.8$ to $> 2\,\mu$m region \citep{2008ApJ...680.1243M}, which is not apparent in astronomical PAH models.  

We also note that the tail,  extending to high $R_{\rm V}$ and low $\alpha_{\rm NIR}$ in Fig.~\ref{fig_NIR_ext1}, cannot be explained by our model. This is probably because the dust along these lines of sight has undergone significant evolution (accretion and coagulation) and thus the constant mass assumption that we adopt here is no longer valid. 

The NIR-MIR optical properties of a-C(:H) materials can also be fit with a two free-parameter Urbach tail model (see paper II). However, in the optEC$_{\rm (s)}$(a) data the observed wavelength-dependent behaviour arises naturally from the physics and no empirical wavelength-dependence is imposed. 

The IR extinction is generally rather invariant but we conclude that variations will arise with compositional and size distribution variations in the ISM, for example, due to the accretion of wide band gap a-C:H material, and also due to particle coagulation in dense regions and erosion in energetic regions. We have yet to fully explore this aspect of our dust model and leave this to targeted studies of particularly well-studied lines of sight.

\subsubsection{The IR absorption bands}

\begin{figure} 
\resizebox{\hsize}{!}{\includegraphics{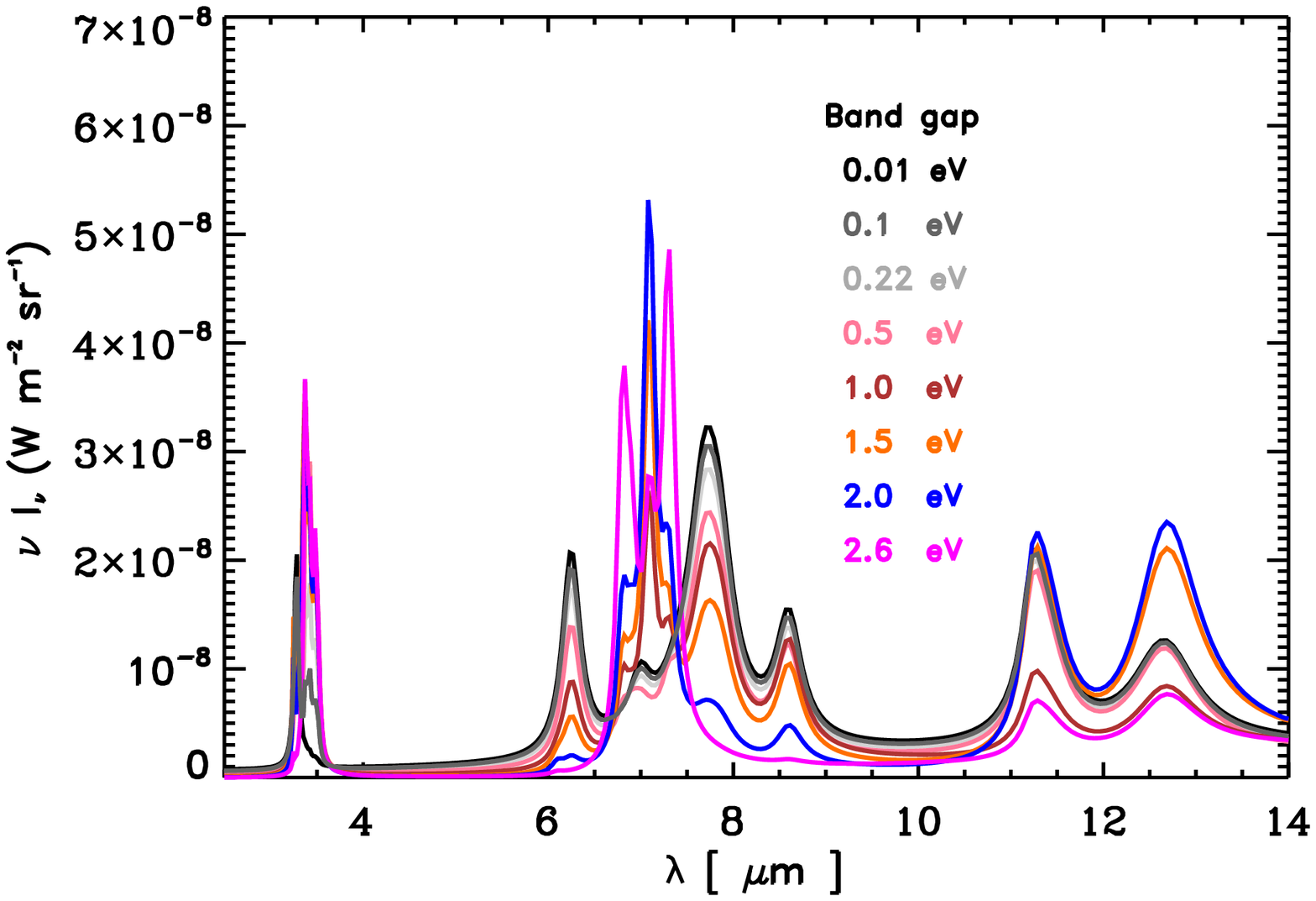}}
\resizebox{\hsize}{!}{\includegraphics{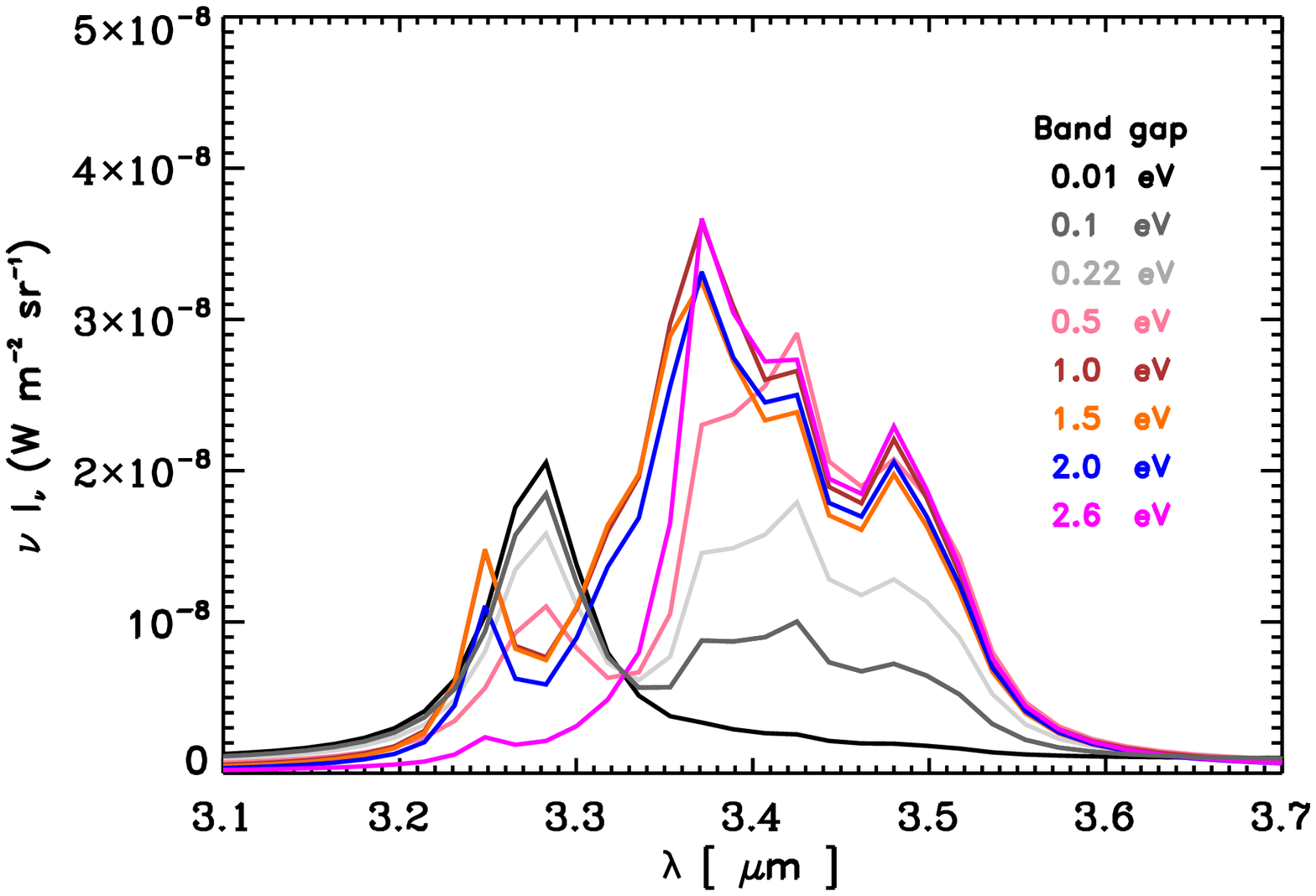}}  
  \caption{The dust model $3-14\,\mu$m and $3.1-3.7\,\mu$m emission band spectra as a function of the a-C(:H) material band gap, $E_{\rm g}$ [eV].}
 \label{fig_dust_results_5}
\end{figure}

As discussed in detail in papers I to III, the optEC$_{\rm (s)}$(a) data is consistent with the profile of the  absorption bands observed towards the Galactic Centre in the $3-4\,\mu$m wavelength region. The predicted absorption band profiles (see Fig.~\ref{fig_dust_3p4_ext}) are similar to those calculated in emission, as shown in Fig.~\ref{fig_dust_results_5}. Using the optEC$_{\rm (s)}$(a) data it is possible to explain, in detail, all of the structure in the observed absorption bands with rather wide band gap ($E_{\rm g} \gtrsim 2.3$\,eV), a-C:H carbonaceous grains with radii $\gtrsim 20$\,nm. As pointed out in Section 5.1 in paper III, a-C nano-particles, with band gaps as low as 0.5\,eV, also appear to have $3-4\,\mu$m IR absorption spectra which match the observed absorption features. However, such small particles are expected to be rapidly UV-photolysed in the ISM and so the good match may simply be fortuitous. 

It is interesting to note that the nature of the bands in the $3-4\,\mu$m region can be used to constrain the emission properties at FIR-mm wavelengths because of the strong optical property coupling at all wavelengths. For example, and as noted above, the absorption band profiles in the $3-4\,\mu$m region indicate the presence of a significant H-rich, wide band gap ($E_{\rm g} \gtrsim 2.3$\,eV), a-C:H dust mass, which ought to be apparent in the long wavelength emission behaviour but this is not the case (see the following section for an explanation).  

%

\subsubsection{The IR emission bands}
\label{sect_IR_mission_bands}

In Fig.~\ref{fig_dust_results_5} we show the IR emission band spectrum in the $2.5-14\,\mu$m (upper panel) and $3.1-3.7\,\mu$m (lower panel) wavelength regions. It is clear, from the lower plot in Fig.~\ref{fig_dust_results_5}, that a-C materials with band gaps in the range $0.1-0.5$\,eV exhibit a 3.3\,$\mu$m aromatic emission band that is always accompanied by a side-band at $\sim 3.4\,\mu$m with a shoulder at $\sim 3.5\,\mu$m. As pointed out in Section 5.1 in paper III, this kind of band structure, as predicted by the optEC$_{\rm (s)}$(a) data, is consistent with the observed emission bands in many regions, {\it e.g.}, across the Orion bar \citep{1997ApJ...474..735S,2001A&A...372..981V}, M17-SW, NGC\,2023 \citep{2001A&A...372..981V} and in M 82 \citep{2012A&A...541A..10Y}, where {\em both} aromatic {\em and} aliphatic CH bands appear in the $3-4\,\mu$m spectral region, even in these energetic environments. 

Our new dust modelling approach, using a-C(:H) optical property data, therefore offers an alternative view to the origin of observed IR emission bands, which are well-matched by astronomical PAH, thermal emission profile models  \citep[{\it e.g.},][]{2001A&A...372..981V,2002A&A...388..639P}. However, in our model it is small ($a \lesssim 0.5$\,nm), 3D, a-C particles that are the carriers of the observed IR emission bands. As proposed in paper III, these $30-40$ C atom, aromatic/aliphatic, cage-like particles \citep{2012ApJ...761...35M}, that we call ``arophatics'', represent the ``end of the road" for a-C(:H) grain evolution because their disintegration products, {\it viz.}, small free-flying, PAH-like species and short hydrocarbon chains, will quickly be destroyed in the diffuse ISM \citep[{\it e.g.},][]{1994ApJ...420..307J,2010A&A...510A..36M,2010A&A...510A..37M,2011A&A...526A..52M,2012A&A...545A.124B}.  The IR emission band-carrying a-C particles are also responsible for the UV bump and FUV extinction, which is consistent with their limited range of variation and the constancy of the IR emission band profiles. Thus, these ``end of the road"-particles, with a narrow range of optical properties, could explain the noted ``stability'' of the FUV extinction carriers in the diffuse ISM \citep{1983ApJ...272..563G}. As shown above, in Section \ref{sect_UV_bump}, a partial de-correlation of the UV extinction components (bump and FUV) and the IR emission bands is predicted by our model. 

The thermal emission from the wide band gap materials observed in absorption (see above) and the low band gap, UV photolysed a-C material component observed in emission should both be evident in the FIR-mm dust emission. However, we find that the predicted dust emissivity at these wavelengths is dominated by the photolysed outer surfaces of the a-C:H/a-C particles, {\it i.e.}, the narrow band gap a-C material ($E_{\rm g} \simeq 0.1 - 0.25$\,eV), that dominates the emission because its emissivity at mm wavelengths is more than four orders of magnitude greater than that for wide band gap a-C(:H) materials ({\it e.g.}, see Fig. 16 in paper II and Fig. 10 in paper III). Thus, and although there may be a significant mass in a-C:H carbonaceous dust, these wide band gap materials are apparently hard to detect and thus there significance has probably been underestimated.

\subsubsection{MIR-FIR-mm emission}
\label{sect_MIRmm_emission}

%
\begin{figure} 
\resizebox{\hsize}{!}{\includegraphics{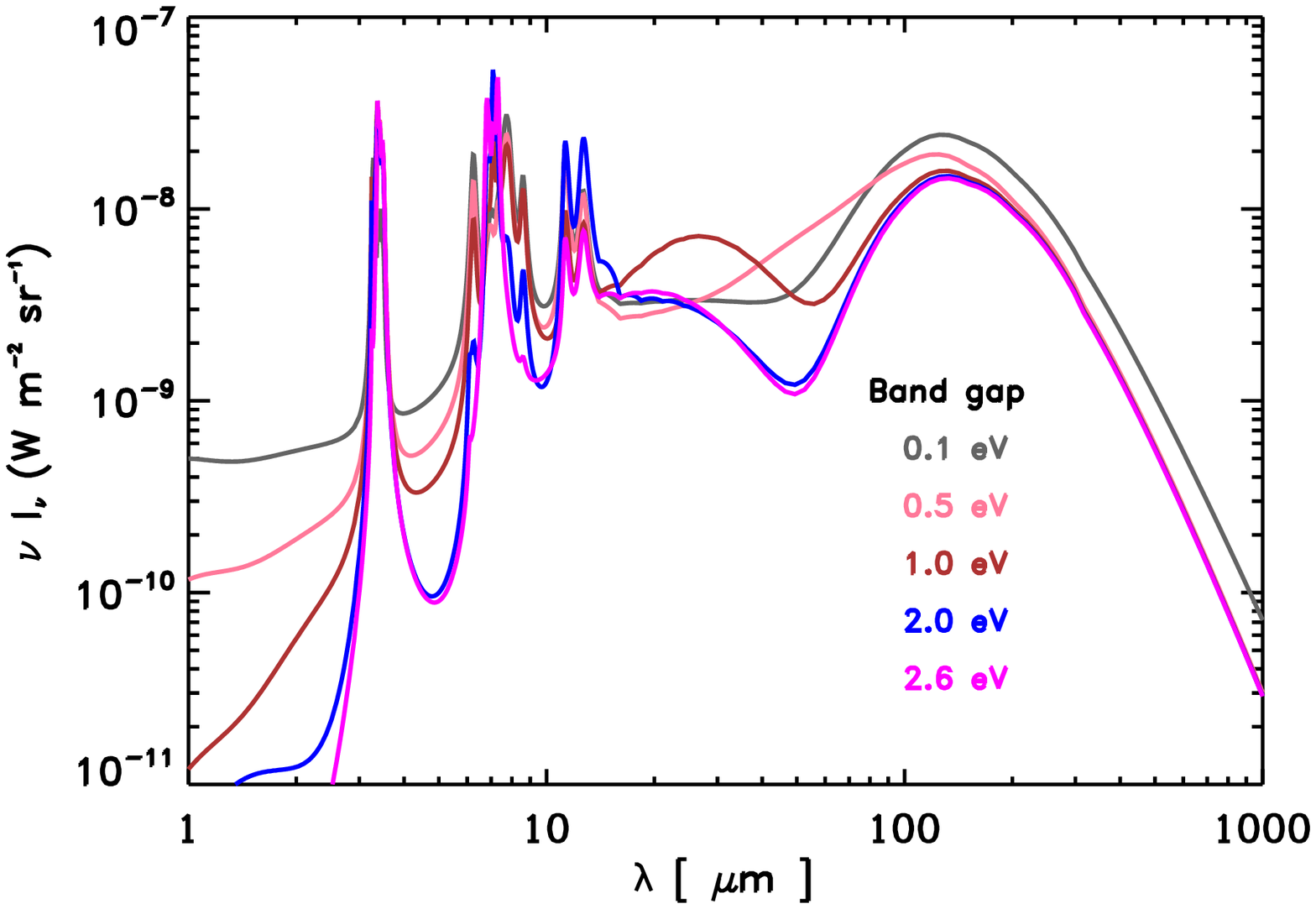}}   
\resizebox{\hsize}{!}{\includegraphics{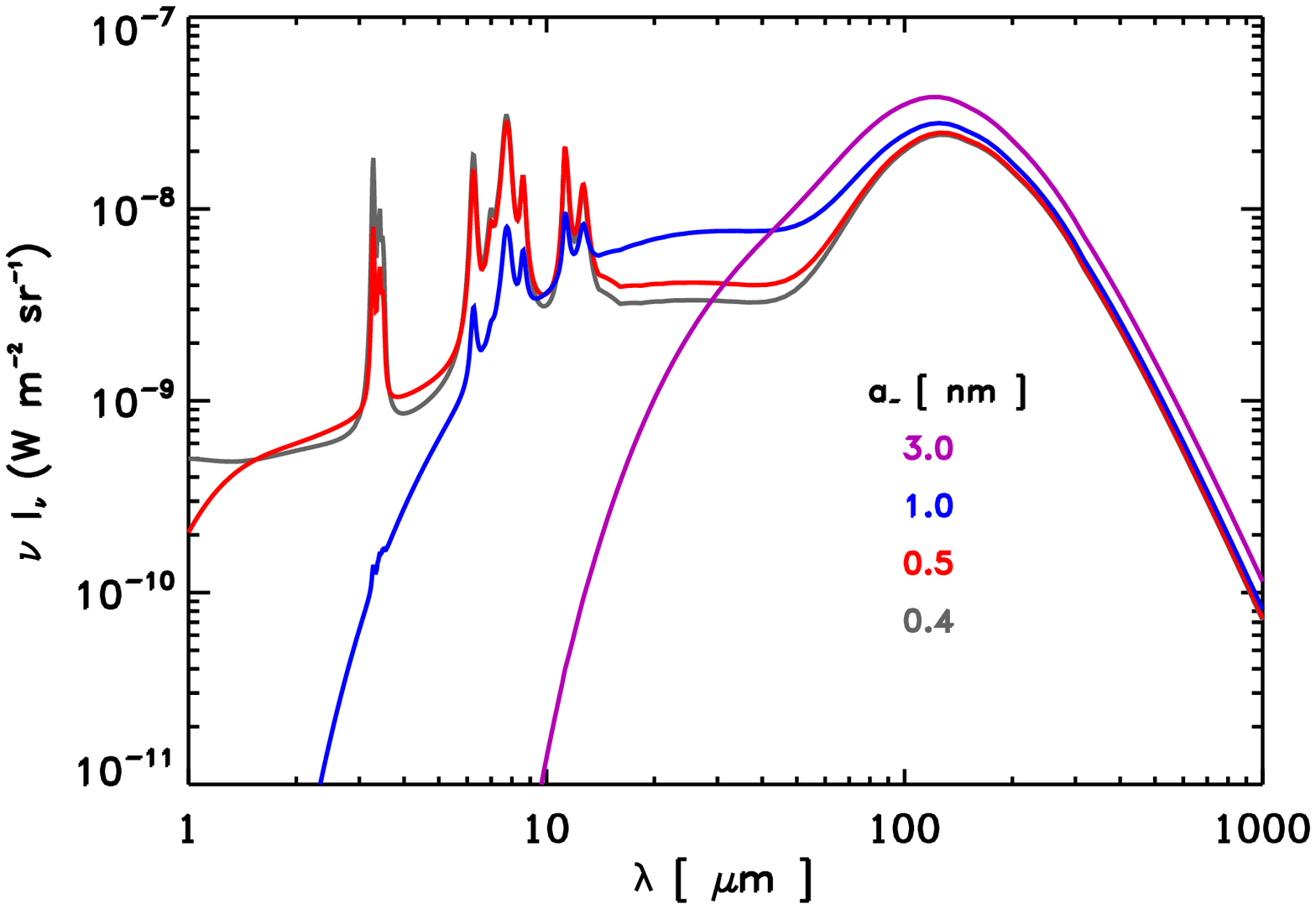}} 
\resizebox{\hsize}{!}{\includegraphics{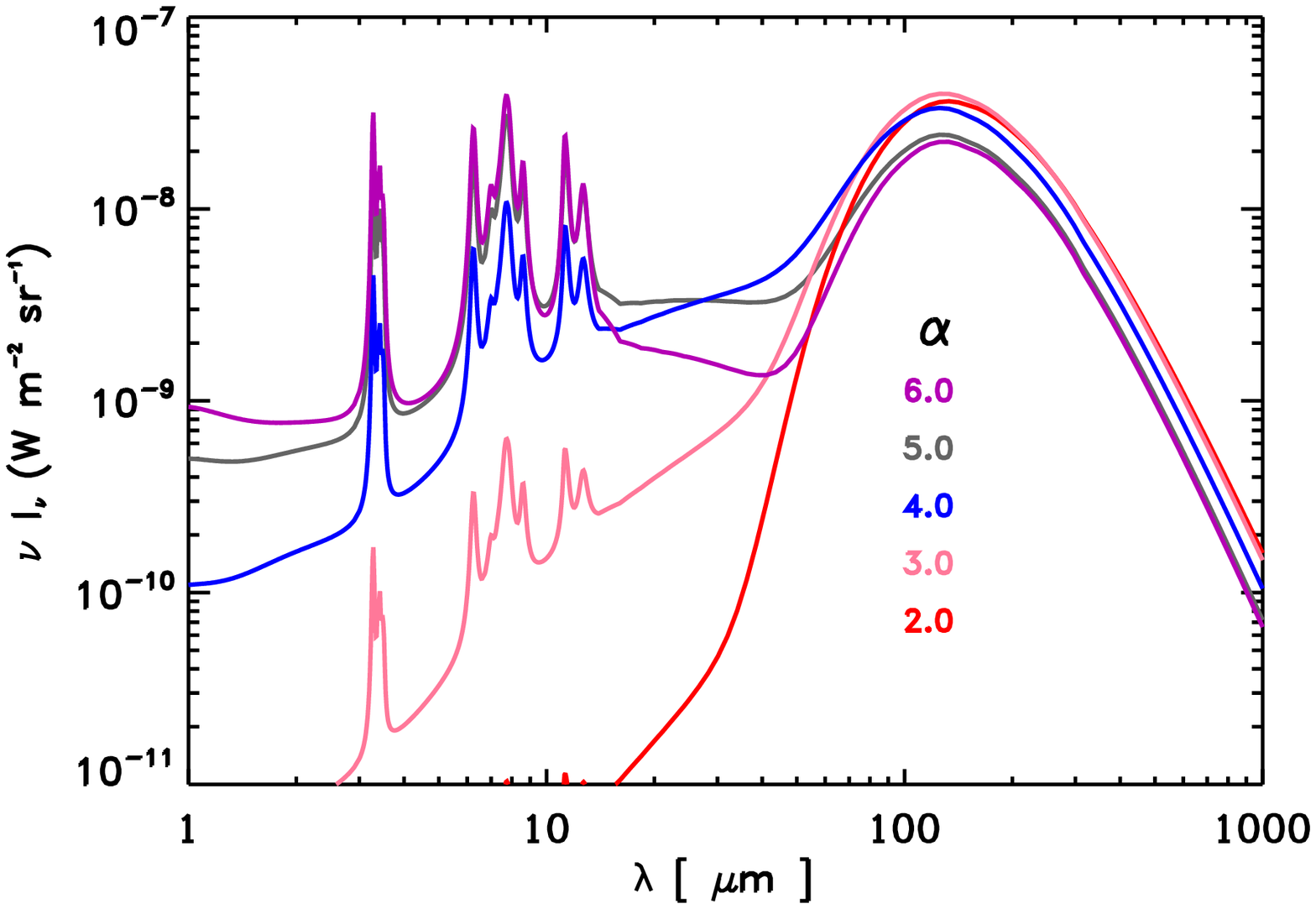}} 
  \caption{The dust model SEDs, for fixed dust mass, as a function of $E_{\rm g}$ (upper panel), $a_-$ (middle panel)  and $\alpha$ (lower panel). The line designations are as per Fig.~\ref{fig_dust_results_2}.}
 \label{fig_dust_results_4}
\end{figure}

In Fig.~\ref{fig_dust_results_4} we show the dust SED as a function of the band gap, $E_{\rm g}$ (upper), of the photo-processed dust components, {\it i.e.}, the small a-C(:H) grains and the outer layer/mantle of the large a-C:H grains. Fig.~\ref{fig_dust_results_4} also shows the SEDs as a function of the minimum grain size, $a_-$ (middle), and the small a-C(:H) grain size distribution power-law index, $\alpha$ (lower). As the band gap increases we note a decrease in the emission at FIR-mm wavelengths, a significant change in the MIR emission to a peak in the $20-30\,\mu$m region and a decrease in the continuum underlying the shortest wavelength emission bands. The decrease in the long wavelength emission is due to the fact that wide band gap a-C:H grains are less emissive, and therefore hotter, and their emission peaks at MIR rather than FIR wavelengths, which explains the broad $20-30\,\mu$m emission peak for $E_{\rm g} = 1.0$\,eV.

As the small a-C grain size distribution $\alpha$ decreases the FIR-mm emission increases and the IR emission bands and MIR emission decrease substantially and eventually disappear. Decreasing $\alpha$ is equivalent to a `flattening' and `narrowing' of the size distribution, which is equivalent to a transfer of mass to the larger particles. It is this that is responsible for the increase in the FIR-mm emission (see Figs.~\ref{fig_dust_dist} and \ref{fig_dust_results_4}), while the loss of the IR emission bands and MIR continuum is simply due to the reduced abundance of their carriers, {\it i.e.}, grains with $a \lesssim 10$\,nm (see Fig.~\ref{fig_dust_dist}). 

%
\begin{figure} 
\resizebox{\hsize}{!}{\includegraphics{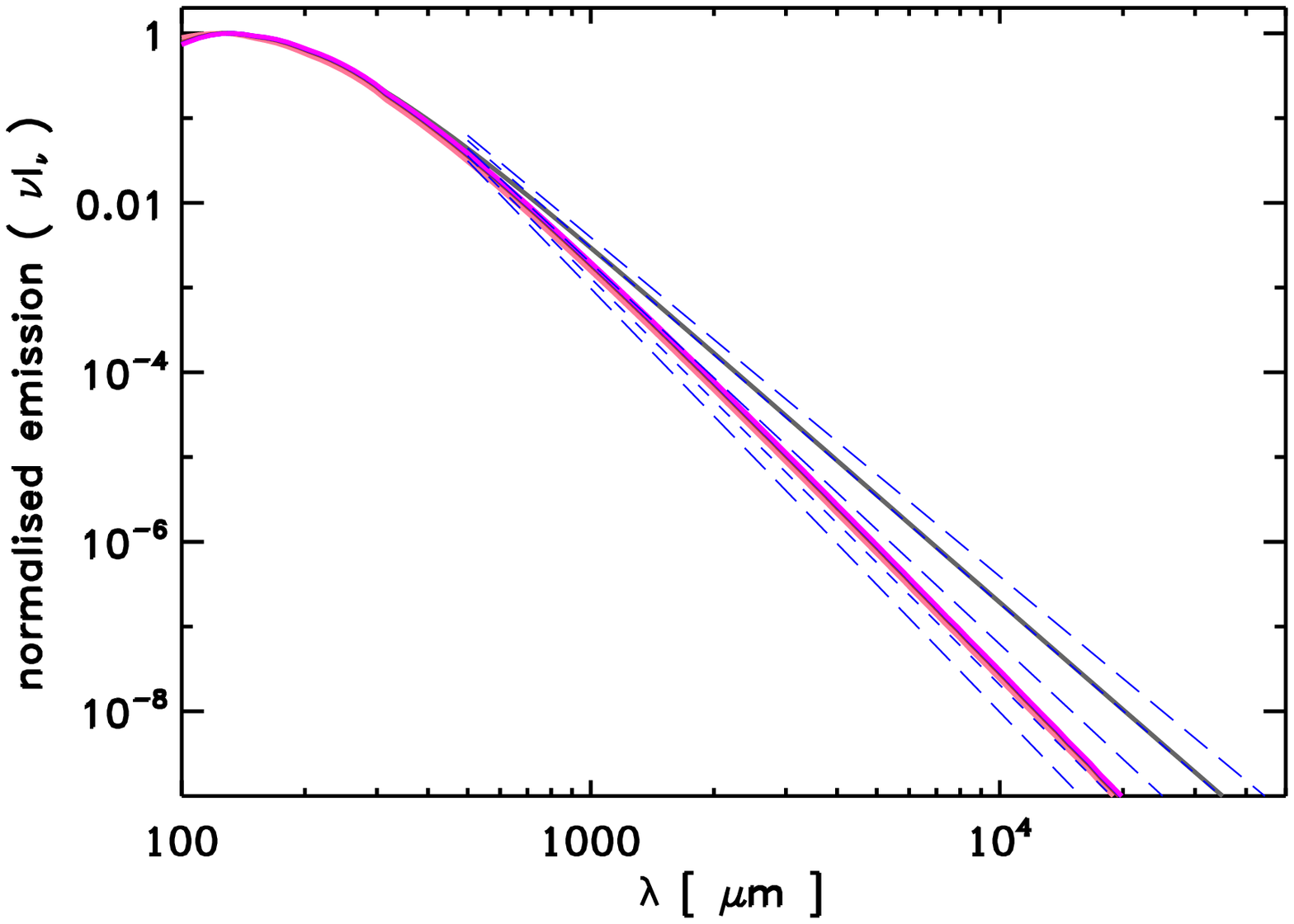}}   
  \caption{The normalised FIR-cm SEDs show that the emissivity slopes segregate into two band-gap groups, one with $\beta \sim 1.3$ for $E_{\rm g} \sim 0$\,eV and the other with $\beta \sim 2$ for $E_{\rm g} \gtrsim 0.5$\,eV. 
  The line designations are as per Fig.~\ref{fig_dust_results_2}. 
  The dashed blue lines indicate slopes of -1.0, -1.3, -1.5, -1.8 and -2.0, from top to bottom.}
 \label{fig_dust_results_4a}
\end{figure}

Fig.~\ref{fig_dust_results_4a} shows a zoom into the FIR-cm wavelength region, where the model SEDs have been normalised to the peak of the emission to allow a direct comparison of the slopes of the emissivity.  This figure shows that the emissivity falls into two well-separated band-gap groups, in effect a two-state system, one with $E_{\rm g}\gtrsim 0.5$\,eV and $\beta \simeq 2$ that `switches' to an $E_{\rm g}\simeq 0$\,eV state with $\beta \simeq 1.3$ as a result of EUV photo-processing. 

\begin{figure} 
 \resizebox{\hsize}{!}{\includegraphics{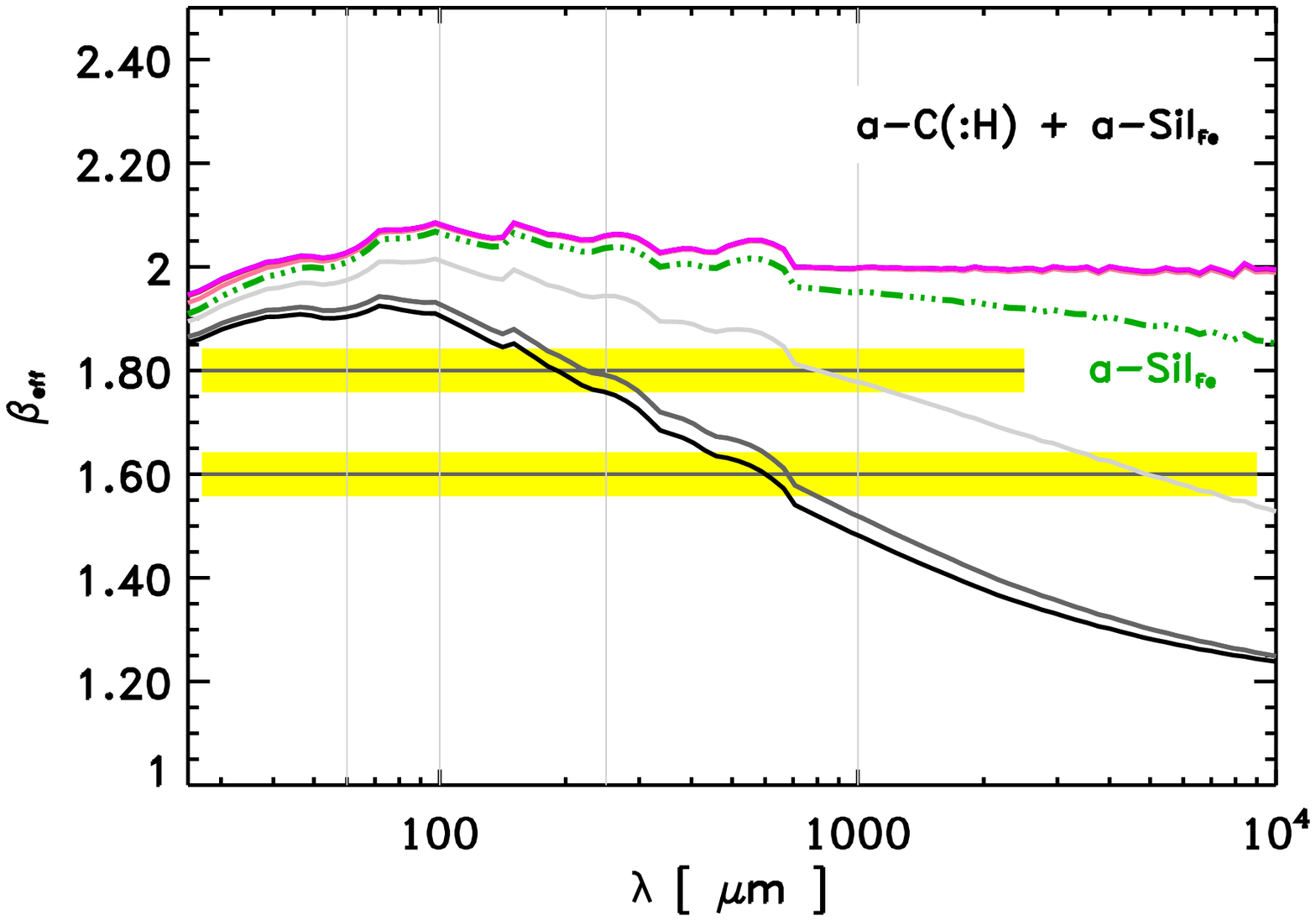}} 
 \vspace*{0.5cm}
 \resizebox{\hsize}{!}{\includegraphics{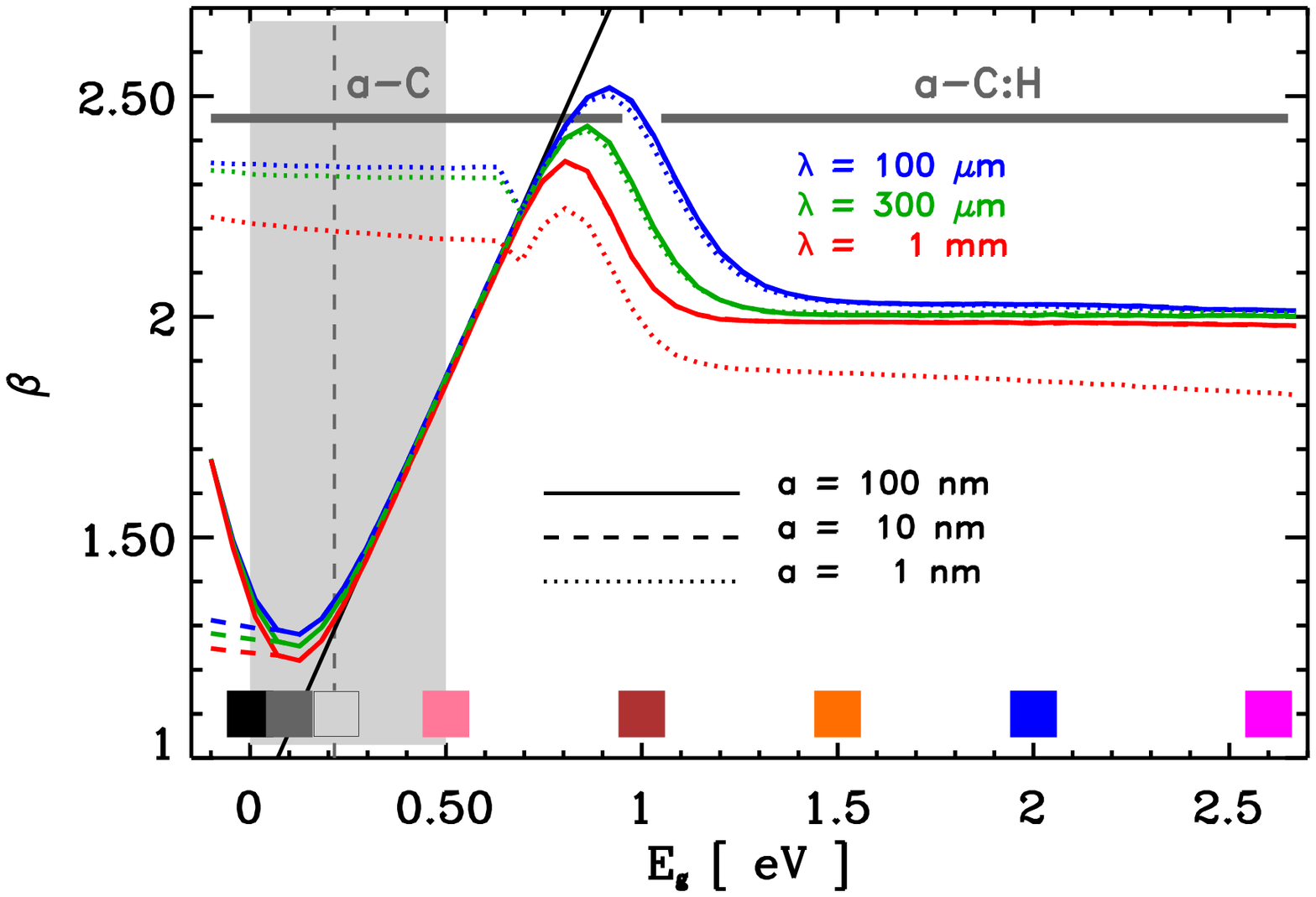}}
  \caption{{\em Upper plot:} $\beta_{\rm eff}$ {\it vs.} $\lambda$, as a function of $E_{\rm g}$, where $\beta_{\rm eff}$ is the effective emissivity slope resulting from a combination of the a-Sil$_{\rm Fe}$/a-C and a-C:H/a-C dust population optical properties. The green triple dot-dashed line shows the data for only the a-Sil$_{\rm Fe}$/a-C grains, which do not depend on the band gap of the thin a-C(:H) coating. The $\lambda$-dependence and variations are therefore only due to the a-C(:H) dust properties. The line colour-coding is as per Fig.~\ref{fig_dust_results_5}. The horizontal grey lines and wide yellow bands show $\beta_{\rm eff} = 1.80 \pm 0.05$ and $1.60 \pm 0.05$. 
  \newline {\em Lower plot:} The emissivity slope, $\beta$, for a-C(:H) dust, for  $\lambda = 100\,\mu$m, 300\,$\mu$m, and 1\,mm, as a function of $E_{\rm g}$, for $ a= 100$, 10 and 1\,nm.  The vertical grey dashed line shows the experimentally-derived `block' on band gap evolution (see paper II). The grey area indicates the range of a-C materials that are consistent with many diffuse ISM dust properties (Section \ref{sect_dust_ext_sed}). The coloured squares at the bottom indicate the band gap colour coding scheme.}
 \label{fig_dust_beta}
\end{figure}

We note that variations in the FIR-mm emissivity slope for a-C(:H) particles are reduced for particles smaller than 3\,nm and that the wavelength-dependence is dominated by the wings of the Drude profiles, {\it i.e.}, $\beta \simeq 2$ for $\lambda \gtrsim 20\,\mu$m (see Fig.~\ref{fig_dust_beta}). However, the emission at these wavelengths is a combination of that from amorphous silicate and carbon grains. Thus, in Fig.~\ref{fig_dust_beta} we show the summed emissivity slopes of all the dust components, $\beta_{\rm eff}$, which is derived directly from the optical properties and is therefore independent of any assumed dust temperature. Fig.~\ref{fig_dust_beta} again reflects the two-state system mentioned above. In detail this figure shows that the switch to the narrow band gap state results in a wavelength-dependent $\beta$ behaviour  {\it i.e.}, $\beta \simeq 1.9$, 1.8, 1.5 and 1.2 at $60-100\,\mu$m, $250\,\mu$m, 1\,mm and 1\,cm, respectively. The EUV processing-driven jump in the optical properties occurs between $E_{\rm g} = 0.1$ and 0.5\,eV and   in Fig.~\ref{fig_dust_beta} we show the results for an intermediate case where $E_{\rm g} = 0.25$\,eV. We note that the optical property jump is not related to the transition between the eRCN and DG models, which occurs at a band gap of $\simeq 1$\,eV.

Fig.~\ref{fig_dust_beta}  shows that the effective emissivity slope, $\beta_{\rm eff}$, varies significantly with wavelength over the FIR-cm range for EUV photo-processed a-C(:H) materials. Thus, an interpretation of the observed interstellar dust emissivity as arising from a single-emissivity-slope material would appear to be too simplistic because it neglects the combination of materials that contribute to the emission at long wavelengths. As Fig.~\ref{fig_dust_model_decomp_std} shows, the FIR emission is dominated by silicates but  at wavelengths longer than a few mm the carbonaceous dust emission is dominant. Carbonaceous and silicate materials make about equal contributions to the emission in the mm region. 

An important prediction of this model arising from the two-state system behaviour is that, at FIR wavelengths ({\it e.g.}, $\approx 60-100\,\mu$m), $\beta_{\rm eff}$ should be  $\sim 2\pm0.1$. However, at longer wavelengths $\beta_{\rm eff}$ must be either $\sim 2$ or $\lesssim 1.8$, with intermediate values  unlikely. In the latter case, $\beta_{\rm eff}$ decreases smoothly from $\sim 1.8$ at $250\,\mu$m to $\sim 1.2-1.3$ at cm wavelengths. 
In general, low (high) values of $\beta_{\rm eff}$ indicate a `mature' aromatic-rich (`young' aliphatic-rich) carbonaceous dust component. We note that for mature materials the $\beta_{\rm eff}$ at $\sim 80-500\,\mu$m wavelengths is dominated by that of the silicate material optical properties, which are not yet well-determined. 
Hence, observed values of $\beta_{\rm eff}$ significantly different from $\sim 2$ at $\lambda \sim 100\,\mu$m (or different from $\sim 1.8$ at $250\,\mu$m) could be an indication that the interstellar silicates are not as we currently model them. 
 
In a future paper we plan to explore the mm-cm dust emission predicted by our model within the context of the latest results from the {\em Planck} mission.

\subsubsection{Band gap evolution and the long wavelength emission}
\label{sect_band_gap_evolution}

Fig.~\ref{fig_dust_beta} shows the band gap and size-dependence of the slope of the a-C(:H) dust emissivity, $\beta$, at $\lambda = 100\,\mu$m, 300\,$\mu$m and 1\,mm. The black line indicates the behaviour of $\beta$ based on the band gap dependence of $\gamma$, the long-wavelength slope of the imaginary part of the refractive index, $k$ (see Eq.~22, paper~II). An upturn in $\beta$, for $E_{\rm g} = -0.1$ to 0.1\,eV, is due to a power-law dependence of $n$ on wavelength and  the peak at $\approx 0.9$\,eV is due to a reduced IR band contribution. Fig.~\ref{fig_dust_beta} shows that a-C(:H), whether in grains or mantles, could have a significant effect on the FIR-mm dust emissivity and that this will be superimposed any effects due to amorphous silicate variations \citep[{\it e.g.},][]{1996ApJ...462.1026A,1998ApJ...496.1058M,2005ApJ...633..272B,2007A&A...468..171M,2011A&A...535A.124C}. 

The data in Fig.~\ref{fig_dust_beta} indicate that in the diffuse ISM large, homogeneous a-C particles ($a \geqslant 10$\,nm, $E_{\rm g} = 0.1-0.2$\,eV) would have a rather flat emissivity slope, $\beta \simeq 1.2-1.3$ (see paper II).  However, particles with radii $> 20$\,nm will exhibit a core material composition that can be different from that of a photo-processed outer mantle (see Footnote \ref{footnote_20nm}). In contrast, small a-C particles ($a \simeq 1$\,nm, $E_{\rm g}({\rm bulk}) = 0.1-0.2$\,eV) will have a steeper emissivity slope, $\beta \simeq 2.2-2.4$. This is determined by the particle size limitation on the aromatic cluster sizes, which elevates the effective band gap to $\gtrsim 0.7$\,eV and suppresses the long wavelength absorptivity (see paper III). Within molecular clouds small particle coagulation and a-C:H mantle accretion will lead to materials with $\beta \simeq 1.8-2.5$ (for $E_{\rm g} = 0.5-2.6$\,eV). However, as discussed above (Section \ref{sect_IR_mission_bands}) a-C:H materials are so much less emissive than a-C that their effects on the FIR-mm emissivity may be hard to observe.

\subsubsection{FIR-to-$A_V$ extinction ratio in the diffuse ISM}

%
\begin{figure} 
\resizebox{\hsize}{!}{\includegraphics{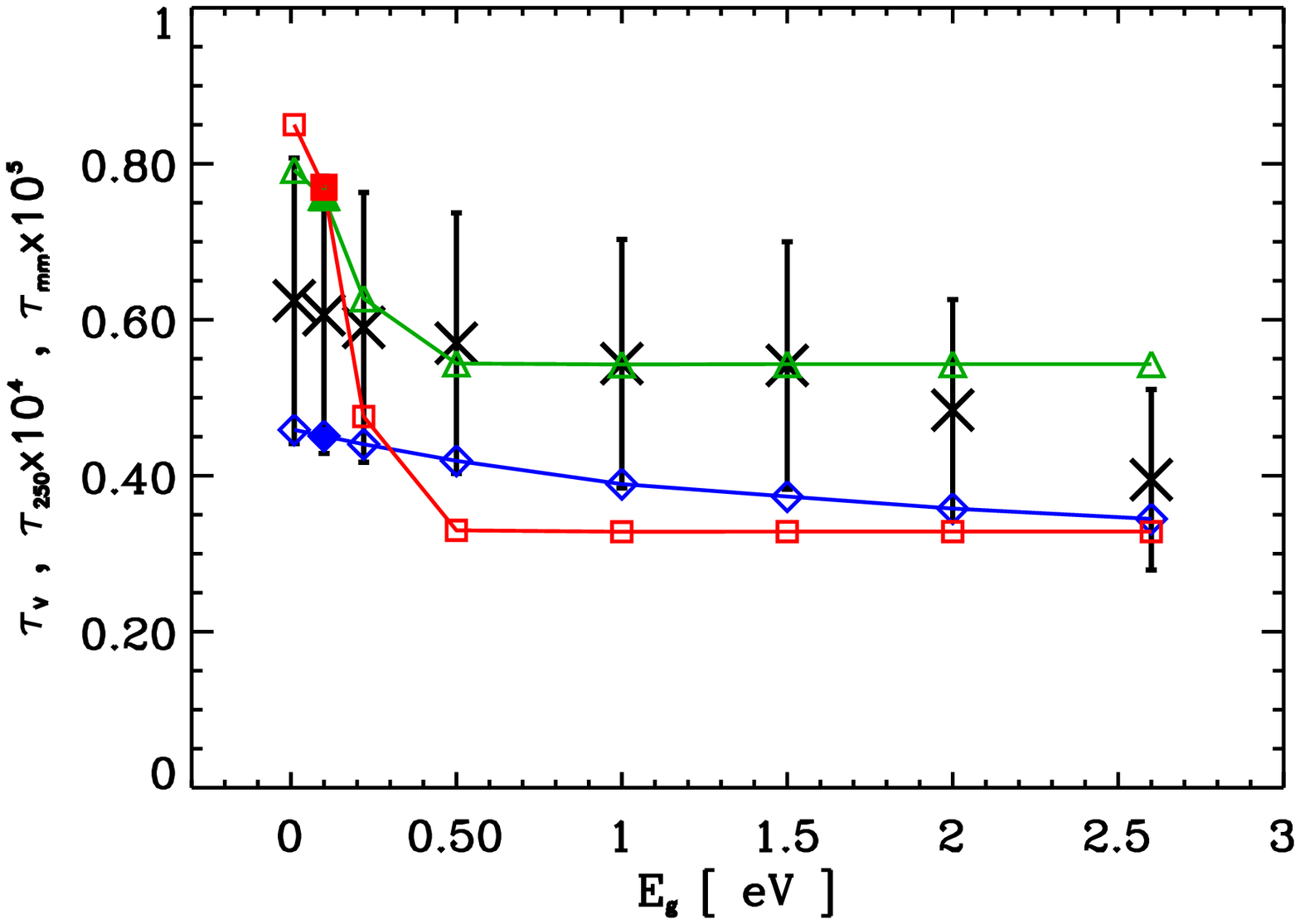}} 
 \vspace*{0.5cm}
\resizebox{\hsize}{!}{\includegraphics{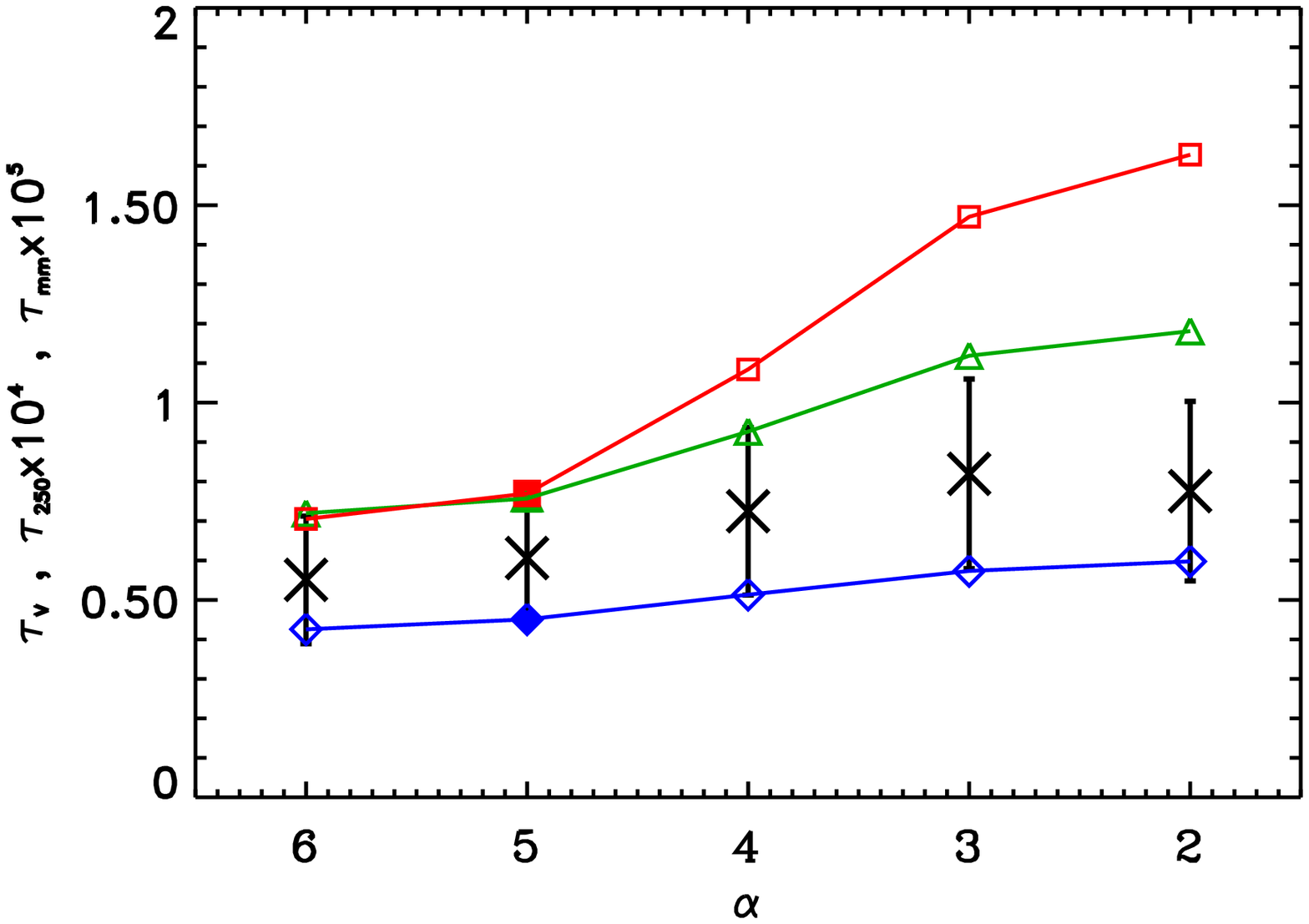}}     
  \caption{The dust model extinction, $\tau_i$, in the V band (blue) and at 250\,$\mu$m (green) and 1\,mm (red) wavelengths for an assumed column density $N_{\rm H} = 10^{21} $\,H\,cm$^{-2}$.   The upper plot shows the variations as a function of the a-C(:H) material band gap, $E_{\rm g}$, and the lower plot as a function of the power-law, $\alpha$, component of the small a-C dust population. The filled data points indicate the values for the best-fit DHGL dust model. The crosses with error bars show $\tau_{250} = 5.8\pm1.7 \times 10^{-4} \,E(B-V)$, from Eq. (\ref{eq_tau_FIR_UV}), multiplied by$10^4 $.}
 \label{fig_dust_results_4b}
\end{figure}

Here we consider the extinction in the diffuse ISM where the dust exhibits only limited variations in temperature and opacity. 

Fig.~\ref{fig_dust_results_4b} shows the model optical depths (opacity) for $N_{\rm H} = 10^{21}$\,H\,cm$^{-2}$ at the V band (blue), $250\,\mu$m (green) and 1\,mm (red) wavelengths as a function of the photo-processed material band gap, $E_{\rm g}$ (for grain radii $< 20$\,nm or 20\,nm thick mantles on larger grains, upper plot), and the small a-C grain power law, $\alpha$ (lower plot). This figure shows that changes in the optical depth are significant for small band gap materials ($E_{\rm g} < 0.5$\,eV) and are more apparent as the wavelength increases and as the power-law index, $\alpha$, decreases. 

Variations in the FIR opacity and in the $E(B-V)$ extinction must, at some level, be coupled because the dust optical properties at these disparate wavelengths vary in tandem ({\it e.g.}, see Fig. 16 in paper II and Figs. $10-13$ in paper III). As shown above,  variations in $E(B-V)$ and in the FIR opacity appear to be significant, while the FUV is practically invariant for fixed dust mass. In order to investigate the inter-dependence between the FIR emissivity and $E(B-V)$  we take the standard extinction-column density relationship from \cite{1978ApJ...224..132B}, $N_{\rm H} = 5.8 \times 10^{21}\,E(B-V)$\,cm$^{-2}$\,mag.$^{-1}$, and combine this with the recent results from the {\em Planck} mission. From a fitting of the FIR to sub-mm dust emissivity in the local diffuse interstellar medium the {\em Planck} data yield $\tau_{250} = 1.0\pm0.3 \times 10^{-25} \, N_{\rm H}$ \citep{2011A&A...536A..24P}. We then find 
\begin{equation}
\tau_{250} = 5.8\pm1.7 \times 10^{-4} \,E(B-V). 
\label{eq_tau_FIR_UV}
\end{equation}
This expression of $\tau_{250}$, calculated using our model results for $E(B-V)$ substituted into Eq.~(\ref{eq_tau_FIR_UV}) and multiplied by$10^4$, is shown in Fig.~\ref{fig_dust_results_4b}, as a function of $E_{\rm g}$ (upper) and $\alpha$ (lower), by the crosses with error bars. It is clear that the model and the recent Planck observational results are in good agreement and that variations in the FIR-mm opacity are rather limited when normalised in this way. 

The major conclusion here is that, for a given dust composition and size distribution, variations in the FIR opacity go hand-in-hand with variations in $E(B-V)$ because of the in-tandem evolution of the dust optical properties at visible and FIR wavelengths. 

\subsubsection{Some further points of note}

As noted above, it appears that many of the distinct dust observables are attributable to carbonaceous matter. 
Further, our use of the optEC$_{\rm (s)}$(a) data indicates that the carriers of the FUV extinction, UV bump, NIR absorption and the IR emission bands are only weakly coupled. 

We also note that there appears to be some degeneracy in the extinction, in that variations in $a_-$ and $\alpha$ often lead to similar extinction curves. For example, our investigations show that increasing $a_-$ from 0.4\,nm to 3\,nm gives essentially the same extinction curve as changing $\alpha$ from $5.0$ to $4.0$ (see Fig. \ref{fig_dust_results_2}). However, a look at the dust SEDs for the different size distributions show very different behaviour in the two cases, with a complete lack of IR emission bands in the case of increasing $a_-$ to 3\,nm  (see Fig. \ref{fig_dust_results_4}). Thus, a combined analysis of the observed extinction {\em and} emission, for a given line of sight, is required in order to constrain the exact dust composition and size distribution. In general, the dust SED is a more powerful constraint on the nature of dust in the ISM than the extinction curve.  

Table~\ref{table_Eg_predictions} summarises the consistency of the model results with the observational constraints. Clearly the un-normalised FUV extinction does not appear to be a strong constraint. However, this table clearly indicates that the IR extinction, and the associated absorption bands, arise from an a-C:H dust component while the bulk of the dust extinction and emission features arise from a-C dust.

\begin{table}
  \caption{Observation-model coherency check: $E_{\rm g}$ characterisation as a function of the $\lambda$-dependent extinction and emission features. The bullets indicate consistency between observation and model.}
\begin{center}
\begin{tabular}{lcccc}
\hline
\hline
                               &                            &                  &                        \\[-0.25cm]
property$\downarrow$ \hspace{0.5cm} $E_{\rm g}$   [ eV ] \ = &  $0-0.25$ & $0.25-0.5$ & $0.5-2.7$               \\
\hline
                               &                            &                  &                                    \\[-0.25cm] 
FUV extinction             & $\bullet$  & $\bullet$     & $\bullet$   \\
UV bump                    & $\bullet$ &               &                              \\ 
vis extinction               & $\bullet$ & $\bullet$ &                                   \\ 
E($\lambda$-V)E(B-V) & $\bullet$ & $\bullet$ &                              \\ 
NIR extinction  variations & $\bullet$ & $\bullet$ & $\bullet$   \\ 
3.4\,$\mu$m abs.        &                            &                  & $\bullet$   \\ 
IR emission bands      & $\bullet$ &                  &                            \\ 
MIR emission              & $\bullet$ &                  &                             \\ 
FIR emission (1.6)       & $\bullet$ &                  &                           \\ 
mm emission (1.6)       & $\bullet$ &                  &                             \\ 
FIR emission (1.8)       &                & $\bullet$ &                        \\ 
mm emission (1.8)       &               & $\bullet$ &                         \\ 
WMAP emission          & $\bullet$ &                  &                      \\ 
\hline
                               &                            &                  &               \\[-0.25cm]  
\end{tabular}     
\end{center}    
  \label{table_Eg_predictions}
\end{table}

\section{Interstellar dust evolution}
\label{sect_dust_variations}

\begin{figure}  
 \resizebox{\hsize}{!}{\includegraphics[angle=2700]{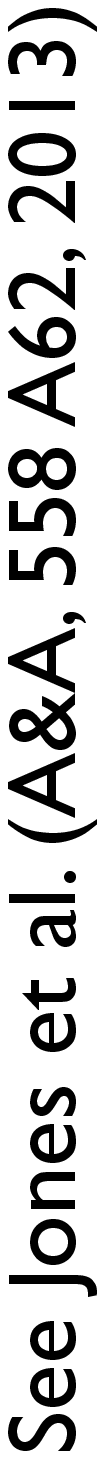}}  
  \caption{A schematic view of the life-cycle of interstellar dust in terms of evolutionary tracks:  
  `parallel-processing' tracks that lead to a smooth evolution are horizontal ({\it i.e.}, accretion and photo-processing) and `cross-talk' tracks that lead to significant mass-transfer between grains sizes ({\it i.e.}, coagulation and disaggregation) are diagonal and vertical. Qualitatively, the red shading reflects the density, $n_{\rm H}$, and the violet shading, on the extreme right, represents high $G_0$ radiation fields.}
 \label{fig_dust_lifecycle}
\end{figure}

Here we outline a scenario for dust evolution as it transits from the dust-forming shells around evolved (asymptotic giant branch, AGB) stars, into the diffuse ISM and then, via cloud collapse, on into the dense star-forming regions of the ISM where dust accretion/(re-)formation is important. It is in star-forming, photon-dominated environments ({\it e.g.}, H{\footnotesize II} regions and PDRs) that dust undergoes significant processing, which plays a key role in determining the course of its subsequent evolution. The constructive (destructive) processes inherent to dust evolution during its lifetime are accretion and coagulation (photo-processing and dis-aggregation/fragmentation).  

A schematic view of the dust evolutionary sequence, as a function of environment, is given in Fig.~\ref{fig_dust_lifecycle}. This figure introduces the idea of interstellar dust evolutionary tracks, which are very different for a-C(:H) and a-Sil$_{\rm Fe}$ grains because of their differing susceptibilities and reactions to processing. 
Observational evidence indicates that the evolutionary processes have a stronger effect on a-C(:H) dust than on amorphous silicate dust \citep[{\it e.g.},][]{2008A&A...492..127S,2011A&A...530A..44J,2012ApJ...760...36P}. 
Nevertheless, the evolutionary tracks are coupled because of dust component mixing, as indicated in Fig.~\ref{fig_dust_lifecycle} by the `parallel-processing' and `cross-talk' pathways. Seemingly, the most important evolutionary tracks for ISM dust are accretion and coagulation in an increasing density ISM (diffuse $\rightarrow$ dense molecular cloud phase change) and photon-driven fragmentation/destruction and disaggregation in  intense radiation field, star-forming environments ({\it i.e.}, H{\footnotesize II} regions and PDRs). 

Dust evolution evidently leads to observable variations in the dust composition, structure and size distribution. 
Here we discuss the major a-C(:H) evolutionary processes in the ISM and how these are consistent with, or can be constrained by, the observational evidence. 

In the following, and within the framework of the schematic dust life-cycle shown in Fig.~\ref{fig_dust_lifecycle}, we explore dust evolution under the assumption of constant dust mass within each dust component. Thus, coagulation and accretion affects are not rigorously-treated because they lead to a net dust mass increase and to mass transfer between the different components. Nevertheless, within the framework of a constant mass assumption and evolving size distributions, we can qualitatively explore the effects of evolution on the dust observables.

\subsection{a-C(:H) UV photo-processing and time-scale issues}
\label{sect_dust_proc_times}

The evolution of the a-C(:H) dust between the aliphatic-rich (a-C:H) and aromatic-rich (a-C) end members is accompanied by a narrowing of the band gap ($E_{\rm g} \gtrsim 2.3$\,eV $\rightarrow \ \simeq 0.1$\,eV) and by photo-fragmentation. This process occurs when a-C:H, accreted in the molecular ISM or formed around evolved stars, is exposed to the ambient radiation field. EUV-UV photon-induced processing (aromatisation) in the diffuse ISM ($G_0 = 1$), and particularly in PDRs ($G_0 \simeq 10-10^4$), on time-scales $\gtrsim 10^6$\,yr$/G_0$ leaves an imprint on the dust extinction and emission \citep{2012cA&A...542A..98J,2012eA&A...545C...3J}. The observable effects of photo-processing (top left and right hand side of Fig. \ref{fig_dust_lifecycle}) are to increase the bump strength and to steepen the NIR-visible extinction (Fig. \ref{fig_dust_results_2}, upper plot), significant modifications to the IR emission band profiles (Fig. \ref{fig_dust_results_5}) and to enhance the MIR-mm emission (Fig. \ref{fig_dust_results_4}, upper plot). 

Here we do not re-visit the time-scale-dependent aspects of photo-processing and photo-fragmentation but leave this until we have a better understanding of how they operate on a-C(:H) nano-particles in intense radiation fields. For example, it is not yet clear whether the aromatisation of a-C:H particles is a direct C$-$H bond photo-dissociation process or whether the transformation is driven by stochastic heating, or a combination of the two.\footnote{Particles with $ a \lesssim 10$\,nm, where the grain heating is predominantly stochastic, in the standard interstellar radiation field \citep[{\it e.g.},][]{1985ApJ...292..494D}, will undergo temperature excursions large enough to drive the dehydrogenation of a-C:H materials but the relevant time-scales are not yet determined. However, we have undertaken a preliminary study of a-C:H dust evolution in planetary nebul\ae\ within the context of fullerene formation \citep{2012ApJ...757...41B,2012ApJ...761...35M}.}  Dust models used in the interpretation of astronomical observations therefore need to be time-dependent in order to characterise the UV photon-induced transformation of a-C(:H) and must take into account its origin and prior history. This kind of study is probably best undertaken on a case-by-case basis, which compares model predictions with the astronomical observations \citep[{\it e.g.},][]{2008A&A...491..797C,2012A&A...541A..19A}. 

\subsection{Accretion}
\label{sect_dust_construct_accn}

Carbon accretion from the gas phase leads to the formation of a-C(:H) mantles on all grains, with material composition depending on the local density and extinction. For example, if the onset of accretion is in the diffuse ISM, where most species are atomic or singly-ionised ({\it e.g.}, Si$^+$, S$^+$, C$^+$), the extinction is low and UV photo-processing still a significant hazard, then a-C mantles will form. However, for intermediate levels of extinction, $A_{\rm V} < 3$, {\it i.e.},  before ice mantle formation, the mantles will be of a-C:H.\footnote{It is perhaps this accreted carbonaceous matter that could help to explain  ``coreshine'' phenomena \citep[][see also paper~II]{2010Sci...329.1622P,2010A&A...511A...9S}.} The onset of accretion is not restricted to C and H and other abundant gas phase atomic species, such as O and N, are also expected to accrete and to leave their observable signatures \citep{2013A&A...555A..39J}. In high extinction regions, {\it i.e.}, $A_{\rm V} > 3 \ (\gtrsim 6)$, the accreted carbon mantles will be submerged under H$_2$O (H$_2$O+CO) ice layers. 

The observable effects of accretion (the `parallel-processing' tracks in the middle of Fig. \ref{fig_dust_lifecycle}) are to increase all grain sizes (Fig. \ref{fig_dust_results_2}, middle plot), to increase the total dust mass and to  steepen the small grain mass distribution (Fig. \ref{fig_dust_results_2}, lower plot). However, small particles ($a < 10$\,nm) will undergo temperature excursions, resulting from stochastic photon absorption, which is likely to inhibit accretion onto the smallest particles in low extinction regions.

\subsection{Coagulation}
\label{sect_dust_construct_coag}

Accretion is a key process in dust evolution, however, it does not occur in isolation but pre-dates and aids in coagulating grains into mixed-composition aggregates in low-velocity grain-grain collisions. The early stages of coagulation, in a quiescent cloud filament in Taurus, were observed and modelled by \cite{2003A&A...398..551S}, where it was shown that the disappearance of small particles is consistent with reduced dust temperatures, enhanced emissivity and their coagulation into and onto larger particles. More recently \cite{2011A&A...528A..96K,2012A&A...548A..61K} found a factor of about 2.7 enhancement in the emissivity at FIR wavelengths, resulting from grain coagulation.  Further, this emissivity gain is rather rapidly achieved, {\it i.e.}, sticking only a few large grains together yields a doubling of the dust emissivity before saturation sets in; additional coagulation then only leads to incremental gains in emissivity. 

The observable effects of coagulation (`cross-talk' tracks in the middle of Fig. \ref{fig_dust_lifecycle}) are a mass re-distribution towards larger grains (smaller grains remain but decrease in abundance, {\it i.e.}, $a_-$ is fixed, see Fig. \ref{fig_dust_dist}). The effects of coagulation on the dust extinction and emission can be mimicked by progressively decreasing the power law index, $\alpha$, of the small a-C(:H) grain population from 5 to 2 as small grains are subsumed into the large grain population (Figs. \ref{fig_dust_results_2} and \ref{fig_dust_results_4}, lower plots). As can be seen the extinction in the FUV flattens, the UV bump broadens and the NIR-visible extinction flattens. In emission these changes manifest as a diminution of the IR emission bands and the MIR continuum accompanied by an increase in the FIR-mm emission. 
 
These evolutionary changes agree well with observations and their interpretation \cite[{\it e.g.},][]{2003A&A...398..551S,2011A&A...528A..96K,2012A&A...548A..61K} and are in general agreement with the evolution of the extinction curve with increasing $R_{\rm V}$ \citep[{\it e.g.},][and Fig.~\ref{fig_dust_results_2} above]{1990ARA&A..28...37M}.

\subsection{Destructive processing --- UV photo-fragmentation}
\label{sect_dust_destruct}

A bottom-up coagulation process appears to be unlikely in the diffuse ISM given the local conditions. In the dust coagulation kernel, $n_{\rm H} \, \sigma \, v_{\rm rel}$, the dust cross-section, $\sigma$, is `fixed' and the grain-grain collision velocities are low ($v_{\rm rel} < 1$\,km\,s$^{-1}$).  Thus, higher densities than those typical of the diffuse ISM are required for efficient coagulation in time-scales $\lesssim 3 \times 10^7$\,yr, the typical star formation time-scale in molecular clouds \cite[{\it e.g.},][]{1989IAUS..135..431M}.  

Dust evolution in the diffuse ISM could be more straight-forwardly be explained by a ``top-down''  scenario, where the major dust growth processes (accretion and coagulation) only occur in dense molecular clouds. The observed extinction and emission variations can then be explained by the de-construction of the accreted/coagulated particles, formed in the dense ISM, as they are subsequently (photo-)processed in H{\footnotesize II} regions and PDRs.  

Consistent with of a ``top-down'' dust evolution scenario is the observation that large $R_{\rm V}$ sight-lines, usually taken to be an indicator of prior dust coagulation, are found in PDRs such as the Orion Trapezium region \citep[{\it e.g.},][]{2007ApJ...663..320F}.   Here we are observing dense molecular cloud dust ablated from a parent cloud and exposed to the intense radiation field of a PDR.  Dust evolution here is driven by EUV-UV photo-processing, which drives off low-volatile mantles, triggers disaggregation and aromatises and photo-fragments a-C(:H) grains and mantles. The dust processing time-scale is generally longer than the local dynamical time-scale and so the dust will therefore not be in equilibrium with the local conditions. 

\begin{figure} 
\resizebox{\hsize}{!}{\includegraphics{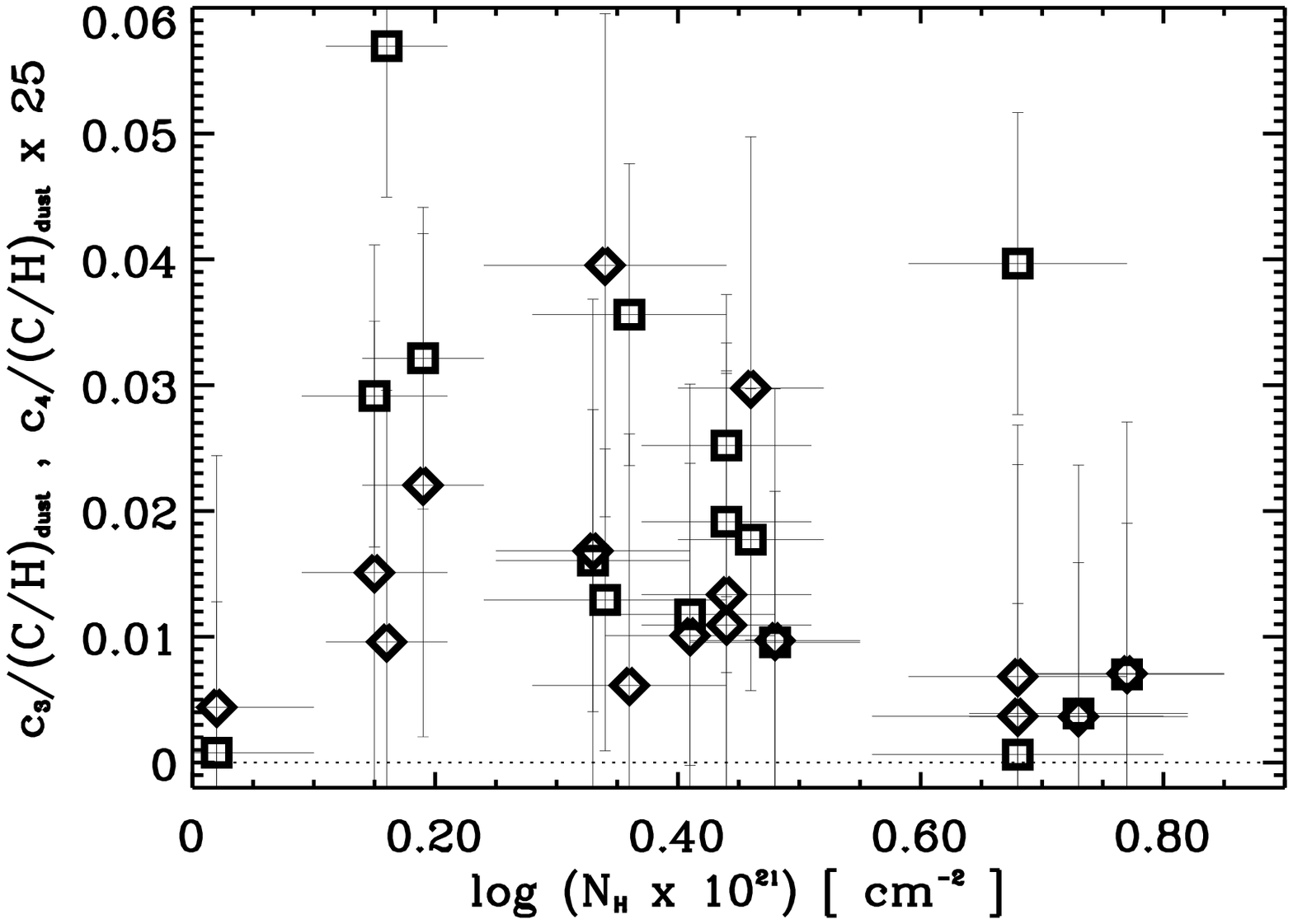}} 
  \caption{The UV bump parameter, c$_3$ (diamonds), and FUV extinction parameter, c$_4 \times 25$ (squares), intensity data normalised by the depletion into dust, (C/H)$_{\rm dust}$, as a function of $N_{\rm H}$. The $y$ value uncertainties are only illustrative. The data are taken from \cite{2012ApJ...760...36P}.}
 \label{fig_UVpropsNH}
\end{figure}

Recent work by \cite{2012ApJ...760...36P} supports the idea that carbonaceous dust undergoes significant  processing in the neutral ISM and shows that the intensity of the UV bump extinction does not correlate with carbon depletion into dust, (C/H)$_{\rm dust}$. This lack of correlation can naturally be explained by our model because (C/H)$_{\rm dust}$ is determined by accretion in dense cloud interiors ({i.e.}, by cloud volume), whereas the intensity of the bump and the FUV extinction reflect the abundance of UV photo-processed small grains in the low density surface regions of clouds. The UV bump and FUV extinction per unit dust mass should therefore decrease with cloud (line of sight) column density. In Fig. \ref{fig_UVpropsNH} we show the UV bump intensity parameter, c$_3$ (diamonds), and FUV extinction parameter, c$_4$ (squares), data from \cite{2012ApJ...760...36P}, normalised by their derived carbon depletion into dust, (C/H)$_{\rm dust}$, and plotted as a function of $N_{\rm H}$. The indicated trends in the UV extinction observations do seem to be consistent with this kind of cloud geometry effect.\footnote{The high FUV extinction line of sight with a weak UV bump, the square at [0.68,0.04], is the HD\,207198 line of sight with the highest (C/H)$_{\rm dust}$ value in the \cite{2012ApJ...760...36P} sample, {\it i.e.}, (C/H)$_{\rm dust} = 395 \pm 61$\,ppm. Strong FUV extinction associated with a weak UV bump is a characteristic of the wide band gap a-C:H materials expected to accrete in the denser regions of the diffuse ISM.} \cite{2012ApJ...760...36P} also note that the FUV extinction appears to show a gradual decrease with decreasing gas density, $n_{\rm H}$, which they interpret as due to the preferential fragmentation of small grains in the diffuse ISM. This observation seemingly lends support to the idea that small carbonaceous dust mass variations are driven by photo-fragmentation in low $n_{\rm H}$ regions. A study by \cite{2012A&A...542A..69P} finds that the photo-fragmentation of very small carbon grains into PAHs occurs in relatively dense, $n_{\rm H} = [100,10^5]$\,cm$^{-3}$, and UV irradiated PDRs, $G_0 = [100,5 \times 10^4]$, where H$_2$ rotational line emission is observed, {\it i.e.}, where $0.5 < G_0/n_{\rm H} < 1.0$. 
This could be observational evidence for H$_2$ formation via a-C(:H) grain photo-processing, as proposed by \cite{2012bA&A...540A...2J}. 

In PDRs EUV/UV photo-processing (a-C:H $\rightarrow$ a-C) progressively destroys the aliphatic component that binds the growing aromatic domains, eventually leading to photo-fragmentation and the release of daughter species \cite[{\it e.g.},][and as discussed in paper III]{2005A&A...435..885P}.\footnote{We note that carbon dust shattering in grain-grain collisions in shock waves results in fragments with a power-law mass distribution index of order $\sim 1.8-2.0$ \citep[{\it e.g.},][]{1996ApJ...469..740J,2008A&A...492..127S}, which is similar to the aromatic cluster mass distribution assumed in paper III and similar to that derived for astronomical PAHs by \cite{2002A&A...388..639P}.} 
Thus, the observable effects of photo-processing in PDRs arise from dust dis-aggregation, a-C:H grain aromatisation and photo-fragmentation, which (re-)generates a power law distribution of small a-C(:H) particles (`cross-talk' tracks on the right of Fig. \ref{fig_dust_lifecycle}), leading to a progressive  `steepening' of the UV extinction (Fig. \ref{fig_dust_results_2}, lower plot), increased MIR emission from nano-particles and decreased FIR-mm emission (Fig. \ref{fig_dust_results_4}, lower plots). This evolutionary scenario seems to be supported by Spitzer and ISO data for the Horsehead Nebula and NGC\,2023 North \citep{2008A&A...491..797C} and Herschel observations of the Orion Bar \citep{2012A&A...541A..19A}, which indicate lower relative abundances of the IR band emitters and the small grain MIR continuum carriers, by up to an order of magnitude compared to the diffuse ISM. 

In a future paper we plan to explore the effects of grain-grain collisional fragmentation and sputtering erosion on the SED in supernova-generated shocks.

\section{Concluding summary}
\label{sect_conclusions}

We propose a new two-material dust model, consisting of amorphous silicate and hydrocarbon dust, a-C(:H), with the  optical properties of the latter determined by the particle size and material band gap. This dust model consists of three dust components: small ($a \simeq 0.4$ to $\sim 100$\,nm) and large a-C(:H) grains, modelled using the optEC$_{\rm (s)}$(a) optical property data, and large carbon-coated, amorphous forsterite-type silicate grains with metallic iron nano-particle inclusions (a-Sil$_{\rm Fe}$). The large grains are assumed to have log-normal size distribution peaking at radii $\simeq 200$\,nm. 

Using the DustEM tool, to calculate the extinction and emission, we find that this model gives a  good fit to astronomical observations of the diffuse ISM extinction, emission and albedo, with only a minimal ``fine-tuning'' of the laboratory-derived data. The fit to these observational data, without the use of ``astronomical'' silicates, graphite or PAHs, is rather remarkable and illustrates the viability of the adopted optEC$_{\rm (s)}$(a) and a-Sil$_{\rm Fe}$ data.

The most innovative aspect of this dust model is that most of the observed extinction and emission, and their evolution, can quite naturally be explained by the size- and composition-dependent optical properties of a-C(:H) dust, with amorphous silicate dust playing only a subordinate role. We propose that the aliphatic-to-aromatic transformation ({\it i.e.}, wide band gap a-C:H $\rightarrow$ narrow band gap a-C) is accompanied by photo-fragmentation. The evolution of the large a-C(:H) grains formed around evolved stars and the mantles formed by accretion in the dense ISM is then driven EUV-UV photo-processing and photo-fragmentation in H{\footnotesize II} regions and PDRs. These processes are likely to be critical for understanding dust evolution in the ISM.  

Our understanding in interstellar dust modelling is currently limited by a lack of suitable, wide-wavelength coverage (EUV-cm) laboratory data for analogues of amorphous silicate and hydrocarbon dust.  
In particular, a major hindrance to the `holistic' dust modelling approach is the current lack of suitable optical property data for analogues of interstellar amorphous silicates, and also the means to validate the optEC$_{\rm (s)}$(a) data. 
A definitive quantitative analysis of astronomical observations must await these data and so some care should currently be exercised in order to avoid over-interpretation. 
Nevertheless, the fact that this model is able to explain so many FUV-mm dust observables, and their inter-correlations, is encouraging.  


\begin{acknowledgements} 
We thank the anonymous referee for a very careful reading of this manuscript and papers I-III, and for many useful and insightful remarks. 
We would like to thank our colleagues at the IAS and elsewhere for many interesting discussions on cosmic dust.
This research was, in part, made possible through the financial support of the Agence National de la Recherche (ANR) through the programs Cold Dust (ANR-07-BLAN-0364-01) and CIMMES (ANR-11-BS56-029-02).
\end{acknowledgements}

\noindent {\it Note added in proof} \\[0.1cm]
In section 4 we should also have made reference to the work of Viktor Zubko, Eli Dwek \& Richard G. Arendt (2004, ApJ, 152:211-249) who present a wide array of PAH, bare-grain and composite (but not core-mantle) dust models that are consistent with the extinction, emission and abundance constraints. In all of the dust models proposed by Zubko et al. (2004) the wavelength-dependence of the FUV extinction is due to the combined effects of  PAHs, carbonaceous/graphite and/or silicate dust components, which appears to be inconsistent with observations (Greenberg \& Chlewicki 1983).


\bibliographystyle{aa} 
\bibliography{biblio_HAC} 



\appendix

\section{Interstellar silicates}
\label{sect_silicate}

\begin{figure} 
 \resizebox{\hsize}{!}{\includegraphics{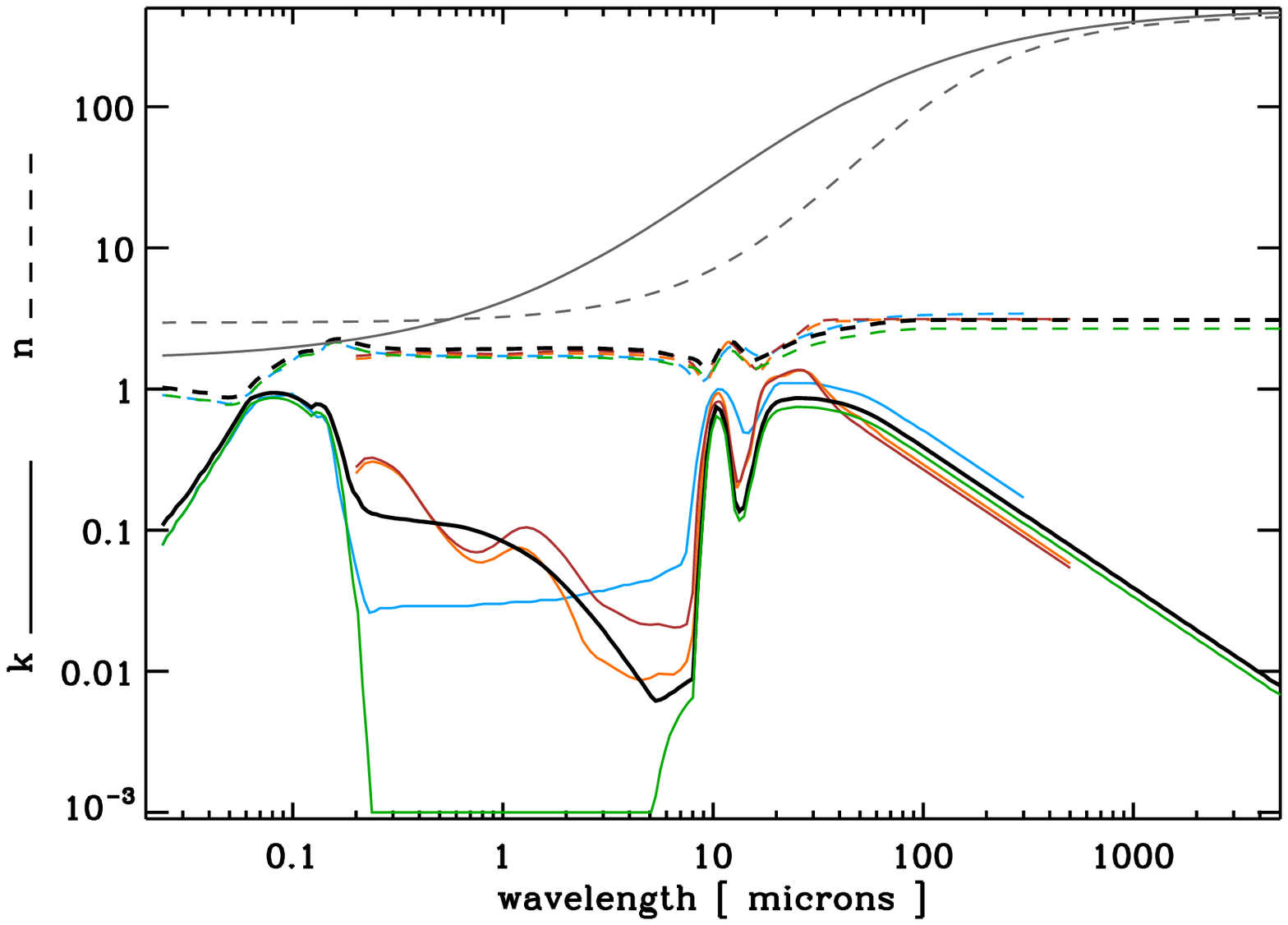}}
  \caption{The black lines show the imaginary (solid) and real (dashed) parts of the complex refractive index of our newly-constructed amorphous silicate/Fe nano-particle material, a-Sil$_{\rm Fe}$. This is a composite constructed using the amorphous forsterite-type silicate data (green) from \cite{1996ApJS..105..401S} and the metallic iron data (grey) from \cite{1983ApOpt..22.1099O,1985ApOpt..24.4493O,1988ApOpt..27.1203O}, see text for details. 
  For comparison we show the experimental data for  iron-containing amorphous olivine-type silicates, MgFeSiO$_4$ (orange) and Mg$_{0.8}$Fe$_{1.2}$SiO$_4$ (brown), from \cite{1995A&A...300..503D} and the astronomical data of \cite{1984ApJ...285...89D} (cobalt blue).}
 \label{fig_new_aSil}
\end{figure}

The optical properties of interstellar silicates have been modelled by empirical fits to observations \citep{1984ApJ...285...89D} and also using laboratory-constrained organic-refractory and amorphous silicate data for core-mantle particles \citep{1996A&A...309..258G}. Both approaches have been shown to give a good fit to the observations. However, these models have a problem in that they require essentially all of the available cosmic iron to be incorporated into the silicate but appear to add it for `free'. In that, they include its atomic fraction and mass but, apparently, neglect its important contribution to the imaginary part of the refractive index in the NIR region, {\it e.g.}, see Fig. 3 in \cite{1984ApJ...285...89D} and Fig.~1 in \cite{1996A&A...309..258G}. However, the inclusion of the cosmic abundance of iron into a silicate is far from `free' but leads to very significant absorption in the NIR \citep[{\it e.g.},][]{2011A&A...526A..68Z} regardless of whether the iron is within the silicate structure or present as metallic nano-particles ({\it e.g.},  see Fig.~\ref{fig_new_aSil}). If the effect of the neglected iron on the optical constants of \cite{1984ApJ...285...89D} and \cite{1996A&A...309..258G} were to be included then the predicted silicate dust temperatures would be hotter than indicated by observations because of the enhanced absorption in the NIR region, where the interstellar radiation field peaks. Thus, the amorphous astronomical silicate data derived by \cite{1984ApJ...285...89D} and \cite{1996A&A...309..258G} are not without significant problems. 

It is clear that, ideally, the recent laboratory data for amorphous silicates by \cite{2011A&A...535A.124C} should be used to constrain the temperature-dependence of its long-wavelength properties. However, the currently-available data does not have a wide enough wavelength coverage to be used in amorphous silicate dust extinction and emission models. 

For amorphous interstellar silicates, and until such time as suitable full wavelength-coverage optical property data are available for interstellar silicate analogue materials, we adopt an approach that allows for a range of silicate metal depletions. We first consider the available laboratory data for amorphous olivine-type silicates with iron incorporated into the silicate structure \cite[Mg:Fe = 1:1 or 1:1.5,][]{1995A&A...300..503D}. Secondly, following the work of \cite{2006A&A...448L...1D} on annealed silicates, where the iron is reduced to metal in the presence of carbon, we consider an amorphous forsterite-type silicate \citep{1996ApJS..105..401S} with $\sim 70$\% of the cosmic iron\footnote{Assuming a cosmic abundance of 32\,ppm for Fe, of which 22\,ppm is incorporated into a silicate matrix as metallic Fe nano-particles or into the silicate as cations. For Mg we require an abundance of 50\,ppm if the Fe is in metal or 67\,ppm if the Fe is present as cations in the silicate (see Section \ref{sect_dust_starter}).} is incorporated in the form of metallic iron \citep{1983ApOpt..22.1099O,1985ApOpt..24.4493O,1988ApOpt..27.1203O} nano-particle inclusions, equivalent to a metal ratio Mg:Fe = 2.3:1. The complex refractive index for the latter amorphous silicate/Fe mix, a-Sil$_{\rm Fe}$,  was calculated using the Maxwell-Garnett effective medium theory (EMT).\footnote{These refractive index data are available from the authors upon request.}
In Fig.~\ref{fig_new_aSil} we show our derived amorphous, Fe nanoparticle-containing, forsterite-type silicate complex refractive index data in comparison with other available silicate data. In the NIR wavelength region ($0.2-8.0\,\mu$m) the a-Sil$_{\rm Fe}$ data are close to those for the amorphous olivine-type silicate, (Mg,Fe)$_2$SiO$_4$, of \cite{1995A&A...300..503D}, as alluded to above, and allow for a range of silicate metal depletions. 

We note that adding $\sim 70$\% of the cosmic iron, as metallic nano-particles, into an amorphous silicate matrix  has about the same effect on the optical properties in the NIR as incorporating the iron into the silicate structure (Fig.~\ref{fig_new_aSil}). Hence, the presence of iron as metallic inclusions or as cations within the silicate structure leads to significant NIR absorption in the $0.3-3\,\mu$m region where the ISRF peaks. 

In conclusion, dust modelling is currently hampered by the lack of laboratory-measured optical constants for appropriate amorphous interstellar silicate analogue materials over a sufficiently wide wavelength range ({\it i.e.}, FUV to mm wavelengths). Further, our uncertain knowledge as to where, and in what solid form, the bulk of the cosmic iron is to be found in interstellar dust is also a severe limitation. 

\section{a-C(:H) material heat capacities}
\label{astro_fudge_Cp}

As proposed in paper~III we assume that the heat capacity of a-C(:H) materials is the linear sum of the abundance-weighted heat capacities of the constituent atoms in the given bonding configurations. We therefore calculate the particle heat capacities using Eq.~(A.3) from paper~III, {\it i.e.},
\[
\frac{C_{\rm V}}{N_{\rm C}} = \frac{1}{(1-X_{\rm H})} \ \times 
\] 
\begin{equation} 
\ \ \ \ \ \ \ \ \ \ \Bigg\{ X_{\rm H}^\prime C_{\rm V}({\rm CH}) \ + \ X_{sp3} \left[ \frac{C_{\rm V}({\rm CC})_{sp2}}{R} + C_{\rm V}({\rm CC})_{sp3} \right] \Bigg\},
\label{eq_Cv_4}
\end{equation}
which includes the  contributions from surface and interior H atoms, $X_{\rm H}^\prime$, $sp^2$ aromatic/olefinic and $sp^3$ aliphatic C atoms, the left, centre and right terms in the lower line of Eq.~(\ref{eq_Cv_4}), respectively. 
For the CH bond heat capacity, $C_{\rm V}({\rm CH})$, and for the aromatic/olefinic C atom heat capacity, $C_{\rm V}({\rm CC})_{sp2}$, we adopt the \cite{2001ApJ...551..807D} approach used to determine the heat capacities of free-flying PAHs in the ISM. However, we are interested in the aromatic domains within solid a-C(:H) and therefore use the number of carbon atoms per aromatic domain rather than the total number of C atoms in the particle in our determination of $C_{\rm V}({\rm CC})_{sp2}$.  For the $sp^3$ component heat capacity, $C_{\rm V}({\rm CC})_{sp3}$,  we adopt the values for polyethylene  \cite[][Fig.~10, p. 18]{polyethyleneCp}.

\section{``Astronomicalisation'' of the optEC$_{\rm(s)}$(a) data}
\label{astro_fudge}

\subsection{IR band profiles and the emission bands}
\label{astro_fudge_IR}

In order to better fit the observed astronomical IR emission and absorption band profile intensities and widths we needed to make only a few, rather minor, modifications to the previously-published optEC$_{\rm(s)}$(a) band profile data \citep{2012aA&A...540A...1J}. These changes are: the re-assignment of four aliphatic C-C bands to aromatic C$\simeq$C, and adjustments to the intensities (widths) of nine (six) bands. The new band profile intensity and width values are shown in boldface in Table~\ref{spectral_bands}; otherwise the data are the same as in Table~2 of paper~I. 

\begin{table*}
\caption{The adopted C-H and C-C band modes for eRCN and DG bulk materials: 
band centre ($\nu_0$), width ($\delta$) and integrated cross-section ($\sigma$).}
\begin{center}
\begin{tabular}{ccccll}
                       &              &               &               &                \\[-0.35cm]
\hline
\hline
                       &                       &              &                &               \\[-0.35cm]
                      &  $\nu_0$                            &  $\delta$           &  $\sigma$         &         Band       &             eRCN            \\
         no.        &  [ cm$^{-1}$ ( $\mu$m ) ]   &  [ cm$^{-1}$ ]   &   [ $\times 10^{-18}$ cm$^{2}$ ]   &        assignment         &   designation  \\[0.05cm]
\hline
                       &                      &            &           &                    \\[-0.2cm]
                       
 & \multicolumn{5}{l}{\underline{C-H stretching modes}}                                  \\[0.1cm]
       1    &    3078  ( 3.25 )      &    22.5    &    1.40   &    $sp^2$ CH$_2$  olefinic,  asy.     &    X$^2_{\rm CH_2}$                 \\ 
       2    &    3050  ( 3.28 )      &    53.1    &    1.50   &    $sp^2$ CH          aromatic            &    X$^2_{\rm CH_{\rm ar}}$       \\
       3    &    3010  ( 3.32 )      &    47.1    &    2.50   &    $sp^2$ CH          olefinic              &    X$^2_{\rm CH}$                      \\
       4    &    2985  ( 3.35 )      &    17.7    &    1.15   &    $sp^2$ CH$_2$  olefinic,  sym.    &    X$^2_{\rm CH_2}$                 \\
       5    &    2960  ( 3.38 )      &    29.3    &{\bf 4.00}&    $sp^3$ CH$_3$  aliphatic, asy.    &    $\phi$ X$^3_{\rm CH_3}$  \\
       6    &    2925  ( 3.42 )      &    28.9    &    3.30   &    $sp^3$ CH$_2$  aliphatic, asy.    &    X$^3_{\rm CH_2}$                    \\
       7    &    2900  ( 3.45 )      &    30.0    &    0.50   &    $sp^3$ CH$_2$  aliphatic            &    X$^3_{\rm CH}$                       \\
       8    &    2882  ( 3.47 )      &    30.2    &    1.00   &    $sp^3$ CH tertiary  aliphatic        &    X$^3_{\rm CH_2}$                   \\
       9    &    2871  ( 3.48 )      &    27.8    &    1.45   &    $sp^3$ CH$_3$  aliphatic, sym.    &    $\phi$ X$^3_{\rm CH_3}$      \\
     10    &    2850  ( 3.51 )      &    41.8    &    2.20   &    $sp^3$ CH$_2$  aliphatic, sym.    &    X$^3_{\rm CH_2}$                 \\[0.2cm]
                       
 & \multicolumn{5}{l}{\underline{C-H bending modes}}                                     \\[0.1cm]
      11    &    1470  ( 6.80 )    &    30.0    &    0.30     &    $sp^3$ CH$_3$  aliphatic, asy.             &    $\phi$ X$^3_{\rm CH_3}$    \\
      12    &    1450  ( 6.90 )    &{\bf 3.0}  &    1.20     &    $sp^3$ CH$_2$  aliphatic                     &    X$^3_{\rm CH_2}$               \\
      13    &    1430  ( 6.99 )    &{\bf 120.0}&{\bf 0.20}&    $sp^2$ CH          aromatic                    &     X$^2_{\rm CH_{\rm ar}}$     \\
      14    &    1410  ( 7.09 )    &    30.0    &    1.00     &    $sp^2$ CH$_2$  olefinic                       &    X$^2_{\rm CH_2}$               \\
      15    &    1400  ( 7.14 )    &    30.0    &    0.10     &    $sp^3$ (CH$_3$)$_3$ aliphatic, sym.   &     X$^3_{\rm (CH_3)_3}$       \\
      16    &    1370  ( 7.30 )    &    30.0    &    0.30     &    $sp^3$ CH$_3$   aliphatic, sym.           &   $\phi$ X$^3_{\rm CH_3}$     \\[0.2cm]
                       
 & \multicolumn{5}{l}{\underline{C-C modes}}                                                  \\[0.1cm]
      17    &    1600  ( 6.25 )    &{\bf 160.0}&{\bf 0.29} &    $sp^2$ C$\simeq$C   aromatic           &    X$_{\rm C\simeq C}$    \\
      18    &    1500  ( 6.67 )    &{\bf 4.0}    &    0.15    &    $sp^2$ C$\simeq$C   aromatic           &    X$_{\rm C\simeq C}$    \\
      19    &    1640  ( 6.10 )    &    40.0     &    0.10    &    $sp^2$ C=C   olefinic                          &    X$_{\rm C=C}$              \\[0.2cm]
                             
 & \multicolumn{5}{l}{\underline{additionally-assumed and estimated C-H modes}}  \\[0.1cm]
      20    &    3020  (   3.31 )    &    50.0  &    0.50      &    $sp^2$ CH   olefinic                          &    X$^2_{\rm CH}$                \\
      21    &      890  ( 11.24 )    &{\bf 8.0} &    0.50      &    $sp^2$ CH   aromatic                       &    X$^2_{\rm CH_{\rm ar}}$   \\
      22    &      880  ( 11.36 )    &    40.0  &    0.50      &    $sp^2$ CH   aromatic                       &    X$^2_{\rm CH_{\rm ar}}$   \\
      23    &      790  ( 12.66 )    &    50.0  &    0.50      &    $sp^2$ CH   aromatic                       &    X$^2_{\rm CH_{\rm ar}}$   \\[0.2cm]
                             
 & \multicolumn{5}{l}{\underline{additionally-assumed and estimated C-C modes}}  \\[0.1cm]
      24    &    1328  (  7.53  )    &{\bf 600.0}&    0.10      &    {\bf $sp^2$ C-C  aromatic}                &    {\bf X$^2_{\rm C\simeq C}$}    \\
      25    &    1300  (  7.69 )     &{\bf 720.0}& {\bf 0.05}  &    {\bf $sp^2$ C-C  aromatic}                &    {\bf X$^2_{\rm C\simeq C}$}    \\
      26    &    1274  (  7.85  )    &{\bf 600.0}& {\bf 0.05}  &    {\bf $sp^2$ C-C  aromatic}                &    {\bf X$^2_{\rm C\simeq C}$}    \\
      27    &    1163  (  8.60  )    &{\bf 270.0}& {\bf 0.05}   &    {\bf $sp^2$ C-C  aromatic}               &    {\bf X$^2_{\rm C\simeq C}$}    \\[0.2cm]
\hline
\hline
                       &                      &                      &                   &        &    \\[-0.35cm]
\end{tabular}
\end{center}
\label{spectral_bands}
\end{table*}

\subsection{The $\pi-\pi^\star$ band and the UV extinction bump}
\label{astro_fudge_UVbump}

We find that the optEC$_{\rm(s)}$(a) data gives a better fit to the observed 217\,nm UV bump when the single aromatic ring ($N_R = 1$), benzene-like clusters are not included in the aromatic cluster distribution summation (Section B.1.1 in paper~III). This was already hinted at in the discussion of the C$_6$ band in Appendix~B of paper~III. There the abundance scaling factor for $N_R = 1$ aromatic species was taken to be 0.08 rather than 1 and the astronomical observation requirements seemingly justify that supposition. Hence, and for particles with $a \leq 30$\,nm, the summation in the denominator of Eq.~(B.5) in paper~III should run from $N_R = 2$ (rather than $N_R = 1$) up to the largest aromatic cluster size, $N_R$(max), which is equivalent to setting the relative abundance for $N_R = 1$ to zero rather than the value of 0.08 asumed in paper~III.

As shown in paper~III, the {\it as-is} use of the optEC$_{\rm(s)}$(a) data predicts a UV bump that is too large by about a factor of two and it was concluded there that we do not yet have a complete understanding of the physics of the likely UV bump carriers. Here we attempt to, in part, fill this gap by adopting an empirical approach. We find that, for nanoparticles with $a < 3$\,nm, a narrowing of the contributing aromatic bands is required to get a reasonable fit to the UV bump width and adopt the following empirical relationship
\begin{equation}
\sigma =   \sigma_0 \times {\rm exp} \left( - \frac{0.3}{N_{\rm R}} \right) \ \ \ \ \ \ \ {\rm for} \ a \leqslant 3\,{\rm nm}, 
\label{eq_UVbump_narrowing}
\end{equation}
where $\sigma_0 = 0.650\, N_{\rm R}^{0.08}$, as per paper~III, and the exponential term gives the aromatic cluster size-dependence.

\subsection{The $\sigma-\sigma^\star$ band and the FUV extinction}
\label{astro_fudge_FUV}

As suggested in paper~III, the $\sigma-\sigma^\star$ band in a-C:H materials should exhibit size-dependent effect, as per the $\pi-\pi^\star$ band (see paper~III for details and also Section~\ref{astro_fudge_UVbump}) but, to date and as far as we are aware, no such effect has been measured for a-C(:H). 

\begin{figure} 
 \resizebox{\hsize}{!}{\includegraphics{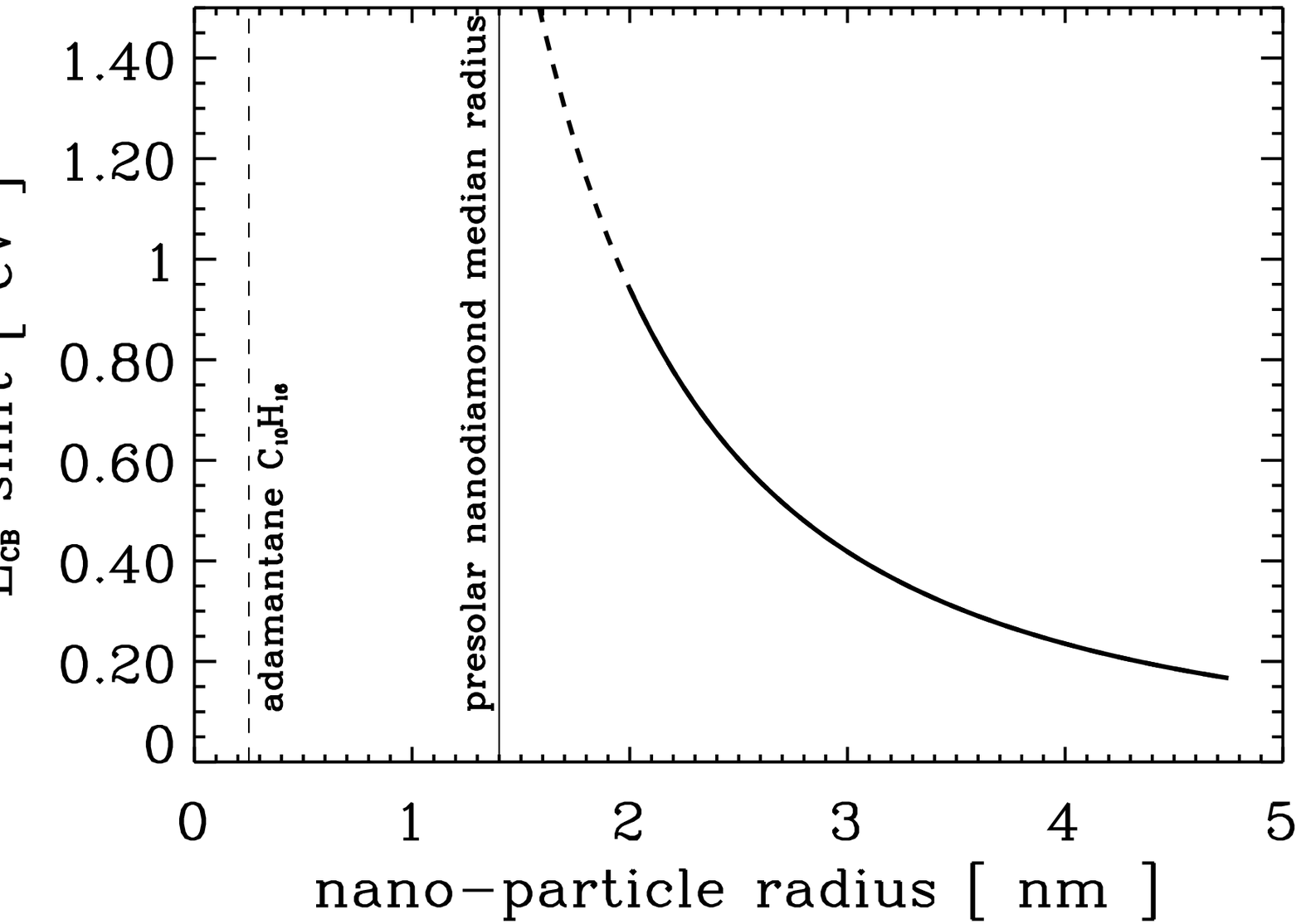}}
 \caption{Shift in the conduction band energy, $E_{\rm CB}$, as a function of radius for nanodiamonds, calculated using Eq.~\ref{eq_deltaECB}. The thick, solid portion of the curve indicates the behaviour supported by the experimental results of \cite{1999PhRvL..82.5377C} and the dashed portion is the extrapolation to smaller sizes using Eq.~\ref{eq_deltaECB}. For reference, the vertical solid line marks the mean radius of the meteoritic pre-solar nanodiamonds ($\sim 1.4$\,nm). The vertical dashed line marks the radius of the smallest diamond-type entity, the molecule adamantane, C$_{10}$H$_{16}$.}
 \label{fig_deltaECB}
\end{figure}

However, the observation of quantum-size effects that can significantly alter the electronic properties of semiconductor materials when they exist as nano-crystals or nanoparticles would have major implications for the optical properties of interstellar dust particles. For example, in diamond, the conduction band and valence band energies ($E_{\rm CB}$ and $E_{\rm VB}$) are, respectively, observed to shift to higher and lower energies as the particle size decreases. This leads to an increase in the band gap and a consequent shift in the absorption edge to shorter wavelengths as the particle size decreases. This effect was well-demonstrated for micro- and nanodiamond particles by \cite{1999PhRvL..82.5377C}. They showed that for diamond particles with radii $< 1.8$\,nm (with $< 4300$ C atoms) $E_{\rm CB}$ does not remain bulk-like. From the results of Chang et al. the shift in the conduction band energy, $\Delta E_{\rm CB}$, can be expressed as,
\begin{equation}
\Delta E_{\rm CB} = \frac{(\pi\, \hbar)^2}{2\, m^\ast\, a^2} 
\label{eq_deltaECB}
\end{equation}
where $m^\ast$ is the electron effective mass ($0.10 m_{\rm e}$, where $m_{\rm e}$ is the electron mass), $\hbar = h/2 \pi$ where $h$ is Planck's constant and $a$ is the particle radius. Fig.~\ref{fig_deltaECB} plots $\Delta E_{\rm CB}$ as a function of radius for nanodiamonds and indicates that the pre-solar, meteoritic nanodiamonds (mean radii $\sim 1.4$\,nm) will clearly have electronic properties that differ markedly from those of bulk diamond. In fact, nanodiamonds may have band gap energies ($E_{\rm CB} - E_{\rm VB}$) shifted by about 1\,eV from the bulk diamond value of 5.5\,eV (226\,nm, $4.4\,\mu$m$^{-1}$), to in excess of 6.4\,eV (190\,nm, $5.3\,\mu$m$^{-1}$). 

Given that the $\sigma-\sigma^\star$ band in a-C(:H) materials is determined by the carbon atom 2$s$ electrons, exactly as in diamond, we use the above approach to determine the size-dependent shift in the band gap. For a-C(:H) materials at nanoparticle sizes we follow \cite{1999PhRvL..82.5377C} and shift the a-C(:H) $\sigma-\sigma^\star$ band peak energy by 
\begin{equation}
E_{\sigma-\sigma^\star}(a) = E_{0,\sigma-\sigma^\star} + \Delta E_{\rm a-C(:H)} = E_{0,\sigma-\sigma^\star} + \frac{(\pi\, \hbar)^2}{2\, m^\ast\, a^2}, 
\label{eq_sigma_shift}
\end{equation}
where we here set $\Delta E_{\rm a-C(:H)} = \Delta E_{\rm CB}$. Recalling that, from paper~III, $E_{0,\sigma-\sigma^\star} = 13.0$\,eV, we limit the band peak position to a maximum of 14\,eV, {\it i.e.}, a maximum band shift of 1\,eV, coherent with that measured in diamond \citep{1999PhRvL..82.5377C}. 

In order to get a reasonable fit to the observed slope and intensity of the FUV extinction we also needed to increase the peak intensity of the intrinsic a-C(:H) $\sigma-\sigma^\star$ band, $I_{0,\sigma-\sigma^\star}$, but only for sub-10\,nm radius particles, {\it i.e.}, 
\begin{equation}
I_{\sigma-\sigma^\star} = I_{0,\sigma-\sigma^\star} \times \bigg\{ 1 + 0.06\, \left[ 10 - \left(\frac{a}{\rm 1\,nm}\right) \right] \bigg\} \ {\rm for} \ a \leqslant 10\,{\rm nm}, 
\label{eq_sigma_mult_factor}
\end{equation}
which corresponds to an increase by  a factor of $\simeq 1.5$, 1.4, 1.3 for $a = 1$, 3 and 5\,nm, respectively.

\subsection{Comparison of nm-sized a-C(:H) particles with PAHs}
\label{astro_fudge_cfPAH}

As per Section 4.1 of paper III, we compare the modified a-C(:H) data with those for astronomical PAHs containing the same number of carbon atoms. Even with the assumed band strength and width modifications, used to give a better match to the astronomical data, it is clear that the discrepancy between the optEC$_{\rm(s)}$(a) data for a-C(:H) and the astronomical PAHs, as pointed out in paper III, is still apparent. In the NIR regions ($\sim 1-5\,\mu$m) we note that laboratory data do indeed show enhanced extinction with respect to interstellar PAH models \cite{2005ApJ...629.1188M,2008ApJ...680.1243M}. Adding more longer wavelength bands, as per the \cite{2007ApJ...657..810D} model, could perhaps bring the optEC$_{\rm(s)}$(a) model data into closer correspondence with other models. However, the exact material/size origin of the bands is not yet completely clear and so this aspect has yet to be fully explored.  

In a recent critical analysis of the interstellar PAH hypothesis \cite{2013ApJ...771....5K} have shown that it is not without rather severe problems. As a response to this they proposed mixed organic and aliphatic nano-particles, MOANs, as a viable carrier of the observed IR emission bands and underlying continua 
\citep{2011Natur.479...80K,2013ApJ...771....5K}. These MOANs are, in essence, part of the same family of materials as a-C(:H). However, the MOANs appear to be much more open and less cross-linked structures than a-C(:H), and require the presence of O, N and S in the structure \citep{2013ApJ...771....5K}, which is not supported by interstellar depletion studies. 

\begin{figure} 
 \resizebox{\hsize}{!}{\includegraphics{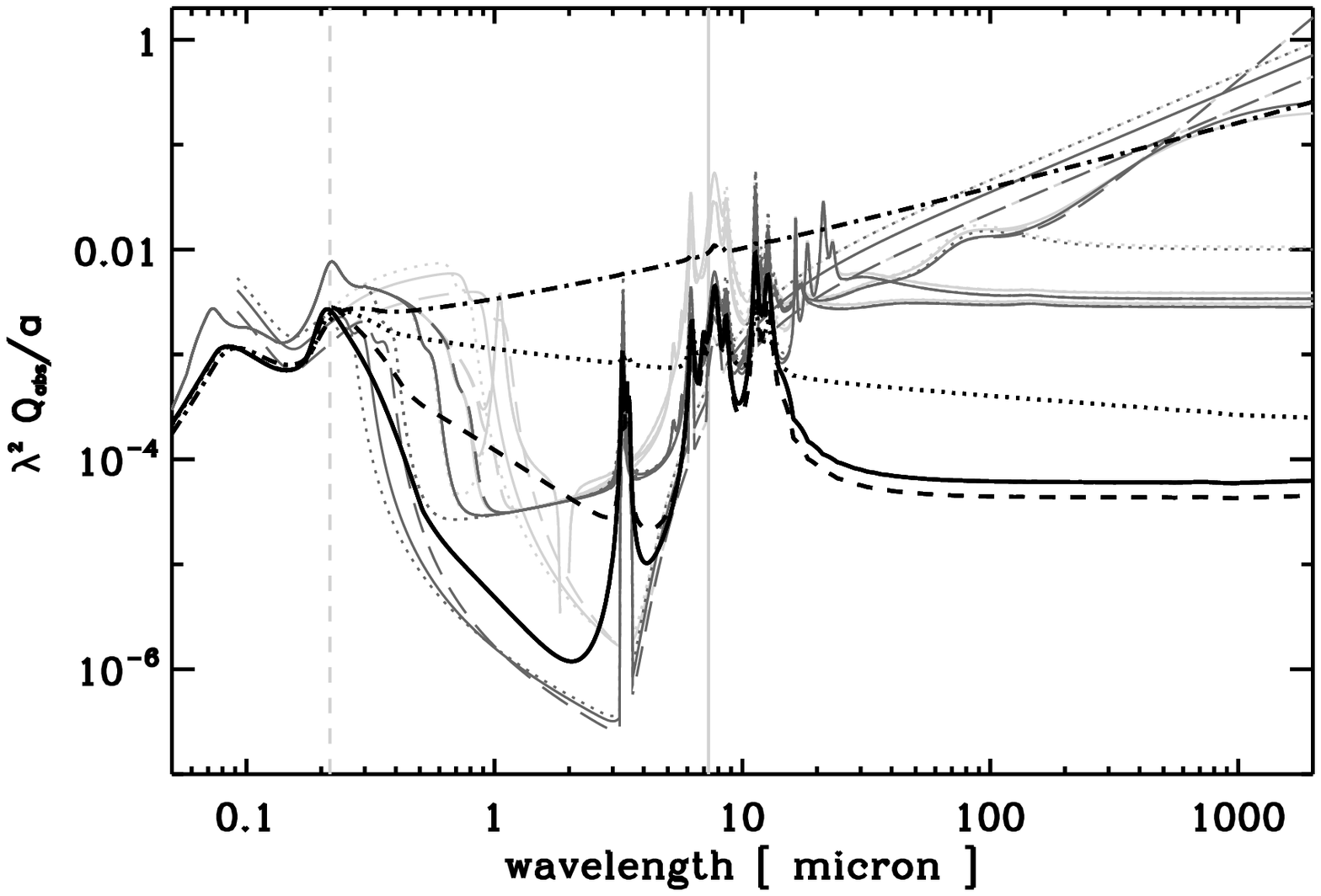}}
 \caption{$\lambda^2$Q$_{\rm abs}$/a for a-C(:H) particles with $E_{\rm g}{\rm (bulk)} = 0.1$\,eV (black lines) and of radii 0.33\,nm (solid), 0.5\,nm (dashed), 1\,nm (dotted) and 3\,nm (dash-dotted).  The data for neutral and cation PAHs, with the same number of C atoms \citep[dark and light grey solid lines, respectively,][]{1990A&A...237..215D,2001ApJ...551..807D,2007ApJ...657..810D,2011A&A...525A.103C} are shown for comparison. 
The dashed, vertical, grey line shows the peak position of the UV bump and the solid, vertical, grey line shows the upper wavelength limit for the well-constrained a-C(:H) IR features.} 
 \label{fig_cf_aCH_PAH}
\end{figure}

\listofobjects

\end{document}